\documentclass[a4paper,11pt]{article}

\usepackage[a4paper,left=2.73cm,right=2.7cm,top=3cm,bottom=3.5cm]{geometry}

\usepackage{amsfonts,amsmath,amssymb,amsthm,tabstackengine}
\usepackage[usenames, dvipsnames]{color}
\usepackage{array}
\usepackage{multirow}
\usepackage{makecell}
\allowdisplaybreaks

\usepackage{color, colortbl,xcolor}
\definecolor{mygray}{gray}{0.90}

\usepackage[colorlinks=true,linktocpage=true,linkcolor=red,citecolor=blue]{hyperref}
\usepackage{graphicx}

\usepackage{comment}

\numberwithin{equation}{section}

\newcommand{\be}{\begin{eqnarray}}
\newcommand{\ee}{\end{eqnarray}}

\begin{document}

\begin{titlepage}

\thispagestyle{empty}

\begin{center}

{\LARGE \textbf{Type~II orientifold flux vacua in 3D}}

\vspace{40pt}
		
{\large  \bf{\'Alvaro Arboleya}}, \,  {\large  \bf{Adolfo Guarino}} \large{ and } {\large  \bf{Matteo Morittu}}
		
\vspace{25pt}

{\normalsize 
Departamento de F\'isica, Universidad de Oviedo,\\
Avda. Federico Garc\'ia Lorca 18, 33007 Oviedo, Spain.}
\\[7mm]

{\normalsize  
Instituto Universitario de Ciencias y Tecnolog\'ias Espaciales de Asturias (ICTEA) \\
Calle de la Independencia 13, 33004 Oviedo, Spain.}
\\[10mm]

\texttt{alvaroarbo@gmail.com}  ,  \, 
\texttt{adolfo.guarino@uniovi.es} ,  \,
\texttt{morittumatteo@uniovi.es}

\vspace{20pt}

\vspace{20pt}
				
\abstract{We initiate a systematic study of type~II orientifold flux vacua in three dimensions including gauge and metric fluxes, O-planes and D-branes. We derive simple flux models (we dub them RSTU-models) that admit a description in terms of gauged supergravities with half-maximal $\,\mathcal{N}=8\,$ supersymmetry in three dimensions. As a landscape appetizer, we present various multi-parametric families of supersymmetric and non-supersymmetric AdS$_{3}$ and Mkw$_{3}$ vacua. Remarkably, negative masses turn out to be always absent in the spectrum of scalar fluctuations, thus making all the vacua perturbatively stable within half-maximal supergravity. We provide examples of non-supersymmetric type~IIB AdS$_{3}$ flux vacua which feature parametrically-controlled scale separation and come along with integer-valued conformal dimensions of the would-be dual CFT$_{2}$ operators. We also comment on the implications of our results in light of the Swampland Program.
}

\end{center}

\end{titlepage}

\tableofcontents

\hrulefill
\vspace{10pt}

\baselineskip 4.55mm

\section{Motivation}

The study of flux compactifications has been and continues being one of the most active research areas in string phenomenology. While most of the effort has been devoted to derive semi-realistic models of particle physics and cosmology in four dimensions (4D), the advent of the so-called Swampland Program \cite{Palti:2019pca} has broadened the interest to other dimensions. As stated in the abstract of \cite{Palti:2019pca}: 

``\textit{The Swampland Program aims to distinguish effective theories which can be completed into quantum gravity in the ultraviolet from those which cannot.}'' \\[-2mm] 

\noindent Since it is not an exclusive program in 4D, we will rather investigate simpler three-dimensional effective theories where some of the conjectures that have been stated in the literature might be easier to test. This motivates us to work within the context of supersymmetric gravity, \textit{i.e.} supergravity, in three dimensions (3D). Supergravity theories in three dimensions are special in that the dreibein and the gravitino field(s) describe non-propagating degrees of freedom. All the massless propagating degrees of freedom are either spin-$0$ scalars or spin-$\frac{1}{2}$ fermions. Still, it is convenient to formulate three-dimensional supergravities in terms of a dreibein and the gravitino fields in order to present the theories in a manifestly covariant form \cite{Marcus:1983hb}.

Flux compactifications down to three dimensions have been considered in the literature, nonetheless, in much less detail than their four-dimensional counterparts (see \cite{Grana:2005jc} for a comprehensive review). Paralleling previous constructions in 4D, examples of $\,\mathcal{N}=1\,$ (two real supercharges) flux models in 3D have been put forward in the context of type~IIA \cite{Farakos:2020phe} and type~IIB \cite{Emelin:2021gzx} orientifold reductions on seven-dimensional internal spaces with $\textrm{G}_{2}$-holonomy and co-calibrated $\textrm{G}_{2}$-structure, respectively. A study of the structure of three-dimensional Minkowski (Mkw$_{3}$) and anti-de Sitter (AdS$_{3}$) flux vacua was also initiated in \cite{Farakos:2020phe,Emelin:2021gzx}. Although an exhaustive classification of vacua was not performed in those references, various general expectations were drawn regarding, for example, the existence of scale-separated AdS$_{3}$ vacua. In the type~IIA models of \cite{Farakos:2020phe} (see also \cite{Farakos:2023nms,Farakos:2023wps}), the AdS$_{3}$ scale could be parametrically decoupled from the Kaluza--Klein (KK) scale while keeping the solution classical and at weak coupling. However, as originally pointed out in \cite{Emelin:2021gzx} and later on reiterated in \cite{VanHemelryck:2022ynr,Apers:2022vfp}, the type IIB AdS$_{3}$ vacua of \cite{Emelin:2021gzx} do not seem to allow for parametrically-controlled scale separation. Another interesting comment regarding the type~IIA AdS$_{3}$ vacua of \cite{Farakos:2020phe} was made in \cite{Apers:2022zjx}. The conformal dimensions of the single-trace operators in the dual CFT$_{2}$'s turn out to be non-integer valued, unlike what was observed in \cite{Conlon:2021cjk,Apers:2022tfm} for the operators in the CFT$_{3}$'s dual to the DGKT AdS$_{4}$ vacua \cite{DeWolfe:2005uu}. In \cite{Apers:2022vfp}, the same non-integer nature was argued to hold also for the conformal dimensions of operators in the CFT$_{2}$'s dual to the type IIB AdS$_{3}$ vacua of \cite{Emelin:2021gzx}. From the results in \cite{Emelin:2021gzx,VanHemelryck:2022ynr,Apers:2022vfp}, type IIB AdS$_{3}$ flux vacua are not expected to be compatible with parametrically-controlled scale separation or to come along with integer-valued conformal dimensions of the would-be dual CFT$_{2}$ operators. Using some simple type~IIB orientifold flux models with a large amount ($\mathcal{N}=8$) of supersymmetry in combination with powerful supergravity and algebraic geometry techniques, we will present potential counterexamples to these expectations.

The small amount of supersymmetry of the $\mathcal{N}=1$ type~II orientifold models in  \cite{Farakos:2020phe,Emelin:2021gzx} allows for intricate systems of intersecting O$p$-planes and D$p$-branes in the compactification scheme. While this enriches the structure of extrema of the scalar potential in the effective 3D supergravity arising upon compactification, it also makes a systematic study of such extrema unfeasible due to the complexity of the extremisation conditions. A systematic and analytic study of AdS$_{3}$ vacua within the context of 3D supergravity with a larger $\,\mathcal{N}=8\,$ supersymmetry (sixteen real supercharges or, equivalently, half-maximal) has been carried out in \cite{Deger:2019tem}. However, and despite the large amount of supersymmetry, a complete classification of extrema of the scalar potential of $\,\mathcal{N}=8\,$ supergravity remains out of computational reach. Still, employing the method put forward in \cite{Dibitetto:2011gm} which exploits the homogeneity of the scalar geometry in half-maximal supergravity, ref.~\cite{Deger:2019tem} succeeded in finding all the AdS$_{3}$ vacua preserving $\,\mathcal{N}=8\,$ supersymmetry. These are AdS$_{3}$ vacua preserving all the supersymmetries of the supergravity theories where they live. However, the string-theoretic embedding (if any) of these 3D supergravities remains generically unknown.

In this work we will combine the two approaches discussed above. We will first identify which $\,\mathcal{N}=8\,$ (half-maximal) supergravities in three dimensions arise from type~II orientifold reductions with background gauge and metric fluxes, as well as spacetime filling O$p$-planes and D$p$-branes. Demanding half-maximal supersymmetry will translate into having a \textit{single type} of coincident O$p$/D$p$ sources, \textit{e.g.}, type~IIA with O$2$/D$2$ or type~IIB with O$5$/D$5$. The formulation of the type~II flux models within the embedding tensor formalism of \cite{Nicolai:2001ac,deWit:2003ja} (see also \cite{Deger:2019tem}) will allow us to carry out a systematic study of vacua using the method of \cite{Dibitetto:2011gm}, to compute the complete mass spectrum within half-maximal supergravity at the various vacua, and also to test some conjectures within the Swampland Program. For instance, we will provide the first examples of parametrically-controlled and scale-separated type IIB AdS$_{3}$ flux vacua which are non-supersymmetric and perturbatively stable within half-maximal supergravity, and that come along with integer-valued conformal dimensions of the would-be dual CFT$_{2}$ operators. Our systematic study of type II orientifold flux vacua in $D=3$ extends/complements previous ones carried out in $D\, \ge 4\,$: see \textit{e.g.} \cite{Dibitetto:2011gm} for $D=4$, \cite{Dibitetto:2019odu} for $D=6$, and \cite{Dibitetto:2012rk,Dibitetto:2015bia} for $D \ge 7$.

The paper is organised as follows. In Section~\ref{sec:N=8_sugra} we describe half-maximal gauged supergravity theories in three dimensions. We present their field content; we introduce the embedding tensor formalism and so the gauging procedure; we explicitly display the bosonic part of the complete Lagrangian and conveniently express its scalar potential in terms of the fermionic mass matrices. In Section~\ref{sec:Type_II_orientifolds} we establish a precise correspondence between single O$p$-plane orientifold reductions of type~II strings in the presence of background fluxes and 3D half-maximal supergravities. In particular, relying on group-theoretical arguments we build a dictionary identifying M/string-theoretic and 3D gauged supergravity quantities, \textit{e.g.} fluxes and embedding tensor components. Equipped with such a well-posed correspondence, in Section~\ref{sec:flux_models&vacua} we prepare the ground for a systematic exploration of the landscape of half-maximal gauged supergravities in three dimensions. More specifically, we discuss two interesting invariant sectors of the full $\,\mathcal{N}=8\,$ 3D theory: a $\mathbb{Z}_2 \times \mathbb{Z}_2$ invariant sector and an SO(3) invariant sector. After connecting the former with the type~IIA with O$2$/O$6$-plane models of \cite{Farakos:2020phe,Farakos:2023nms,Farakos:2023wps}, we study the SO(3) invariant sector (we dub it RSTU-model) for the type~IIB with an O$5$-plane duality frame. We investigate its vacuum structure and comment on the properties of the (surprisingly) rich structure of flux vacua that we find. In Section~\ref{sec:Conclusions} we draw our conclusions and discuss future research lines in light of the Swampland Program. A couple of technical appendices are also included.

\section{A crash course on half-maximal 3D supergravity}
\label{sec:N=8_sugra}

The study of all possible half-maximal ($\,\mathcal{N}=8\,$) supergravities in three dimensions (3D) was initiated in \cite{Nicolai:2001ac,deWit:2003ja} and more recently completed in \cite{Deger:2019tem}. Starting from the ungauged theory of \cite{Marcus:1983hb}, which features an $\textrm{SO}(8,n)$ global symmetry, the most general $\,\mathcal{N}=8\,$ supergravity in 3D was constructed by applying the so-called gauging procedure. Upon this procedure, a subgroup $\textrm{G} \subset \textrm{SO}(8,n)$ is promoted from global to local. The theory then becomes a gauged supergravity, and a scalar potential and scalar-dependent mass terms for the various fermionic fields are generated by the gauging. The field content of the half-maximal gauged supergravities consists, first of all, of the $\,\mathcal{N}=8\,$ supergravity multiplet which is composed of the driebein $e_{\mu}{}^a$ and eight gravitini $\psi_{\mu}{}^{\cal A}$ with ${\mathcal{A} = 1,\ldots,8}$. The latter transform in the spinorial representation of the R-symmetry group $\textrm{SO}(8)_{\textrm{R}}$. In addition, the theory can also include $n$ matter multiplets. Each matter multiplet is composed of eight real scalars $\phi^{\mathcal{I}}$ (spin-$0$) and eight two-component Majorana fermions $\chi^{\dot{\mathcal{A}}}$ (spin-$\frac{1}{2}$) with $\mathcal{I}=1,\ldots,8$ and $\dot{\mathcal{A}}=1,\ldots,8$ being associated with the vectorial and conjugate-spinorial representations of $\textrm{SO}(8)_{\textrm{R}}$. In this work we will focus on the case $n=8$. This is the largest value for which the duality group of three-dimensional half-maximal supergravity can be embedded into the one of maximal ($\,\mathcal{N}=16\,$) supergravity, \textit{i.e.} $\textrm{SO}(8,8) \subset \textrm{E}_{8(8)}$. In this case the scalars in the theory serve as coordinates in the coset space
\begin{equation}
\label{scalar_geometry_N=8}
{\mathcal{M}}_{\textrm{scal}} = \frac{{\rm SO}(8,8)}{{\rm SO}(8) \times {\rm SO}(8)}  \ ,
\end{equation}
which can be viewed as a maximal subspace of the scalar geometry $\textrm{E}_{8(8)}/\textrm{SO}(16)$ of the maximal theory \cite{Marcus:1983hb}: the scalar geometry in (\ref{scalar_geometry_N=8}) comprises $64$ scalars which are half of the $128$ scalars in the $\textrm{E}_{8(8)}/\textrm{SO}(16)$ scalar geometry of the maximal theory.

The complete Lagrangian of the $\,\mathcal{N}=8\,$ gauged supergravities in 3D can be found in \cite{Deger:2019tem}. (We closely follow the notation and conventions therein). Its bosonic part, which will be the relevant one for this work, is\footnote{The term $- \frac{1}{32} \, e \, D_\mu M_{MN} \, D^{\mu} M^{MN}$ in \eqref{L_bosonic} precisely corresponds to the contribution $+ \frac{1}{4} \, e \, P_{\mu}^{Ir} P^{\mu \, Ir}$ in eq.~(2.12) of \cite{Deger:2019tem}.}
\begin{equation}
\label{L_bosonic}
\mathcal{L}_{\textrm{bos}} = - \frac{1}{4} \, e \,  R - \frac{1}{32} \, e \, D_\mu M_{MN} \, D^{\mu} M^{MN} - e \, V + \mathcal{L}_{\textrm{CS}} \ .
\end{equation}
The first term in (\ref{L_bosonic}) is the standard Einstein--Hilbert term for the driebein field. The second term is a kinetic term for the scalar fields in the theory. Using the standard coset construction, the scalar-dependent matrix $M_{MN}$ entering (\ref{L_bosonic}) is obtained from a coset representative
\begin{equation}
\label{scalar_coset_N=8}
\mathcal{V}_{M}{}^{\underline{P}}(\phi) \in \frac{\textrm{SO}(8,8)}{\textrm{SO}(8) \times \textrm{SO}(8)}  
\end{equation}
as $M_{MN} = \mathcal{V}_{M}{}^{\underline{P}} \,\, \mathcal{V}_{N}{}^{\underline{Q}} \,\, \delta_{\underline{PQ}}$, where $M$ and $\underline{M}$ denote fundamental (vectorial) indices of $\textrm{SO}(8,8)$ and $\textrm{SO}(8) \times \textrm{SO}(8)$, respectively.\footnote{In our conventions, the coset representative $\mathcal{V}_{M}{}^{\underline{N}}(\phi)$ transforms under global $\textrm{SO}(8,8)$ transformations from the left and under local $\textrm{SO}(8) \times \textrm{SO}(8)$ transformations from the right (underlined index).} The third term in (\ref{L_bosonic}) is a scalar potential whose explicit expression will be given in a moment. Finally, the last term in (\ref{L_bosonic}) denotes a topological term for the vector fields $A_{\mu}{}^{MN}=A_{\mu}{}^{[MN]}$ in the theory which formally transform in the $\bf{120}$ (adjoint) irreducible representation (irrep) of $\textrm{SO}(8,8)$. Note that in three dimensions vectors are dual to scalars and therefore do not carry an independent dynamics (see \cite{Deger:2019tem} for more details).

\subsection{Gaugings, embedding tensor and quadratic constraints}

The gaugings of half-maximal supergravity are encoded in the so-called embedding tensor
\begin{equation}
\label{ET_N=8}
\Theta_{MN|PQ} = \theta_{MNPQ} + 2 \left( \eta_{M[P} \theta_{Q]N} - \eta_{N[P} \theta_{Q]M} \right) + 2 \eta_{M[P} \eta_{Q]N} \theta \ ,
\end{equation}
which consists of three irreducible representations (irreps) of the duality group $\textrm{SO}(8,8)$: $\theta_{MNPQ}=\theta_{[MNPQ]} \in \bf{1820}$,  $\theta_{MN}=\theta_{(MN)} \in \bf{135}$ (with $\theta_{M}{}^M=0$) and $\theta \in \bf{1}$. The $\eta_{MN}$ matrix in (\ref{ET_N=8}) is the non-degenerate $\textrm{SO}(8,8)$ invariant matrix that is used to raise/lower vectorial indices. In this work we will alternate between two different basis for $\textrm{SO}(8,8)$: the light-cone (LC) basis and the Lorentzian (Ltz) basis. The $\,\eta_{MN}\,$ matrix is
\begin{equation}
\label{eta_matrices}
\eta_{MN} \Big|_{\textrm{LC}} = \left( \begin{matrix}
    0 & \mathbb{I} \\
    \mathbb{I} & 0
\end{matrix} \right) 
\hspace{10mm} \textrm{ and } \hspace{10mm}
\eta_{MN}\Big|_{\textrm{Ltz}} = \left( \begin{matrix}
    -\mathbb{I} & 0 \\
    0 & \mathbb{I}
\end{matrix} \right) \ ,
\end{equation}
in the two different basis. While the LC basis will be relevant to establish the dictionary between the embedding tensor and the various type~II flux parameters, the Lorentzian basis will be more adequate to compute the fermionic mass terms and, from them, the scalar potential in (\ref{L_bosonic}). Both basis are related by the $\textrm{SO}(16)$ rotation
\begin{equation}
\label{LV_vs_Ltz}
\eta_{MN}\Big|_{\textrm{Ltz (LC)}} = R_{M}{}^{P} \, R_{N}{}^{Q}  \,\,  \eta_{PQ}\Big|_{\textrm{LC (Ltz)}}
\hspace{10mm} \textrm{ with } \hspace{10mm}
R_{M}{}^{N} = \frac{1}{\sqrt{2}} 
\left( \begin{matrix}
    -\mathbb{I} & \mathbb{I} \\
    \mathbb{I} & \mathbb{I}
\end{matrix} \right) \ .
\end{equation}

The subgroup $\textrm{G} \subset \textrm{SO}(8,8)$ of the duality group that is gauged (and then becomes a local symmetry of the theory) upon the gauging procedure has generators of the form
\begin{equation}
\label{XMN_tensor}
X_{MN} = \Theta_{MN|PQ} \, L^{PQ} \ ,
\end{equation}
with 
\begin{equation}
\label{Generators_SO(8,8)}
\left( L^{PQ} \right)_R{}^T = 2 \, \delta_R^{[P} \, \eta^{Q]T}
\hspace{10mm} \textrm{ and } \hspace{10mm}
\left( L^{PQ} \right)_{RS}{}^{TU} = 8 \, \delta_{[R}^{[P} \,  \eta^{Q][T} \, \delta_{S]}^{U]}\ ,
\end{equation}
being the generators of $\textrm{SO}(8,8)$ in the vectorial and adjoint representations, respectively. Using (\ref{Generators_SO(8,8)}), the gauge group generators (\ref{XMN_tensor}) are
\begin{equation}
X_{MNR}{}^{T} \equiv \Theta_{MN|PQ} \, \left( L^{PQ} \right)_R{}^T  
\hspace{10mm} \textrm{ and } \hspace{10mm}
X_{MNRS}{}^{TU}  \equiv \Theta_{MN|PQ} \, \left( L^{PQ} \right)_{RS}{}^{TU}  \ ,
\end{equation}
in the fundamental (vectorial) and adjoint representations, respectively. The fundamental representation enters the covariant derivatives of the scalars in (\ref{L_bosonic}), namely,
\begin{equation}
D_{\mu} M_{MN} = \partial_{\mu} M_{MN} - 2 \, g \, A_{\mu}{}^{PQ} \, X_{PQ(M}{}^{R} M_{N)R} \ ,
\end{equation}
where $g$ denotes the gauge coupling in the supergravity. The adjoint representation determines the gauge brackets
\begin{equation}
\label{brackets_N=8}
\left[ X_{MN} , X_{PQ}  \right] = -\frac12 X_{MNPQ}{}^{RS} \, X_{RS} \ ,
\end{equation}
and closure of the gauge brackets (\ref{brackets_N=8}) imposes a set of quadratic constraints (QC's) of the form
\begin{equation}
\label{QC_N=8}
X_{RSM}{}^{K} \Theta_{KN|PQ} + X_{RSN}{}^{K} \Theta_{MK|PQ} + X_{RSP}{}^{K} \Theta_{MN|KQ} + X_{RSQ}{}^{K} \Theta_{MN|PK} = 0 \ .
\end{equation}
Each embedding tensor configuration solving the QC's in (\ref{QC_N=8}) gives rise to a consistent half-maximal gauged supergravity in three dimensions. Lastly, as discussed in Section~$6$ of \cite{Deger:2019tem}, the set of QC's in (\ref{QC_N=8}) must be supplemented with two additional ones, namely
\begin{equation}
\label{QC_N=16_extra}
\begin{array}{rcl}
48 \, \theta \, \theta_{MN} + \theta_{M}{}^{PQR} \, \theta_{NPQR} - \frac{1}{16} \, \eta_{MN} \, \theta^{PQRS} \, \theta_{PQRS} &=& 0 \ , \\[2mm]
\theta_{M_{1}M_{2}M_{3}M_{4}} \,\, \theta_{M_{5}M_{6}M_{7}M_{8}}  \,\, [\Gamma^{M_{1} \ldots M_{8}}]_{\mathcal{\dot{A} \, \dot{B}}} &=& 0 \ ,
\end{array}
\end{equation}
where $[\Gamma^{M_{1} \ldots M_{8}}]_{\mathcal{\dot{A} \, \dot{B}}}$ denotes the (self-dual) eight-fold antisymmetric product of $\Gamma$-matrices of $\textrm{SO}(8,8)$, for a half-maximal supergravity to be embeddable into the maximal theory. We will come back to these additional constraints when analysing gaugings of half-maximal supergravity arising from M-theory/type~II/type~I flux compactifications.

\subsection{Fermion masses and scalar potential}

The gauging procedure generates scalar-dependent mass terms for the various fermionic fields in the theory \cite{Deger:2019tem}. These take the form
\begin{equation}
\label{L_mass_fermi}
\mathcal{L}_{\textrm{mass}} = \frac{1}{2} \, e \, A_{1}^{\mathcal{A}\mathcal{B}} \, \bar{\psi}_{\mu}{}^{\mathcal{A}} \gamma^{\mu\nu} \psi_{\nu}{}^{\mathcal{B}} + i \, e \, A_2^{\mathcal{A \, \dot{B}} s} \, \bar{\chi}^{\dot{\mathcal{B}}s} \, \gamma^{\mu} \, \psi_{\mu}^{\mathcal{A}} +  \frac{1}{2} \, e \, A_3^{{\cal \dot A} r {\cal \dot B} s} \,  \bar{\chi}^{\dot{\mathcal{A}}r}  \, \chi^{\dot{\mathcal{B}}s} \ ,
\end{equation}
where the index $r=1, \ldots, n$ runs over the $n=8$ matter multiplets and the indices $\mathcal{A}$, $\dot{\mathcal{A}}$ refer, together with $\mathcal{I}$, to the (triality-related) spinorial, conjugate-spinorial and vectorial representations of $\textrm{SO}(8)_{\textrm{R}}$. To obtain the various fermionic mass terms in (\ref{L_mass_fermi}) we will proceed as follows. Using the Lorentzian basis in which $\,\eta_{MN}\,$ takes the block-diagonal form in (\ref{eta_matrices}) and the $\textrm{SO}(8,8)$ vectorial index splits as $M=(\mathcal{I},r)$, we will start by building the coset representative as
\begin{equation}
\label{coset_rep_Ltz}
\mathcal{V}_{M}{}^{\underline{N}}(\phi) = e^{\phi_{\mathcal{I}r} \, \left(L^{\mathcal{I}r}\right)_{M}{}^{\underline{N}}} \ ,
\end{equation}
where $\phi_{\mathcal{I}r}$ are the $64$ physical scalars associated with the non-compact generators $L^{\mathcal{I}r}$. Note that, amongst the $L^{MN}$ generators in (\ref{Generators_SO(8,8)}), those of type $L^{\mathcal{IJ}}$ and $L^{rs}$ span the compact $\textrm{SO}(8) \times \textrm{SO}(8) \subset \textrm{SO}(8,8)$ in the denominator of (\ref{scalar_coset_N=8}) and, therefore, have no physical scalars associated. Equipped with the coset representative (\ref{coset_rep_Ltz}) in the Lorentzian basis, we will dress up the embedding tensor components in (\ref{ET_N=8}) to obtain the so-called $T$-tensor components\footnote{Importantly, the non-singlet embedding tensor components $\theta_{MNPQ}$ and $\theta_{MN}$ must be given in the Lorentzian basis too for the contractions in (\ref{T-tensor_components}) to be well-defined.}
\begin{equation}
\label{T-tensor_components}
\begin{array}{rcl}
T_{\underline{MNPQ}} &=&  (\mathcal{V}^{-1})_{\underline{M}}{}^R \, (\mathcal{V}^{-1})_{\underline{N}}{}^S \, (\mathcal{V}^{-1})_{\underline{P}}{}^T \, (\mathcal{V}^{-1})_{\underline{Q}}{}^U \, \theta_{RSTU} \ , \\[2mm]
T_{\underline{MN}} &=&  (\mathcal{V}^{-1})_{\underline{M}}{}^R \, (\mathcal{V}^{-1})_{\underline{N}}{}^S \,  \theta_{RS} \ , \\[2mm]
T &=& \theta \ .
\end{array}
\end{equation}
Finally, the fermionic mass terms in (\ref{L_mass_fermi}) can be extracted from (\ref{T-tensor_components}) upon taking suitable projections into irreps of the R-symmetry group using $\textrm{SO}(8)_{\textrm{R}}$ invariant $\gamma^{(p)}$-forms (see Appendix~\ref{App:Gamma_conventions}). The result is
\begin{equation}
\label{Fermion-Shifts}
\begin{array}{rcl}
A_1^{\mathcal{AB}} &=& -\frac{1}{48} \, [\gamma^{\mathcal{IJKL}}]_{\cal A B} \,\, T_{\underline{\mathcal{IJKL}}} - \frac14 \delta^{\cal A B} \left( T_{\underline{\mathcal{II}}} - 4T \right) \ , \\[3mm]
A_2^{\mathcal{A \, \dot{B}} s}  &=& -\frac{1}{12} \,  [\gamma^{\mathcal{IJK}}]_\mathcal{A\dot{B}} \,\, T_{\underline{\mathcal{IJK}s}} - \frac{1}{2} [\gamma^{\mathcal{I}}]_{\mathcal{A\dot{B}}} \,\, T_{\underline{\mathcal{I} s}} \ , \\[3mm]
A_3^{{\cal \dot A} r {\cal \dot B} s} &=& \frac{1}{48} \, \delta^{rs} [\gamma^{\mathcal{IJKL}}]_{{\cal \dot A}{\cal \dot B}}\,\, T_{\underline{\mathcal{IJKL}}} + \frac{1}{2} \,[\gamma^{\mathcal{IJ}}]_{{\cal \dot A}{\cal \dot B}} \,\, T_{\underline{\mathcal{IJ}rs}} -2 \, \delta^{{\cal \dot A}{\cal \dot B}} \left[ \delta^{rs} \Big( T  - \frac{1}{8} T_{\underline{qq}} \Big) +  T_{\underline{rs}} \right] \ .
\end{array}
\end{equation}
In terms of the above fermionic mass terms, the scalar potential of the gauged supergravity can be expressed as
\begin{equation}
\label{V_N=8}
V(\phi) = 
-\frac{1}{4} A_1^{{\cal A B}} A_1^{{\cal A B}} + \frac{1}{8} A_2^{{\cal A \, {\dot B}} s} A_2^{{\cal A \, {\dot B}} s} \ .
\end{equation}

Selecting a particular embedding tensor configuration $\Theta_{MN|PQ}$ in (\ref{ET_N=8}) that solves the QC's in (\ref{QC_N=8}) amounts to select a particular gauged supergravity, namely, a particular theory. Then, the extremisation of the corresponding scalar potential (\ref{V_N=8}) yields the structure of maximally symmetric vacua, \textit{i.e.} Anti de Sitter (AdS$_{3}$), Minkowski (Mkw$_{3}$) or de Sitter (dS$_{3}$), of such a theory. Once a vacuum has been found, the spectrum of fermions around that vacuum is obtained directly from (\ref{Fermion-Shifts}). The spectrum of vectors and scalars can be computed from the mass matrices\footnote{These mass matrices were obtained assuming that the vacuum solution is placed at the origin of moduli space, namely, at $\mathcal{V}_{M}{}^{\underline{N}}(0)=\mathbb{I}$. Note that this does not imply a lack of generality in the context of 3D gauged supergravity since the coset space (\ref{scalar_coset_N=8}) is a homogeneous space and, therefore, a vacuum at $\mathcal{V}_{M}{}^{\underline{N}}(\phi)\neq\mathbb{I}$ can be brought to the origin of moduli space by acting with a $\textrm{SO}(8,8)$ transformation. Also notice that the original embedding tensor $\Theta_{MN|PQ}$ yielding the vacuum outside the origin must be transformed accordingly for the $T$-tensor (\ref{T-tensor_components}) and, therefore, the scalar potential (\ref{V_N=8}) and the mass spectra, to remain invariant.} in eqs.~$(2.22)-(2.23)$ of \cite{Deger:2019tem}.

\section{Type~II orientifolds}
\label{sec:Type_II_orientifolds}

Type~II orientifolds have been studied in \cite{Koerber:2007hd} within the context of supersymmetric compactifications. The orientifold action is a composition of both worldsheet and target space transformations. In particular, a worldsheet orientation-reversal parity transformation $\Omega_{P}$, a worldsheet fermion number projector for left-moving fermions $(-1)^{F_{L}}$ (which is not always needed \cite{Koerber:2007hd}) and an internal target space involution $\sigma$. The internal involution $\sigma$ may leave certain submanifolds of the internal space invariant. The product of one of such invariant submanifolds with the external spacetime is referred to as an orientifold O$p$-plane, where $p$ denotes the total number of spatial dimensions filled by the orientifold plane.

O$p$-planes have negative tension, are electrically (magnetically) charged under a $C_{(p+1)}$ ($C_{(7-p)}$) gauge potential and do not have an associated dynamics. Being charged objects, they enter charge conservation conditions known as tadpole cancellation conditions (we refer the reader to \cite{Koerber:2007hd} for more details) of the form
\begin{equation}
\label{Tadpole_p-form}
D F_{(8-p)} + \beta_{8-p} \, H_{(1)} \wedge F_{(8-p)}  -  H_{(3)} \wedge F_{(6-p)} = J_{\textrm{O}p/\textrm{D}p} \ ,
\end{equation}
where $\,\beta_{8-p}\,$ is a constant, $\,D \equiv d+\omega\,$ is a nilpotent $(D^2=0)$ twisted exterior derivative on the internal space (see the discussion on twisted tori below) and $J_{\textrm{O}p/\textrm{D}p}$ denotes the net current of charged O$p$-plane/D$p$-brane sources in the smeared limit. Similarly to (\ref{Tadpole_p-form}), and in the absence of NS$5$-branes, there is a condition for the NS-NS three-form flux of the form
\begin{equation}
\label{Bianchi_H3-form}
D H_{(3)}  + \alpha \, H_{(1)} \wedge H_{(3)}  = 0  \ ,
\end{equation}
with $\,\alpha\,$ being a constant. The parameters $\beta_{p}$ in (\ref{Tadpole_p-form}) are not independent. In particular, one has that $\,\beta_{p} = \beta - \frac{4-p}{2}\alpha\,$ in the type~IIA context and $\,\beta_{p} = \beta - \frac{5-p}{2}\alpha\,$ in the type~IIB one for arbitrary parameters $\,\alpha\,$ and $\,\beta\,$. Then, as we will show, the matching between the tadpole conditions (\ref{Tadpole_p-form})-(\ref{Bianchi_H3-form}) and the quadratic constraints (\ref{QC_N=8}) of the gauged supergravity requires an $(\alpha,\beta)$-dependent embedding of the dilaton flux $\,H_{(1)}\,$ into the embedding tensor. The attentive reader might have noticed the unconventional terms $\,H_{(1)} \wedge F_{(8-p)}\,$ and $\,H_{(1)} \wedge H_{(3)}\,$ in (\ref{Tadpole_p-form}) and (\ref{Bianchi_H3-form}), respectively. Starting from a standard formulation of type II supergravity in the string frame, these terms can be generated by performing non-standard field redefinitions $\,F_{(p)} \rightarrow e^{-\beta_{p} \Phi} F_{(p)}\,$ and $\,H_{(3)} \rightarrow e^{-\alpha \Phi} H_{(3)}\,$ involving different powers of the type~II dilaton $\,e^{\Phi}$. The need for this unconventional frame stems from the fact that, when considering type~IIA models, the metric fluxes $\,\omega\,$ and the dilaton flux $\,H_{(1)}\,$ must be treated on equal footing as they both come from metric fluxes in 11D. The parameters $\,\alpha\,$ and $\,\beta\,$ will determine the embedding of $\,H_{(1)}\,$ inside the embedding tensor of the three-dimensional gauged supergravities.\footnote{It will also occur that the matching between the tadpole conditions (\ref{Tadpole_p-form}) and (\ref{Bianchi_H3-form}) and the quadratic constraints of the gauged supergravity (\ref{QC_N=8}) will allow for additional parameters, we denote them $\,\gamma\,$ and $\,\delta$, in the embedding of the dilaton flux $\,H_{(1)}\,$ inside the embedding tensor. This ambiguity would ultimately be fixed by performing a full-fledged dimensional reduction of type II supergravity in presence of $H_{(1)}$.} If $\,H_{(1)}=0\,$ one can set $\,\alpha=\beta=0\,$ and recover the standard string frame.

\subsection*{Twisted tori: metric and gauge fluxes}

We will consider type II orientifold reductions on a seven-dimensional group manifold $\textrm{G}$ commonly referred to as a twisted torus \cite{Scherk:1979zr}. Introducing a basis of one-forms $\,\eta^{m}$ ($m=1,\ldots,7$) specified by a coordinate dependent twist matrix $\,U^{m}{}_{n}(y) \in \textrm{G}$
\begin{equation}
\label{eta_basis}
\eta^{m} = U^{m}{}_{n}(y) \, dy^{n} \ ,
\end{equation}
one can proceed and expand the (constant) internal components of the background gauge fluxes in the basis (\ref{eta_basis}): $H_{(3)} = \tfrac{1}{3!} \,H_{mnp} \, \eta^{m} \wedge \eta^{n} \wedge \eta^{p}$, etc. Due to the twist in (\ref{eta_basis}), the one-forms $\,\eta^{m}\,$ are no longer closed but obey the structure equation
\begin{equation}
\label{structure_equation}
d\eta^{p} + \tfrac{1}{2} \, \omega_{mn}{}^{p} \, \eta^{m} \wedge \eta^{n} = 0   \ ,
\end{equation}
with $\,\omega_{mn}{}^{p}=(U^{-1})_{m}{}^{r} (U^{-1})_{n}{}^{s} \left( \partial_{r} U^{p}{}_{s} - \partial_{s} U^{p}{}_{r}  \right)\,$ being the structure constants -- commonly referred to as metric fluxes -- of the Lie algebra 
\begin{equation}
\label{X_commutators}
[X_m, X_n] = \omega_{mn}{}^{p} \, X_p \ ,
\end{equation}
spanned by the isometry generators $\,X_m = (U^{-1})_m{}^n \, \partial_{n}$. The Jabobi identity for the algebra (\ref{X_commutators}) amounts to the integrability condition for (\ref{structure_equation}). The compactness of the group manifold $\,\textrm{G}\,$ can be established by direct evaluation of the Killing--Cartan (KC) metric $\,K_{mn} = \omega_{m p}{}^{q} \, \omega_{n q}{}^{p}\,$. If $\,\textrm{G}\,$ is compact, then the internal space can be taken to be the group manifold. If $\,\textrm{G}\,$ is non-compact, then the internal space is locally isomorphic to the group manifold but global aspects must be carefully addressed. We refer the reader to refs~\cite{Hull:2005hk,Hull:2006tp} for more details on this and related issues.

The tadpole cancellation conditions in (\ref{Tadpole_p-form}) and (\ref{Bianchi_H3-form}) involve the metric fluxes $\,\omega\,$ specifying the twisted $\mathbb{T}^{7}$ torus, as well as gauge fluxes $\,H_{(3)}\,$ and $\,F_{(p)}$, and a dilaton flux $\,H_{(1)}$. However, the coexistence of metric $\,\omega\,$ and one-form $\,H_{(1)}\,$ or $\,F_{(1)}\,$ fluxes in a compactification scheme poses some issues. On the one hand, metric fluxes $\,\omega\,$ arise from Scherk--Schwarz (SS) reductions on twisted tori \cite{Scherk:1979zr}. These fluxes have a geometric origin and can be thought of as operators that map an internal $p$-form into a ($p+1$)-form, \textit{e.g.} $(DF)_{mnpq}=\partial_{[m} F_{npq]} + \omega_{[mn}{}^{r} F_{|r|pq]}$, and specify the twist of fields (gauge potentials) with legs along the internal space. Since the dilaton $\,\Phi\,$ is a scalar (no legs along the internal space), a dilaton flux $\,H_{(1)}\,$ cannot be generated by the ordinary SS procedure. The same holds for a flux $\,F_{(1)}$. On the other hand, $\,H_{(1)}\,$ and $\,F_{(1)}\,$ background fluxes are valid ingredients in a standard flux compactification on an ordinary $\,\mathbb{T}^{7}$ torus. Therefore, we are left with two possible well-defined scenarios:

\begin{itemize}

\item[$i)$] \textbf{Ordinary Scherk--Schwarz reduction}. In this scenario one has
\begin{equation}
\omega_{mn}{}^{p} \neq 0 
\hspace{10mm} \textrm{ and } \hspace{10mm}
H_{(1)}=F_{(1)}=0 \ .
\end{equation}
Since the metric fluxes $\,\omega\,$ are identified with the structure constants of the Lie algebra (\ref{X_commutators}), they must obey standard Jacobi identities of the form
\begin{equation}
\label{Jacobi_identity}
\omega_{[mn}{}^{r} \, \omega_{p]r}{}^{q} = 0 \ ,
\end{equation}
together with an additional ``unimodularity'' or trace condition \cite{Scherk:1979zr}
\begin{equation}
\label{Zero_Trace}
\omega_{m r}{}^{r} = 0 \ .
\end{equation}
In ordinary SS reductions, the unimodularity condition (\ref{Zero_Trace}) ensures consistency of the truncation of the higher-dimensional action. For non-unimodular gaugings there is still a consistent truncation of the equations of motion but, in general, not of the action \cite{Bergshoeff:2003ri,Hull:2005hk}.\footnote{An example of a non-unimodular gauging still describing a consistent truncation at the level of the action can be found in \cite{Derendinger:2007xp}. However, this counterexample to the general statement includes an additional duality twist (see \textit{e.g.} \cite{Dabholkar:2002sy}) that introduces global pathologies in the higher-dimensional background.}
\\

\item[$ii)$] \textbf{Standard flux compactification}. In this scenario one has
\begin{equation}
\omega_{mn}{}^{p} = 0 
\hspace{10mm} \textrm{ and } \hspace{10mm}
H_{(1)} \,,\, F_{(1)} \neq 0 \ .
\end{equation}

\end{itemize}

In what follows we will consider all the fluxes at the same time when discussing the fluxes/embedding tensor dictionary, quadratic constraints on fluxes, etc., but always bearing in mind that one of the two scenarios above must hold for a given three-dimensional gauged supergravity to describe a truncation of a type~II supergravity action. One could of course adopt a bottom-up approach and activate \textit{any} embedding tensor component associated with a (generically non-unimodular) metric flux $\,\omega\,$ as well as gauge fluxes $\,H_{(1)}$, $\,H_{(3)}\,$ and $\,F_{(p)}$ provided the quadratic constraints (\ref{brackets_N=8}) hold. However, the uplift of the three-dimensional gauged supergravity described by such an embedding tensor would generically deviate from an ordinary SS reduction or a standard flux compactification. These more exotic scenarios are beyond the scope of this work.

\subsection{Warming up: M-theory fluxes/embedding tensor dictionary}
\label{sec:M-theory}

\begin{table}[t]
\begin{center}
\renewcommand{\arraystretch}{1.7}
\begin{tabular}{|c|c|c|c|c|}
\hline
&  $\textrm{E}_{8(8)}$ / Maximal & ${\rm SO(8,8)}$ & Half-Maximal &  ${\rm SL(8)} \times \mathbb{R}_1$  \\ 
\hline  
\hline
 & -- & $\bf{16}$ &  --  & ${\bf{8'}}_{-1} \oplus {\bf{8}}_{+1}$  \\ 
\hline  
\hline
\multirow{2}{*}{\rotatebox{90}{Scalars\,\,}}    & \multirow{2}{*}{$\bf{248}$}     & $\bf{120}$ & $L^{MN}$ & ${\bf 28'}_{-2} \oplus {\color{RoyalBlue}{{\bf (63+1)}_{0}}} \oplus {\color{Orange}{{\bf 28}_{+2}}}$ \\ \cline{3-5}
&  &  ${\bf 128'}$ & -- & ${\bf 8}_{-3} \oplus {\bf 56}_{-1} \oplus {\color{Orange}{{\bf 56'}_{+1}}} \oplus \boxed{{\bf 8'}_{+3}}$ \\[0.5mm]
\hline
\hline
\multirow{9}{*}{\rotatebox{90}{Embedding Tensor / Fluxes}} &  ${\bf 1}$   & ${\bf 1}$ & $\theta$ & ${\bf 1}_{0} $ \\
\cline{2-5}
&  \multirow{8}{*}{${\bf 3875}$} & ${\bf 135}$ & $\theta_{MN}$ & ${\bf 36'}_{-2} \oplus {\bf 63}_{0} \oplus  {\bf 36}_{+2}$ \\ \cline{3-5}
&   & \multirow{3}{*}{${\bf 1820}$ } & \multirow{3}{*}{$\theta_{MNPQ}$} & ${\bf 70}_{-4} \oplus {\bf 28'}_{-2} \oplus {\bf 420'}_{-2}$ \\ 
&   & & & ${\bf 720}_{0} \oplus {\bf 63}_0 \oplus {\bf 1}_0$ \\
&   & & & ${\color{Orange}{{\bf 70}_{+4}}} \oplus {\bf 28}_{+2} \oplus {\bf 420}_{+2}$ \\ \cline{3-5}
&   & \multirow{4}{*}{${\bf 1920'}$} & \multirow{4}{*}{--} & ${\bf 8'}_{-5}  \oplus {\bf 216}_{-3} \oplus {\bf 8}_{-3}$ \\
&   & & & ${\bf 56}_{-1} \oplus {\bf 504}_{-1} \oplus  {\bf 168}_{-1}$ \\
&   & & & ${\bf 56'}_{+1} \oplus {\bf 504'}_{+1} \oplus  {\bf 168'}_{+1}$ \\
&   & & & ${\color{Orange}{{\bf 8}_{+5}}} \oplus  {{\color{RoyalBlue}{\bf 216'}_{+3}}} \oplus {\color{RoyalBlue}{{\bf 8'}_{+3}}}$ \\
     \hline
\end{tabular}
\caption{M-theory branching rules for the embedding chain $\textrm{E}_{8(8)} \supset {\rm SO(8,8)} \supset {\rm SL(8)} \times \mathbb{R}_1$. The subscripts in the third column indicate $\mathbb{R}_1$-charges. We have highlighted the scalars ${\color{RoyalBlue}{e_{A}{}^{B}}}$, ${\color{Orange}{C_{(3)}}}$ and ${\color{Orange}{C_{(6)}}}$ in (\ref{scalars_SL(8)}) as well as their associated fluxes ${\color{RoyalBlue}{\omega_{AB}{}^{C}}}$, ${\color{Orange}{G_{(4)}}}$ and ${\color{Orange}{G_{(7)}}}$ in (\ref{fluxes_SL(8)}). Also the physical internal derivatives $\partial_{A}$ have been put in a box for their quick identification. Note that the spinorial representations $\bf{128'}$ and $\bf{1920'}$ of $\textrm{SO}(8,8)$ give rise to $\textrm{SL}(8) \times \mathbb{R}_{1}$ irreps with an odd $\mathbb{R}_{1}$-charge. These irreps are projected out when truncating maximal to half-maximal supergravity and are marked with $``-"$ as they do not give rise to scalars or embedding tensor deformations in half-maximal supergravity.} 
\label{Table:E8/SO(8,8)/SL8}
\end{center}
\end{table}

In the absence of sources, the dimensional reduction of 11D supergravity on $\mathbb{T}^{8}$ yields the ungauged 3D supergravity with $\textrm{E}_{8(8)}$ global U-duality symmetry \cite{Marcus:1983hb}. This $\textrm{E}_{8(8)}$ symmetry group becomes manifest when 11D supergravity is reformulated in the form of an exceptional field theory living in a generalised $(3+248)$-dimensional spacetime with an extended set of internal coordinates in the $\bf{248}$ (adjoint) irrep of $\textrm{E}_{8(8)}$ \cite{Hohm:2014fxa}. The eight physical internal coordinates of 11D supergravity are then part of the ${\bf{248}} \in \textrm{E}_{8(8)}$. On the other hand, standard Kaluza--Klein (KK) reduction of 11D supergravity on $\mathbb{T}^{8}$ comes along with a global $\textrm{GL}(8) = \textrm{SL}(8) \times \mathbb{R}_{1}$ symmetry descending from the ordinary internal diffeomorphisms. This implies that the eight physical internal coordinates of 11D supergravity $y^{A}$, with $A=1,\ldots,8$, must transform as an eight-dimensional irrep of $\textrm{SL}(8)$. A quick inspection of Table~\ref{Table:E8/SO(8,8)/SL8} then confirms that, up to a conventional choice, one has that
\begin{equation}
\label{partial_SL(8)}
y^{A} \in {\bf{8}}_{-3} 
\hspace{10mm} \textrm{ and } \hspace{10mm}
\partial_{A} \in {\bf{8}'}_{+3} \ .
\end{equation}

After having group-theoretically identified the physical derivatives $\partial_{A}$, we proceed to identify the internal components of the elfbein $e_{A}{}^{B}$ as well as the three-form $C_{(3)} \equiv C_{ABC}=C_{[ABC]}$ and six-form $C_{(6)} \equiv C_{A_{1}\ldots A_{6}}=C_{[A_{1}\ldots A_{6}]}$ gauge potentials of 11D supergravity. An inspection of Table~\ref{Table:E8/SO(8,8)/SL8} reveals that 
\begin{equation}
\label{scalars_SL(8)}
e_{A}{}^{B} \in ({\bf 63 + 1})_{0}
\hspace{5mm} , \hspace{5mm}
C_{(3)} \in {\bf{56}'}_{+1}
\hspace{5mm} \textrm{ and } \hspace{5mm}
C_{(6)} \in {\bf{28}}_{+2} \ ,
\end{equation}
where $28$ compact scalars must be subtracted from $e_{A}{}^{B}$ upon gauge-fixing of the internal $\textrm{SO}(8)$ local symmetry, namely, $e_{A}{}^{B} \in \textrm{GL}(8)/\textrm{SO}(8)$. Lastly, metric and gauge fluxes are constructed by applying exterior derivatives on the scalars in (\ref{scalars_SL(8)}). This yields
\begin{equation}
\label{fluxes_SL(8)}
\omega_{AB}{}^{C} \in ({\bf 216' + 8'})_{+3}
\hspace{5mm} , \hspace{5mm}
G_{(4)} \in {\bf{70}}_{+4}
\hspace{5mm} \textrm{ and } \hspace{5mm}
G_{(7)} \in {\bf{8}}_{+5} \ .
\end{equation}
Importantly, bosonic (spinorial) irreps of $\textrm{SO}(8,8)$ give rise to ${\rm SL(8)} \times \mathbb{R}_1$ irreps with an even (odd) $\mathbb{R}_{1}$-charge and are kept (projected out) when truncating maximal to half-maximal supergravity (see Table~\ref{Table:E8/SO(8,8)/SL8}). Therefore, since the physical internal derivatives in (\ref{partial_SL(8)}) are spinorial in this sense ($\mathbb{R}_{1}$-charge $+3$), only the spinorial scalars $C_{(3)}$ ($\mathbb{R}_{1}$-charge $+1$) in (\ref{scalars_SL(8)}) will produce a bosonic flux $G_{(4)}$ ($\mathbb{R}_{1}$-charge $+4$) in (\ref{fluxes_SL(8)}) suitable to enter the embedding tensor of half-maximal supergravity. Summarising, the half-maximal supergravities obtained from M-theory will contain physical scalars and gauge fluxes of the form
\begin{equation}
\label{Fields&Fluxes_bosonic_M-theory}
\begin{array}{lcl}
\textrm{Scalars} & : & e_{A}{}^{B} \in \frac{\textrm{GL}(8)}{\textrm{SO}(8)}
\hspace{3mm} , \hspace{3mm}
C_{(6)} \ , \\[2mm]
\textrm{Fluxes} & : & 
G_{(4)} \ .
\end{array}
\end{equation}
This simple spinorial/bosonic grading will help us later when discussing more complicated flux models arising from type~II orientifold reductions. 

Before moving to discuss specific algebraic properties of the M-theory gaugings, let us perform a precise counting of bosonic and spinorial scalars in the M-theory context. As we have already seen, bosonic scalars arise from  $e_{A}{}^{B} \in \textrm{GL}(8)/\textrm{SO}(8)$ and $C_{(6)}$ adding up to $64$ scalars. However, there are only $56$ spinorial scalars arising from $C_{(3)}$. The difference is explained by the $8$ additional spinorial scalars dual to the vectors $e_{\mu}{}^{A}$. All together, there are $64$ bosonic and $64$ spinorial scalars as required by the $\textrm{E}_{8(8)}/\textrm{SO}(16)$ scalar geometry of the maximal theory. 

Since half-maximal supergravity does not contain spinorial irreps of $\textrm{SO}(8,8)$, only the gauge flux $G_{(4)}$ and the $64$ bosonic scalars $e_{A}{}^{B} \in \textrm{GL}(8)/\textrm{SO}(8)$ and $C_{(6)}$ (all of them with an even $\mathbb{R}_{1}$-charge) are present in the theory. The resulting gauging in 3D has a simple embedding tensor (\ref{ET_N=8}) specified by the non-zero components
\begin{equation}
\label{ET_M-theory}
\theta^{ABCD} = \tfrac{1}{4!} \, \varepsilon^{ABCDEFGH} G_{EFGH}  \ ,
\end{equation}
where we have made the $\textrm{SO}(8,8)$ index splitting $T_{M}=(T_{A},T^{A})$ in light-cone coordinates. This is nothing but the branching ${\bf{16}} \rightarrow {\bf{8'}}_{-1} \oplus {\bf{8}}_{+1}$ under $\textrm{SO}(8,8) \supset \textrm{SL}(8) \times \mathbb{R}_{1}$ in Table~\ref{Table:E8/SO(8,8)/SL8}. The embedding tensor (\ref{ET_M-theory}) satisfies the QC's in (\ref{QC_N=8}) and specifies an Abelian gauge group of dimension $28$ for the non-zero generators $X^{AB} = \theta^{ABCD} \, L_{CD}$ in the light-cone basis.\footnote{In the light-cone basis there are two maximal Abelian subgroups spanned by $L^{AB}$ and $L_{AB}$, respectively.}

\begin{table}[t!]
\begin{center}
\scalebox{0.78}{
\renewcommand{\arraystretch}{1.7}
\begin{tabular}{|c|c|c||c|}
     \hline 
     & Half-Maximal & ${\rm SL(8)} \times \mathbb{R}_1$ &  ${\rm SL(7)} \times \mathbb{R}_2 \times \mathbb{R}_1$ \\ 
     \hline \hline
     & -- & ${\bf{8'}}_{-1} \oplus {\bf{8}}_{+1}$ &  $[{\bf{7'}}_{(-1,-1)} \oplus {\bf{1}}_{(+7,-1)}] \oplus [{\bf{7}}_{(+1,+1)} \oplus {\bf{1}}_{(-7,+1)}]$ \\
     \hline\hline  
     \multirow{4}{*}{\rotatebox{90}{Scalars\,\,}} &  \multirow{2}{*}{$L^{MN}$}  & \multirow{2}{*}{${\bf 28'}_{-2} \oplus {\color{RoyalBlue}{{\bf (63+1)}_{0}}}  \oplus {\color{Orange}{{\bf 28}_{+2}}}$} & $[{\bf 21'}_{(-2,-2)} \oplus {\bf 7'}_{(+6,-2)}] \oplus [{\color{BrickRed}{{\bf 7'}_{(-8,0)}}} \oplus {\color{RoyalBlue}{{\bf (48+1)}_{(0,0)}}} \oplus {\bf 7}_{(+8,0)}] $ 
     \\
     & &   & ${\color{RoyalBlue}{{\bf 1}_{(0,0)}}}  \oplus [{\color{ForestGreen}{{\bf 7}_{(-6,+2)}}} \oplus {\color{BrickRed}{{\bf 21}_{(+2,+2)}}}]$
     \\\cline{2-4}
     & \multirow{2}{*}{--} &  \multirow{2}{*}{${\bf 8}_{-3} \oplus {\bf 56}_{-1} \oplus {\color{Orange}{{{\bf 56'}_{+1}}}} \oplus \boxed{{\bf 8'}_{+3}}$} & $ [{\bf 7}_{(+1,-3)} \oplus {\bf 1}_{(-7,-3)}] \oplus [{\bf 35}_{(+3,-1)} \oplus {\bf 21}_{(-5,-1)}] $
     \\
     & &  $ $ & $[{\color{BrickRed}{{\bf 35'}_{(-3,+1)}}} \oplus {\color{ForestGreen}{{\bf 21'}_{(+5,+1)}}}] \oplus [\boxed{{\bf 7'}_{(-1,+3)}} \oplus {\bf 1}_{(+7,+3)}]$ 
     \\[2mm]
     \hline \hline
      \multirow{20}{*}{\rotatebox{90}{Embedding Tensor / Fluxes}} & $\theta$ &  ${\bf 1}_0$ & ${\bf 1}_{(0,0)}$  
     \\\cline{2-4}
     & \multirow{3}{*}{$\theta_{MN}$} &  \multirow{3}{*}{${\bf 36'}_{-2} \oplus {\bf 63}_{0} \oplus  {\bf 36}_{+2}$} & $[{\bf 28'}_{(-2,-2)} \oplus {\bf 7'}_{(+6,-2)} \oplus {\bf 1}_{(+14,-2)}]$
     \\
     & &  $ $ & $[{\bf 7'}_{(-8,0)} \oplus {\bf 48}_{(0,0)} \oplus {\bf 1}_{(0,0)} \oplus {\bf 7}_{(+8,0)}]$
     \\
     & &  $ $ & $[{\bf 28}_{(+2,+2)} \oplus {\bf 7}_{(-6,+2)} \oplus {\color{magenta}{{\bf 1}_{(-14,+2)}}}]$
     \\\cline{2-4}
     & \multirow{6}{*}{$\theta_{MNPQ}$}  &  \multirow{2}{*}{${\bf 70}_{-4} \oplus {\bf 28'}_{-2} \oplus {\bf 420'}_{-2}$} & $[{\bf 35}_{(-4,-4)} \oplus {\bf 35'}_{(+4,-4)}] \oplus [{\bf 21'}_{(-2,-2)} \oplus {\bf 7'}_{(+6,-2)}]$
     \\
     & &  $ $ & $[{\bf 35'}_{(-10,-2)} \oplus {\bf 21'}_{(-2,-2)} \oplus {\bf 224'}_{(-2,-2)} \oplus {\bf 140'}_{(+6,-2)}]$
     \\ \cline{3-4}
     & &  \multirow{2}{*}{${\bf 720}_{0} \oplus {\bf 63}_0 \oplus {\bf 1}_0$} & $[{\bf 140'}_{(-8,0)} \oplus {\bf 392}_{(0,0)} \oplus {\bf 48}_{(0,0)} \oplus {\bf 140}_{(+8,0)}]$
     \\
     & &  $ $ & $[{\bf 7'}_{(-8,0)} \oplus {\bf 48}_{(0,0)} \oplus {\bf 1}_{(0,0)} \oplus {\bf 7}_{(+8,0)}] \oplus {\bf 1}_{(0,0)}$
     \\ \cline{3-4}
     & & \multirow{2}{*}{${\color{Orange}{{\bf 70}_{+4}}} \oplus {\bf 28}_{+2} \oplus {\bf 420}_{+2}$} & $[{\color{BrickRed}{{\bf 35}_{(-4,+4)}}} \oplus {\color{ForestGreen}{{\bf 35'}_{(+4,+4)}}}] \oplus [{\bf 7}_{(-6,+2)} \oplus {\bf 21}_{(+2,+2)}]$
     \\
     & &  $ $ & $[{\bf 35}_{(+10,+2)} \oplus {\bf 21}_{(+2,+2)} \oplus {\bf 224}_{(+2,+2)} \oplus {\bf 140}_{(-6,+2)}]$
     \\\cline{2-4}
     & \multirow{10}{*}{--}  &  \multirow{2}{*}{${\bf 8'}_{-5} \oplus {\bf 8}_{-3} \oplus {\bf 216}_{-3}$} & $[{\bf 7'}_{(-1,-5)} \oplus {\bf 1}_{(+7,-5)}] \oplus [{\bf 1}_{(-7,-3)} \oplus {\bf 7}_{(+1,-3)}]$
     \\
     & &  $ $ & $[{\bf 48}_{(-7,-3)} \oplus {\bf 140}_{(+1,-3)} \oplus {\bf 7}_{(+1,-3)} \oplus {\bf 21}_{(+9,-3)}]$
     \\ \cline{3-4}
     & &  \multirow{3}{*}{${\bf 56}_{-1} \oplus {\bf 504}_{-1} \oplus  {\bf 168}_{-1}$} & $[{\bf 21}_{(-5,-1)} \oplus {\bf 35}_{(+3,-1)}]$
     \\
     & &  $ $ & $[{\bf 224}_{(-5,-1)} \oplus {\bf 210}_{(+3,-1)} \oplus {\bf 35}_{(+3,-1)} \oplus {\bf 35'}_{(+11,-1)}]$
     \\
     & &  $ $ & $[{\bf 7}_{(-13,-1)} \oplus {\bf 21}_{(-5,-1)} \oplus {\bf 28}_{(-5,-1)} \oplus {\bf 112}_{(+3,-1)}]$
     \\ \cline{3-4}
     & &  \multirow{3}{*}{${\bf 56'}_{+1} \oplus {\bf 504'}_{+1} \oplus  {\bf 168'}_{+1}$} & $[{\bf 21'}_{(+5,+1)} \oplus {\bf 35'}_{(-3,+1)}]$
     \\
     & &  $ $ & $[{\bf 224'}_{(+5,+1)} \oplus {\bf 210'}_{(-3,+1)} \oplus {\bf 35'}_{(-3,+1)} \oplus {\bf 35}_{(-11,+1)}]$
     \\
     & &  $ $ & $[{\bf 7'}_{(+13,+1)} \oplus {\bf 21'}_{(+5,+1)} \oplus {\bf 28'}_{(+5,+1)} \oplus {\bf 112'}_{(-3,+1)}]$
     \\ \cline{3-4}
     & &  \multirow{2}{*}{${\color{Orange}{{\bf 8}_{+5}}} \oplus {\color{RoyalBlue}{{\bf 8'}_{+3}}} \oplus {\color{RoyalBlue}{{\bf 216'}_{+3}}}$} & $[{\color{BrickRed}{{\bf 7}_{(+1,+5)}}} \oplus {\color{ForestGreen}{{\bf 1}_{(-7,+5)}}}] \oplus [{\color{RoyalBlue}{{\bf 7'}_{(-1,+3)}}} \oplus {\bf 1}_{(+7,+3)}]$
     \\
     & &  $ $ & $[{\bf 48}_{(+7,+3)} \oplus {\color{RoyalBlue}{{\bf 140'}_{(-1,+3)}}} \oplus {\color{RoyalBlue}{{\bf 7'}_{(-1,+3)}}} \oplus {\color{BrickRed}{{\bf 21'}_{(-9,+3)}}}]$
     \\
     \hline
\end{tabular}
}
\caption{Type~IIA with O$2$-plane branching rules for the embedding ${\rm SL(7)} \times \mathbb{R}_2 \times \mathbb{R}_1 \subset {\rm SL(8)} \times \mathbb{R}_1$. The subscripts in the third column indicate $(\mathbb{R}_2,\mathbb{R}_1)$-charges. We have highlighted the scalars ${\color{RoyalBlue}{e_{m}{}^{n}}}$, ${\color{RoyalBlue}{\Phi}}$, ${\color{ForestGreen}{B_{(2)}}}$, ${\color{ForestGreen}{B_{(6)}}}$ and ${\color{BrickRed}{C_{(p)}}}$ in (\ref{scalars_SL(7)_IIA}), as well as their associated fluxes in (\ref{fluxes_SL(7)_IIA}). The Romans mass parameter ${\color{magenta}{F_{(0)} \in {\bf 1}_{(-14,+2)}}}$ has also been highlighted and the physical internal derivatives $\partial_{m}$ have been put in a box for their quick identification. This table should be understood as a continuation of Table~\ref{Table:E8/SO(8,8)/SL8}.} 
\label{Table:SL8-SL7}
\end{center}
\end{table}

\subsection{Type~IIA with O$2$-planes}
\label{sec:O2-plane}

The M-theory reduction on $\mathbb{T}^{8}=\mathbb{T}^{7} \times \mathbb{S}^{1}$ can be reinterpreted as a (massless) type~IIA reduction on $\mathbb{T}^{7}$ with the coordinate along the $\mathbb{S}^{1}$ being the M-theory coordinate. The $\textrm{GL}(8)$ diffeomorphisms on $\mathbb{T}^{8}$ are then reduced to the $\textrm{GL}(7)$ diffeomorphisms on $\mathbb{T}^{7}$ and coordinates split as $y^{A} = (y^{m},y^{8})$ with $m=1,\ldots,7$. The relevant branching rules for the M-theory $\Leftrightarrow$ type~IIA with O2 correspondence are summarised in Table~\ref{Table:SL8-SL7}.

The seven physical coordinates and their associated derivatives follow from the branching rules in Table~\ref{Table:SL8-SL7}. They are identified with 
\begin{equation}
\label{partial_SL(7)}
y^{m} \in {\bf{7}}_{(+1,-3)} 
\hspace{10mm} \textrm{ and } \hspace{10mm}
\partial_{m} \in {\bf{7}'}_{(-1,+3)} \ .
\end{equation}
On the other hand, the type~IIA dilaton $\Phi$, the internal components of the zehnbein $e_{m}{}^{n}$ and the various type~IIA gauge potentials $B_{(2)}$, $B_{(6)}$ and $C_{(1)}$, $C_{(3)}$, $C_{(5)}$ are identified as
\begin{equation}
\label{scalars_SL(7)_IIA}
\begin{array}{c}
e_{m}{}^{n} \in ({\bf 48 + 1})_{(0,0)}
\hspace{5mm} , \hspace{5mm}
C_{(1)} \in {\bf{7}'}_{(-8,0)}
\hspace{5mm} , \hspace{5mm}
\Phi \in {\bf{1}}_{(0,0)} \ , \\[2mm]
C_{(3)} \in {\bf 35'}_{(-3,+1)}
\hspace{4mm} , \hspace{4mm}
B_{(2)} \in {\bf 21'}_{(+5,+1)}
\hspace{4mm} , \hspace{4mm}
B_{(6)} \in {\bf 7}_{(-6,+2)}
\hspace{4mm} , \hspace{4mm}
C_{(5)} \in {\bf 21}_{(+2,+2)} \ ,
\end{array}
\end{equation}
where $21$ compact scalars must be subtracted from $e_{m}{}^{n}$ upon gauge-fixing of the internal $\textrm{SO}(7)$ local symmetry, namely, $e_{m}{}^{n} \in \textrm{GL}(7)/\textrm{SO}(7)$. While the bosonic scalars in (\ref{scalars_SL(7)_IIA}) correctly add up to $64$, the number of the spinorial ones turns out to be $56$. The $7+1$ missing spinorial scalars are dual to the three-dimensional vectors $e_{\mu}{}^{n}$ and $C_{\mu}$.\footnote{Since we are not including $C_{(7)}$ in the democratic formulation of type~IIA supergravity, the vector $C_{\mu}$ that would be captured by a purely internal $C_{(7)}$ must be taken into account explicitly.} The metric and gauge fluxes obtained by applying exterior derivatives on the scalars in (\ref{scalars_SL(7)_IIA}) are
\begin{equation}
\label{fluxes_SL(7)_IIA}
\begin{array}{c}
\omega_{mn}{}^{p} \in ({\bf 140' + 7'})_{(-1,+3)}
\hspace{5mm} , \hspace{5mm}
F_{(2)} \in {\bf{21}'}_{(-9,+3)}
\hspace{5mm} , \hspace{5mm}
H_{(1)} \in {\bf{7'}}_{(-1,+3)} \ , \\[2mm]
F_{(4)} \in {\bf 35}_{(-4,+4)}
\hspace{4mm} , \hspace{4mm}
H_{(3)} \in {\bf 35'}_{(+4,+4)}
\hspace{4mm} , \hspace{4mm}
H_{(7)} \in {\bf 1}_{(-7,+5)}
\hspace{4mm} , \hspace{4mm}
F_{(6)} \in {\bf 7}_{(+1,+5)} \ .
\end{array}
\end{equation}
Finally, the massive version of type~IIA supergravity admits an additional deformation parameter known as the Romans mass $F_{(0)}$ \cite{Romans:1985tz}. This parameter does not follow from a gauge potential and it is group-theoretically identified with
\begin{equation}
F_{(0)} \in {\bf 1}_{(-14,+2)} \ . 
\end{equation}

Again, since half-maximal supergravity does not contain spinorial irreps of $\textrm{SO}(8,8)$, we must keep scalars and fluxes with an even $\mathbb{R}_{1}$-charge. The result is that only the $64$ physical scalars and gauge fluxes
\begin{equation}
\label{Fields&Fluxes_bosonic_IIA}
\begin{array}{lcl}
\textrm{Scalars} & : & e_{m}{}^{n} \in \frac{\textrm{GL}(7)}{\textrm{SO}(7)}
\hspace{3mm} , \hspace{3mm}
C_{(1)}
\hspace{3mm} , \hspace{3mm}
\Phi
\hspace{3mm} , \hspace{3mm}
B_{(6)}
\hspace{3mm} , \hspace{3mm}
C_{(5)} \ , \\[2mm]
\textrm{Fluxes} & : & 
H_{(3)}
\hspace{3mm} , \hspace{3mm}
F_{(4)} 
\hspace{3mm} , \hspace{3mm}
F_{(0)} \ ,
\end{array}
\end{equation}
are present in the theory. These fluxes induce a gauging in 3D specified by the embedding tensor components
\begin{equation}
\label{ET_IIA-theory}
\theta^{mnpq} = \tfrac{1}{3!} \, \varepsilon^{mnpqrst} H_{rst}
\hspace{5mm} , \hspace{5mm}
\theta^{mnp8} = \tfrac{1}{4!} \, \varepsilon^{mnpqrst} F_{qrst}
\hspace{5mm} , \hspace{5mm}
\theta^{88} = F_{(0)} \ ,
\end{equation}
provided the QC's in (\ref{QC_N=8}) hold. An explicit computation of (\ref{QC_N=8}) for the embedding tensor (\ref{ET_IIA-theory}) yields
\begin{equation}
\label{QC_IIA_O2}
F_{(0)} \, H_{(3)} = 0 \ ,
\end{equation}
so $F_{(0)}$ and $H_{(3)}$ can not be simultaneously turned on in a compactification preserving half-maximal supersymmetry in 3D. The embedding tensor components in (\ref{ET_IIA-theory}) result in the gauge brackets 
\begin{equation}
\begin{array}{rcl}
\left[ X^{mn},  X^{8p} \right] &=& \tfrac{2}{3!} \, \epsilon^{mnpqrst} \, H_{qrs} \, X^{8}{}_{t}  \ , \\[2mm]
\left[ X^{8m} ,  X^{8n} \right] &=& 2\left(\tfrac{1}{4!} \, \epsilon^{mnpqrst} \, F_{pqrs} \, X^{8}{}_{t} - F_{(0)} \, X^{mn}\right)\ ,
\end{array}
\end{equation}
which generically involve the $21+7+7$ generators $\left\lbrace  X^{mn} \,,\, X^{8p} \,,\, X^{8}{}_{t} \right\rbrace$. Solving (\ref{QC_IIA_O2}) by setting $F_{(0)}=0$ trivialises $X^{8}{}_{t}=0$ and the remaining algebra spanned by $\left\lbrace  X^{mn} \,,\, X^{8p} \right\rbrace$ reduces to the $28$-dimensional Abelian gauging of M-theory. On the contrary, solving (\ref{QC_IIA_O2}) by setting $H_{(3)}=0$ does not trivialise any generator and yields a $2$-step nilpotent algebra of dimension $35$ with a $28$-dimensional center.

\subsubsection*{Orientifold interpretation and O$2$-planes}

The half-maximal supergravity models that we have just derived can be realised more string-theoretically in terms of massive type~IIA orientifold reductions including O$2$-planes/D$2$-branes. Three-dimensional Lorentz invariance requires these sources to be located at
\begin{equation}
\label{O2_location}
\begin{array}{lll|lc|lc|lc|c}
x^{0} & x^{1} & x^{2} & y^{1}  & y^{2} & y^{3}  & y^{4}& y^{5} & y^{6} & y^{7}   \\
\hline
\times & \times &\times & &  &  &  &  &  &  
\end{array}   
\end{equation}
thus filling the 3D external spacetime completely. From a group-theoretical viewpoint, the orientifold action $\mathcal{O}_{\mathbb{Z}_{2}}$ in the string theory side is precisely the $\mathbb{Z}_{2}$ symmetry that truncates from maximal to half-maximal supergravity. More specifically, this $\mathbb{Z}_{2}$ projects out those fields and fluxes sitting in spinorial representations of $\textrm{SO}(8,8)$ when branching $\textrm{E}_{8(8)} \supset \textrm{SO}(8,8)$ (see Table~\ref{Table:E8/SO(8,8)/SL8}).

\begin{table}[t]
\begin{center}
\renewcommand{\arraystretch}{1.7}
\begin{tabular}{|c|c|c|c|c|c|c|c|c|}
\hline
Fields &    $e_{n}{}^{p}$ &  $C_{(1)}$ &  $\Phi$ & $C_{(3)}$ &   $B_{(2)}$ & $B_{(6)}$ & $C_{(5)}$ \\
\hline
$\Omega_{P}$ & $+$  & $-$  & $+$ & $+$ & $-$ & $+$ &  $-$  \\
\hline
$\sigma_{\textrm{O}2}$  & $+$  & $-$ & $+$ & $-$ & $+$ & $+$ & $-$    \\
\hline
$\mathcal{O}_{\mathbb{Z}_{2}}$  & $+$  & $+$ & $+$ & $-$ &  $-$ & $+$ & $+$  \\
\hline\hline
Fluxes &    $\omega_{mn}{}^{p}$ &  $F_{(2)}$ &  $H_{(1)}$ & $F_{(4)}$ &   $H_{(3)}$ & $H_{(7)}$ & $F_{(6)}$ \\
\hline  
$\mathcal{O}_{\mathbb{Z}_{2}}$  & $-$  & $-$ & $-$ & $+$ & $+$ & $-$ & $-$  \\
\hline
\end{tabular}
\caption{Grading of type~IIA fields and fluxes under the O$2$-plane orientifold action $\mathcal{O}_{\mathbb{Z}_{2}} = \Omega_{P} \, \sigma_{\textrm{O}2}$. The Romans mass parameter $F_{(0)}$ is not included in the table since it does not originate from a gauge potential.} 
\label{Table:O2_fields_fluxes}
\end{center}
\end{table}

The location of the O$2$-plane in (\ref{O2_location}) is compatible with the $\textrm{SL}(7)$ covariance of the massive type~IIA models. The internal target space involution $\sigma_{\textrm{O}2}$ reflects all the coordinates (and derivatives) on $\mathbb{T}^{7}$ transverse to the O$2$-plane, namely,
\begin{equation}
\label{sigma_O2}
\begin{array}{cc}
\sigma_{\textrm{O}2} : & \hspace{5mm}  y^{m} \,\, \rightarrow\,\,  - \, y^{m} \\[2mm]
 & \hspace{5mm}  \partial_{m} \,\, \rightarrow\,\,  - \, \partial_{m}
\end{array}
\hspace{10mm} \textrm{ with } \hspace{10mm} m = 1 ,\ldots, 7 \ .
\end{equation}
The full orientifold action $\mathcal{O}_{\mathbb{Z}_{2}} = \Omega_{P} \, \sigma_{\textrm{O}2}$ acts on the various type~IIA fields and fluxes as summarised in Table~\ref{Table:O2_fields_fluxes}. Observe that the set of $\mathcal{O}_{\mathbb{Z}_{2}}$-even fields and fluxes in Table~\ref{Table:O2_fields_fluxes} precisely matches the ones in (\ref{Fields&Fluxes_bosonic_IIA}).

The string-theoretic interpretation of the half-maximal supergravity models also allows for a better understanding of the QC's in (\ref{QC_IIA_O2}). The orientifold $\mathcal{O}_{\mathbb{Z}_{2}}$ we are considering here allows for O$2$-planes (and D$2$-branes). As a result, the flux combination $H_{(3)} \wedge F_{(4)}$ entering (\ref{Tadpole_p-form}) can be used to cancel a tadpole for $C_{(5)}$ and, therefore, its value is totally unrestricted from the 3D supergravity perspective. In other words, O$2$-planes/D$2$-branes are compatible with half-maximal supersymmetry in 3D. However, once O$2$-planes/D$2$-branes are present, no additional sources can be added to the compactification scheme without causing a further breaking of supersymmetry. In light of this, the QC in (\ref{QC_IIA_O2}) is nothing but the absence of (a net charge of) O$6$-planes/D$6$-branes. Note also that the cancellation of a potential tadpole for $C_{(3)}$ due to O$4$-planes/D$4$-branes would require $H_{(3)} \wedge F_{(2)} \neq 0$, but this is simply not possible since the orientifold action $\mathcal{O}_{\mathbb{Z}_{2}}$ we are considering projects out $F_{(2)}$ (see Table~\ref{Table:O2_fields_fluxes}). Finally, let us observe that less than half-maximally supersymmetric models including simultaneously O$2$-planes/D$2$-branes and O$6$-planes/D$6$-branes (thus violating the condition (\ref{QC_IIA_O2})) have been investigated in \cite{Farakos:2020phe} and \cite{Farakos:2023nms,Farakos:2023wps}.

\subsection{Type~IIB with O$3$-planes}
\label{sec:O3-plane}

\begin{table}[t!]
\begin{center}
\scalebox{0.76}{
\renewcommand{\arraystretch}{1.7}
\begin{tabular}{|c|c|c|}
\hline 
& Half-Maximal &  ${\rm SL(6)} \times \mathbb{R}_3 \times \mathbb{R}_2 \times \mathbb{R}_1$ \\ 
\hline \hline
 & -- & $ [({\bf{6'}}_{(-1,-1,-1)} \oplus {\bf{1}}_{(+6,-1,-1)})  \oplus {\bf{1}}_{(0,+7,-1)}] \oplus [({\bf{6}}_{(+1,+1,+1)} \oplus {\bf{1}}_{(-6,+1,+1)})  \oplus {\bf{1}}_{(0,-7,+1)}]$ \\
\hline \hline
\multirow{10}{*}{\rotatebox{90}{Scalars}} & \multirow{4}{*}{$L^{MN}$} & $[({\bf 15'}_{(-2,-2,-2)} \oplus {\bf 6'}_{(+5,-2,-2)}) \oplus ({\bf 6'}_{(-1,+6,-2)} \oplus {\bf 1}_{(+6,+6,-2)})]$ \\
 & & ${\color{RoyalBlue}{{\bf 1}_{(0,0,0)}}} \oplus [({\color{ForestGreen}{{\bf 6'}_{(-7,0,0)}}} \oplus {\color{RoyalBlue}{({\bf 35+1})_{(0,0,0)}}}
 \oplus {\bf 6}_{(+7,0,0)}) \oplus {\color{RoyalBlue}{{\bf 1}_{(0,0,0)}}}$ \\
& & $({\color{BrickRed}{{\bf 6'}_{(-1,-8,0)}}} \oplus {\color{BrickRed}{{\bf 1}_{(+6,-8,0)}}}) \oplus ({\bf 6}_{(+1,+8,0)} \oplus {\bf 1}_{(-6,+8,0)})]$ \\ 
&  & $[({\color{ForestGreen}{{\bf 6}_{(+1,-6,+2)}}} \oplus \boxed{{\bf 1}_{(-6,-6,+2)}}) \oplus ({\color{BrickRed}{{\bf 15}_{(+2,+2,+2)}}} \oplus {\color{BrickRed}{{\bf 6}_{(-5,+2,+2)}}} )]$ \\[2mm]
\cline{2-3}
 & \multirow{6}{*}{--} & $[({\bf 6}_{(+1,+1,-3)} \oplus {\bf 1}_{(-6,+1,-3)}) \oplus {\bf 1}_{(0,-7,-3)}]$ \\
& &  $[({\bf 20}_{(+3,+3,-1)} \oplus {\bf 15}_{(-4,+3,-1)}) \oplus ( {\bf 15}_{(+2,-5,-1)} \oplus {\color{RoyalBlue}{{\bf 6}_{(-5,-5,-1)}}}]$ \\
& &  $[ ({\color{ForestGreen}{{\bf 15'}_{(-2,+5,+1)}}} \oplus {\color{RoyalBlue}{{\bf 6'}_{(+5,+5,+1)}}}) \oplus ({\color{BrickRed}{{\bf 20}_{(-3,-3,+1)}}} \oplus {\color{BrickRed}{{\bf 15'}_{(+4,-3,+1)}}}) ]$ \\
& & $[(  \boxed{ {\bf 6'}_{(-1,-1,+3)}} \oplus {\color{ForestGreen}{{\bf 1}_{(+6,-1,+3)}}}) \oplus {\color{BrickRed}{{\bf 1}_{(0,+7,+3)}}}]$ \\[2mm]
\hline
\end{tabular}}
\caption{Type~IIB with O$3$-plane branching rules for the embedding $\textrm{SL}(6) \times \mathbb{R}_3 \times \mathbb{R}_2 \times \mathbb{R}_1 \subset {\rm SL(7)} \times \mathbb{R}_2 \times \mathbb{R}_1$. The subscripts in the third column indicate $(\mathbb{R}_3,\mathbb{R}_2,\mathbb{R}_1)$-charges. We have highlighted the scalars ${\color{RoyalBlue}{e_{m}{}^{n}}}$, ${\color{RoyalBlue}{\Phi}}$, ${\color{ForestGreen}{B_{(2)}}}$, ${\color{ForestGreen}{B_{(6)}}}$ and ${\color{BrickRed}{C_{(p)}}}$ listed in (\ref{scalars_SL(6)_IIB_even})-(\ref{scalars_SL(6)_IIB_odd}). The physical internal derivatives $\tilde{\partial}_{7} \equiv \partial^{7}$ and $\partial_{\mathsf{m}}$ have been put in a box for their quick identification too. This table should be understood as a continuation of Table~\ref{Table:SL8-SL7}.}
\label{Table:SL7-SL6_IIB}
\end{center}
\end{table}

Type~IIB reductions on a (twisted) $\mathbb{T}^{7}$ down to half-maximal supergravity in 3D can be studied similarly to the M-theory and massive type~IIA cases. However, type~IIB reductions come along with an additional complication: there is no orientifold action $\mathcal{O}_{\mathbb{Z}_{2}}$ treating the seven internal coordinates on $\mathbb{T}^{7}$ on equal footing. 

In this section we will consider the orientifold action induced by an O$3$-plane filling the external spacetime and a single one-cycle inside
\begin{equation}
\label{T7_factorisation}
\mathbb{T}^{7} = \mathbb{T}_{1}^{2} \times \mathbb{T}_{2}^{2} \times \mathbb{T}_{3}^{2} \times \mathbb{S}^{1}  \ .
\end{equation}
Taking as a starting point the O$2$-plane of the type~IIA models in the previous section and applying a T-duality along one coordinate -- we denote that coordinate as $y^{7}$ which specifies the $\mathbb{S}^{1}$ in (\ref{T7_factorisation}) -- one is left with an O$3$-plane in the type~IIB theory with a dual type~IIB coordinate $\tilde{y}^{7} \equiv y_{7}$ (notice the change of position of the index). The presence of the O$3$-plane breaks $\textrm{SL}(7)$ covariance down to a subgroup $\textrm{SL}(6) \subset \textrm{SL}(7)$. The relevant branching rules for the type~IIA with O2 $\Leftrightarrow$ type~IIB with O3 correspondence are summarised in Table~\ref{Table:SL7-SL6_IIB}. The original type~IIA internal derivatives $\partial_{m} \in {\bf{7}'}_{(-1,+3)} \in \textrm{SL}(7) \times \mathbb{R}_{2} \times \mathbb{R}_{1}$ in (\ref{partial_SL(8)}) branch under $\textrm{SL}(6) \times \mathbb{R}_{3} \times \mathbb{R}_{2} \times \mathbb{R}_{1}$ as
\begin{equation}
\label{partial_SL(6)}
\begin{array}{rccccc}
{\bf{7}'}_{(-1,+3)} &\rightarrow& {\bf 6'}_{(-1,-1,+3)}  &\oplus& {\bf 1}_{(+6,-1,+3)} & , \\[2mm]
\partial_{m} &\rightarrow&  \partial_{\mathsf{m}} &\oplus& \partial_{7} & .
\end{array}
\end{equation}
Performing a T-duality along the type~IIA coordinate $y^{7}$ replaces $\partial_{7} \rightarrow \tilde{\partial}_{7}$ and leaves $\partial_{\mathsf{m}}$ unaffected. As a result, the physical derivatives in the new type~IIB duality frame differ from the original type~IIA ones and are now identified as
\begin{equation}
\label{partial_IIB_O3}
\tilde{\partial}_{7} \equiv \partial^{7} \in {\bf 1}_{(-6,-6,+2)} 
\hspace{10mm} \textrm{ and } \hspace{10mm}
\partial_{\mathsf{m}} \in {\bf 6'}_{(-1,-1,+3)} \ .
\end{equation}
Unlike for the type~IIA models in the previous section, now the physical derivatives in the type~IIB models have a mixed nature:  $\tilde{\partial}_{7}$ is bosonic ($\mathbb{R}_{1}$ charge $+2$) and $\partial_{\mathsf{m}}$ are spinorial ($\mathbb{R}_{1}$ charge $+3$).

\begin{table}[t]
\begin{center}
\renewcommand{\arraystretch}{1.8}
\begin{tabular}{|c|c|c|}
\hline
Fluxes &  Flux components  &  Embedding tensor   \\
\hline
\hline
\multirow{2}{*}{$\omega$} & $\omega_{7\mathsf{m}}{}^{\mathsf{n}} \in ({\bf{35}}+{\bf{1}})_{(-6,-6,+2)}$ & $\theta^{7\mathsf{n}8}{}_{\mathsf{m}}$  \\
\cline{2-3}
 & $\omega_{\mathsf{m}\mathsf{n}}{}^{7} \in {\bf{15'}}_{(+4,+4,+4)} $  &  $\theta^{\mathsf{p}\mathsf{q}\mathsf{r}\mathsf{s}}$   \\
\hline
\hline
$H_{(1)}$ & $H_{7} \in {\bf{1}}_{(-6,-6,+2)}$  &  $\frac{1}{3\gamma-\alpha+\beta}\theta^{78}$, $\frac{1}{\gamma}\theta^{7\mathsf{m}8}{}{}{}_{\mathsf{m}}$ \\
\hline
\hline
$F_{(1)}$ & $F_{7} \in {\bf{1}}_{(0,-14,+2)}$  &  $\theta^{88} $ \\
\hline
\hline
$H_{(3)}$ & $H_{\mathsf{m}\mathsf{n}\mathsf{p}} \in {\bf{20}}_{(-3,+4,+4)}$  & $-\theta^{\mathsf{q}\mathsf{r}\mathsf{s}7}$  \\
\hline
\hline
$F_{(3)}$ & $F_{\mathsf{m}\mathsf{n}\mathsf{p}} \in {\bf{20}}_{(+3,-4,+4)}$ &  $ \theta^{\mathsf{q}\mathsf{r}\mathsf{s}8}$ \\
\hline
\hline
$F_{(5)}$ & $F_{\mathsf{m}\mathsf{n}\mathsf{p}\mathsf{q}7} \in {\bf{15}}_{(-4,-4,+4)}$ &  $\theta^{\mathsf{r}\mathsf{s}78} $ \\
\hline
\end{tabular}
\caption{Type~IIB fluxes in the O$3$-plane duality frame of half-maximal supergravity and their identification with embedding tensor components. }
\label{Table:O3_fluxes}
\end{center}
\end{table}

Let us move to group-theoretically identify the internal components of the various type~IIB fields. A detailed analysis of bosonic scalars yields
\begin{equation}
\label{scalars_SL(6)_IIB_even}
\begin{array}{rclcrclcrlcl}
e_{\mathsf m}{}^{\mathsf n} &\in& {(\bf{35}+\bf{1})}_{(0,0,0)} & , &  C_{\mathsf{m}\mathsf{n}\mathsf{p}\mathsf{q}\mathsf{r}7} &\in& {\bf{6}}_{(-5,+2,+2)} & , & C_{(0)} &\in& {\bf{1}}_{(+6,-8,0)} & ,
 \\[2mm]
e_{7}{}^{7} &\in& {\bf{1}}_{(0,0,0)} & , &   C_{\mathsf{m}\mathsf{n}\mathsf{p}\mathsf{q}} &\in& {\bf{15}}_{(+2,+2,+2)} & , & B_{\mathsf{m}7} &\in& {\bf{6'}}_{(-7,0,0)} & , \\[2mm]
\Phi &\in& {\bf{1}}_{(0,0,0)}  &  , & C_{\mathsf{m}7} &\in& {\bf{6'}}_{(-1,-8,0)} & , & B_{\mathsf{m}\mathsf{n}\mathsf{p}\mathsf{q}\mathsf{r}7} &\in& {\bf{6}}_{(+1,-6,+2)}  & ,
\end{array}
\end{equation}
where $15$ compact scalars must be subtracted from $e_{\mathsf{m}}{}^{\mathsf{n}}$ upon gauge-fixing of the internal $\textrm{SO}(6)$ local symmetry, namely, $e_{\mathsf{m}}{}^{\mathsf{n}} \in \textrm{GL}(6)/\textrm{SO}(6)$. Note also that there are $64-1=63$ bosonic scalars in (\ref{scalars_SL(6)_IIB_even}), the missing one being dual to the vector field $e_{\mu}{}^{7}$. The spinorial scalars are given by
\begin{equation}
\label{scalars_SL(6)_IIB_odd}
\begin{array}{rcllrclcrclc}
e_{\mathsf{m}}{}^{7} &\in& {\bf{6'}}_{(+5,+5,+1)} & , & e_{7}{}^{\mathsf{m}} &\in& {\bf{6}}_{(-5,-5,-1)} & , & C_{\mathsf{m}\mathsf{n}\mathsf{p}7} &\in& {\bf{20}}_{(-3,-3,+1)} & ,
 \\[2mm]
B_{\mathsf{m}\mathsf{n}} &\in& {\bf{15'}}_{(-2,+5,+1)} & , & B_{\mathsf{m}\mathsf{n}\mathsf{p}\mathsf{q}\mathsf{r}\mathsf{s}} &\in& {\bf{1}}_{(+6,-1,+3)} & , & C_{\mathsf{m}\mathsf{n}} &\in& {\bf{15'}}_{(+4,-3,+1)} & , \\[2mm]
C_{\mathsf{m}\mathsf{n}\mathsf{p}\mathsf{q}\mathsf{r}\mathsf{s}} &\in& {\bf{1}}_{(0,+7,+3)} & ,
\end{array}
\end{equation}
adding up to $64-6=58$. The six missing spinorial scalars are dual to the vector fields $e_{\mu}{}^{\mathsf{m}}$. 
Finally, acting with the partial derivatives in (\ref{partial_IIB_O3}) on the scalar fields (\ref{scalars_SL(6)_IIB_even})-(\ref{scalars_SL(6)_IIB_odd}) produces the set of bosonic fluxes listed in Table~\ref{Table:O3_fluxes} together with additional spinorial ones which are projected out of the half-maximal theory.

\subsubsection*{Orientifold interpretation and O$3$-planes}

\begin{table}[t]
\begin{center}
\renewcommand{\arraystretch}{1.7}
\begin{tabular}{|c|c|c|c|c|}
\hline
Fields & $\Omega_{P} $& $(-1)^{F_{L}}$   &  $\sigma_{\textrm{O}3}$ & $\mathcal{O}_{\mathbb{Z}_{2}}$  \\
\hline
\hline
$e_{\mathsf{m}}{}^{\mathsf{n}} \, , \, e_{7}{}^{7}$ & \multirow{2}{*}{$+$}  & \multirow{2}{*}{$+$}  & $+$  & $+$  \\
\cline{1-1}\cline{4-5}
$e_{\mathsf{m}}{}^{7} \, , \, e_{7}{}^{\mathsf{m}}$ &  & & $-$  & $-$ \\
\hline
\hline
$\Phi$ & $+$ & $+$ & $+$  & $+$ \\
\hline
\hline
$B_{\mathsf{m}\mathsf{n}}$ & \multirow{2}{*}{$-$}  & \multirow{2}{*}{$+$}  & $+$  & $-$  \\
\cline{1-1}\cline{4-5}
$B_{\mathsf{m}7}$ &  &  & $-$  & $+$ \\
\hline
\hline
$B_{\mathsf{m}\mathsf{n}\mathsf{p}\mathsf{q}\mathsf{r}\mathsf{s}}$ & \multirow{2}{*}{$-$}  & \multirow{2}{*}{$+$}  & $+$  & $-$  \\
\cline{1-1}\cline{4-5}
$B_{\mathsf{m}\mathsf{n}\mathsf{p}\mathsf{q}\mathsf{r}7}$ &  &  & $-$  & $+$ \\
\hline
\end{tabular}
\hspace{5mm}
\begin{tabular}{|c|c|c|c|c|}
\hline
Fields & $\Omega_{P} $ & $ (-1)^{F_{L}}$   &  $\sigma_{\textrm{O}3}$ & $\mathcal{O}_{\mathbb{Z}_{2}}$  \\
\hline
\hline
$C_{\mathsf{m}\mathsf{n}\mathsf{p}\mathsf{q}}$ & \multirow{2}{*}{$-$}  & \multirow{2}{*}{$-$}  & $+$  & $+$  \\
\cline{1-1}\cline{4-5}
$C_{\mathsf{m}\mathsf{n}\mathsf{p}7}$ &  &  & $-$  & $-$ \\
\hline
\hline
$C_{(0)}$ & $-$ & $-$ & $+$  & $+$ \\
\hline
\hline
$C_{\mathsf{m}\mathsf{n}}$ & \multirow{2}{*}{$+$}  & \multirow{2}{*}{$-$}  & $+$  & $-$  \\
\cline{1-1}\cline{4-5}
$C_{\mathsf{m}7}$ &  & &  $-$  & $+$ \\
\hline
\hline
$C_{\mathsf{m}\mathsf{n}\mathsf{p}\mathsf{q}\mathsf{r}\mathsf{s}}$ & \multirow{2}{*}{$+$}  & \multirow{2}{*}{$-$}  & $+$  & $-$  \\
\cline{1-1}\cline{4-5}
$C_{\mathsf{m}\mathsf{n}\mathsf{p}\mathsf{q}\mathsf{r}7}$ &  &  & $-$  & $+$ \\
\hline
\end{tabular}
\caption{Grading of the various type~IIB fields under the O$3$-plane orientifold action ${\mathcal{O}_{\mathbb{Z}_{2}} = \Omega_{P} \, (-1)^{F_{L}}\, \sigma_{\textrm{O}3}}$. The $\mathcal{O}_{\mathbb{Z}_{2}}$-even fields match the ones in (\ref{scalars_SL(6)_IIB_even}), whereas the $\mathcal{O}_{\mathbb{Z}_{2}}$-odd fields match the ones in (\ref{scalars_SL(6)_IIB_odd}).} 
\label{Table:O3_fields_fluxes}
\end{center}
\end{table}

The type~IIB flux models that we have just described can be string-theoretically interpreted as type~IIB orientifold reductions including O$3$-planes (and D$3$-branes) placed as
\begin{equation}
\label{O3_location}
\begin{array}{lll|lc|lc|lc|c}
x^{0} & x^{1} & x^{2} & y^{1}  & y^{2} & y^{3}  & y^{4}& y^{5} & y^{6} & \tilde{y}^{7}   \\
\hline
\times & \times &\times & & &  & &  &  &  \times
\end{array}   
\end{equation}
The location of the O$3$-plane in (\ref{O3_location}) is consistent with the $\textrm{SL}(6)$ covariance of the type~IIB models. The internal target space involution $\sigma_{\textrm{O}3}$ now reflects the six coordinates (and derivatives) on the $\mathbb{T}^{6} = \mathbb{T}_{1}^{2} \times \mathbb{T}_{2}^{2} \times \mathbb{T}_{3}^{2}$ in (\ref{T7_factorisation}) transverse to the O$3$-plane, namely,
\begin{equation}
\label{sigma_O3}
\begin{array}{cc}
\sigma_{\textrm{O}3} : & \hspace{5mm}  \tilde{y}^{7} \,\, \rightarrow\,\, \tilde{y}^{7} \hspace{5mm} , \hspace{5mm} y^{\mathsf{m}} \,\, \rightarrow\,\,  - \, y^{\mathsf{m}} \\[2mm]
 & \hspace{5mm}  \tilde{\partial}_{7} \,\, \rightarrow\,\, \tilde{\partial}_{7} \hspace{5mm} , \hspace{5mm}   \partial_{\mathsf{m}} \,\, \rightarrow\,\,  - \, \partial_{\mathsf{m}}
\end{array}
\hspace{6mm} \textrm{ with } \hspace{6mm} 
\mathsf{m} = 1,\ldots, 6 \ ,
\end{equation}
and the full orientifold action $\mathcal{O}_{\mathbb{Z}_{2}} = \Omega_{P} \,(-1)^{F_{L}}\, \sigma_{\textrm{O}3}$ acts on the various type~IIB fields and fluxes as summarised in Table~\ref{Table:O3_fields_fluxes}. Note that the $\mathcal{O}_{\mathbb{Z}_{2}}$-even and $\mathcal{O}_{\mathbb{Z}_{2}}$-odd fields in Table~\ref{Table:O3_fields_fluxes} consistently match the ones in (\ref{scalars_SL(6)_IIB_even}) and (\ref{scalars_SL(6)_IIB_odd}), respectively. The same matching holds at the level of the fluxes.

The string-theoretic realisation of the half-maximal supergravity models again allows us to better understand the QC's restricting the type~IIB fluxes. These include the nilpotency condition $(D^2=0)$ of the $D = d + \omega$ twisted exterior derivative on the internal space, as well as sourceless Bianchi identities of the form $DH_{(1)}=0$, $DF_{(1)} = 0$, $DH_{(3)} +\alpha  H_{(1)}\wedge H_{(3)}= 0$ and $DF_{(3)} +\beta_{3} H_{(1)} \wedge F_{(3)} - H_{(3)} \wedge F_{(1)} = 0$. These follow from the total absence of $7$-branes (the first two), NS$5$-branes (the third one) and $5$-branes (the fourth one) in the compactification scheme. More relevant are the Bianchi identities involving O$3$/D$3$-sources in (\ref{Tadpole_p-form}), namely, $DF_{(5)} - H_{(3)} \wedge F_{(3)} = J_{\textrm{O}3/\textrm{D}3}$. Since the orientifold action we are considering here is generated by O$3$-planes that extend along the internal direction $\tilde{y}^{7}$, we don't expect any QC in the supergravity restricting the number of such sources. Indeed, an explicit computation of the QC's in (\ref{QC_N=8}) using the embedding tensor/flux correspondence in Table~\ref{Table:O3_fluxes} shows that 
\begin{equation}
\left. DF_{(5)} - H_{(3)} \wedge F_{(3)} \, \right|_{dy^{\mathsf{m}_1} \wedge \cdots \wedge \,dy^{\mathsf{m}_6} } = \textrm{unrestricted} \ ,
\end{equation}
while any other component must vanish. In other words, O$3$/D$3$-sources threading the submanifold whose Poincar\'e dual is $dy^{\mathsf{m}_1} \wedge \cdots \wedge \,dy^{\mathsf{m}_6}$ are compatible with the half-maximal supersymmetry of the type~IIB flux models.

\subsection{Type~IIA with O$4$-planes}
\label{sec:O4-plane}

\begin{table}[t!]
\begin{center}
\scalebox{0.71}{
\renewcommand{\arraystretch}{1.7}
\begin{tabular}{|c|c|c|}
\hline 
& Half-Maximal &  ${\rm SL(5)} \times {\rm SL(2)} \times \mathbb{R}_3 \times \mathbb{R}_2 \times \mathbb{R}_1$ \\ 
\hline \hline
 & -- & $ [({\bf{5'}},{\bf{1}})_{(-2,-1,-1)} \oplus ({\bf{1}},{\bf{2}})_{(+5,-1,-1)}  \oplus ({\bf{1},\bf{1}})_{(0,+7,-1)}] \oplus [({\bf{5}},{\bf{1}})_{(+2,+1,+1)} \oplus ({\bf{1}},{\bf{2}})_{(-5,+1,+1)}  \oplus ({\bf{1},\bf{1}})_{(0,-7,+1)}]$ \\
\hline \hline
\multirow{10}{*}{\rotatebox{90}{Scalars}} & \multirow{4}{*}{$L^{MN}$} & $[({\bf (10',1)}_{(-4,-2,-2)} \oplus {\bf (5',2)}_{(+3,-2,-2)} \oplus {\bf (1,1)}_{(+10,-2,-2)}) \oplus ({\bf (5',1)}_{(-2,+6,-2)} \oplus {\bf (1,2)}_{(+5,+6,-2)})]$ \\
 & & ${\color{RoyalBlue}{{\bf (1,1)}_{(0,0,0)}}} \oplus [( {\color{ForestGreen}{{\bf (5',2)}_{(-7,0,0)}}} \oplus {\color{RoyalBlue}{{\bf (24+1,1)}_{(0,0,0)}}}  \oplus {\color{RoyalBlue}{{\bf (1,3+1)}_{(0,0,0)}}} \oplus {\bf (5,2)}_{(+7,0,0)} )$ \\
& & $({\color{BrickRed}{{\bf (5',1)}_{(-2,-8,0)}}} \oplus {\color{BrickRed}{{\bf (1,2)}_{(+5,-8,0)}}}) \oplus ({\bf (5,1)}_{(+2,+8,0)} \oplus {\bf (1,2)}_{(-5,+8,0)})]$ \\ 
&  & $[({\color{ForestGreen}{{\bf (5,1)}_{(+2,-6,+2)}}} \oplus \boxed{{\bf (1,2)}_{(-5,-6,+2)}}) \oplus ({\color{BrickRed}{{\bf (10,1)}_{(+4,+2,+2)}}} \oplus {\color{BrickRed}{{\bf (5,2)}_{(-3,+2,+2)}}} \oplus {\bf (1,1)}_{(-10,+2,+2)})]$ \\[2mm]
\cline{2-3}
 & \multirow{6}{*}{--} & $[({\bf (5,1)}_{(+2,+1,-3)} \oplus {\bf (1,2)}_{(-5,+1,-3)}) \oplus {\bf (1,1)}_{(0,-7,-3)}]$ \\
& &  $[({\bf (5,1)}_{(-8,+3,-1)} \oplus {\bf (10,2)}_{(-1,+3,-1)} \oplus{\bf (10',1)}_{(+6,+3,-1)})$ \\
& &  $( {\color{RoyalBlue}{{\bf (5,2)}_{(-3,-5,-1)}}} \oplus {\bf (10,1)}_{(+4,-5,-1)} \oplus {\color{ForestGreen}{{\bf (1,1)}_{(-10,-5,-1)}}} )]$ \\
& &  $[( {\color{RoyalBlue}{{\bf (5',2)}_{(+3,+5,+1)}}} \oplus {\color{ForestGreen}{{\bf (10',1)}_{(-4,+5,+1)}}} \oplus {\bf (1,1)}_{(+10,+5,+1)} )$ \\
& &  $({\color{BrickRed}{{\bf (5',1)}_{(+8,-3,+1)}}} \oplus {\color{BrickRed}{{\bf (10',2)}_{(+1,-3,+1)}}} \oplus  {\color{BrickRed}{{\bf (10,1)}_{(-6,-3,+1)}}} )]$ \\
& & $[( \boxed{{\bf (5',1)}_{(-2,-1,+3)}} \oplus {\color{ForestGreen}{{\bf (1,2)}_{(+5,-1,+3)}}}) \oplus {\color{BrickRed}{{\bf (1,1)}_{(0,+7,+3)}}}]$ \\[2mm]
\hline
\end{tabular}}
\caption{Type~IIA with O$4$-plane branching rules for the embedding $\textrm{SL}(5) \times \textrm{SL}(2) \times \mathbb{R}_3 \times \mathbb{R}_2 \times \mathbb{R}_1 \subset {\rm SL(7)} \times \mathbb{R}_2 \times \mathbb{R}_1$. The subscripts in the third column indicate $(\mathbb{R}_3,\mathbb{R}_2,\mathbb{R}_1)$-charges. We have highlighted the scalars ${\color{RoyalBlue}{e_{m}{}^{n}}}$, ${\color{RoyalBlue}{\Phi}}$, ${\color{ForestGreen}{B_{(2)}}}$, ${\color{ForestGreen}{B_{(6)}}}$ and ${\color{BrickRed}{C_{(p)}}}$ listed in (\ref{scalars_SL(5)xSL(2)_IIA_even})-(\ref{scalars_SL(5)xSL(2)_IIA_odd}). The physical internal derivatives $\tilde{\partial}_{\mathtt{i}} \equiv \partial^{\mathtt{i}}$ and $\partial_{\hat{\mathtt{a}}}$ have been put in a box for their quick identification too. This table should be understood as a continuation of Table~\ref{Table:SL8-SL7}.}
\label{Table:SL7-SL2xSL5_IIA}
\end{center}
\end{table}

\begin{table}[t]
\begin{center}
\renewcommand{\arraystretch}{1.8}
\begin{tabular}{|c|c|c|}
\hline
Fluxes &  Flux components  &  Embedding tensor   \\
\hline
\hline
\multirow{3}{*}{$\omega$} & $\omega_{\mathtt{i}\mathtt{j}}{}^{\mathtt{j}} \in ({\bf{1},\bf{2}})_{(-5,-6,+2)}$ & $\theta^{\mathtt{i}\mathtt{j}8}{_{\mathtt{j}}}$  \\
\cline{2-3}
 & $\omega_{\mathtt{i}\hat{\mathtt{a}}}{}^{\hat{\mathtt{b}}} \in ({\bf{24}+\bf{1},\bf{2}})_{(-5,-6,+2)} $  &  $-\theta^{\mathtt{i}\hat{\mathtt{b}}8}{}_{\hat{\mathtt{a}}}$   \\
 \cline{2-3}
 & $\omega_{\hat{\mathtt{a}}\hat{\mathtt{b}}}{}^{\mathtt{i}} \in ({\bf{10'},\bf{2}})_{(+1,+4,+4)} $  &  $\theta^{\hat{\mathtt{c}}\hat{\mathtt{d}}\hat{\mathtt{e}}\mathtt{j}}$   \\
\hline
\hline
$H_{(1)}$ & $H_{\mathtt{i}} \in ({\bf{1},\bf{2}})_{(-5,-6,+2)}$  &  $ \frac{1}{2\gamma+\delta-\beta}\theta^{\mathtt{i}8},\frac{1}{\delta}\theta^{\mathtt{i}\mathtt{j}8}{_{\mathtt{j}}},\frac{1}{\gamma}\theta^{\mathtt{i}\hat{\mathtt{a}}8}{}_{\hat{\mathtt{a}}}$ \\
\hline
\multirow{2}{*}{$H_{(3)}$} & $H_{\mathtt{i}\mathtt{j}\hat{\mathtt{a}}} \in ({\bf{5'},\bf{1}})_{(-12,-6,+2)}$  & $\theta^{\mathtt{i}\mathtt{j}8}{}_{\hat{\mathtt{a}}} $  \\
\cline{2-3}
& $H_{\hat{\mathtt{a}}\hat{\mathtt{b}}\hat{\mathtt{c}}} \in ({\bf{10},\bf{1}})_{(-6,+4,+4)} $  &  $\theta^{\hat{\mathtt{d}}\hat{\mathtt{e}}\mathtt{i}\mathtt{j}}$\\
\hline
\hline
\multirow{2}{*}{$F_{(2)}$} & $F_{\mathtt{i}\mathtt{j}} \in ({\bf{1},\bf{1}})_{(0,-14,+2)}$ &  $ \theta^{88}$ \\
\cline{2-3}
& $F_{\hat{\mathtt{a}}\hat{\mathtt{b}}} \in ({\bf{10'},\bf{1}})_{(+6,-4,+4)}$ &  $ -\theta^{\hat{\mathtt{c}}\hat{\mathtt{d}}\hat{\mathtt{e}}8}$ \\
\hline
$F_{(4)}$ & $F_{\mathtt{i}\hat{\mathtt{a}}\hat{\mathtt{b}}\hat{\mathtt{c}}} \in ({\bf{10},\bf{2}})_{(-1,-4,+4)}$ &  $\theta^{\mathtt{i}\hat{\mathtt{d}}\hat{\mathtt{e}}8} $ \\
\hline
$F_{(6)}$ & $F_{\mathtt{i}\mathtt{j}\hat{\mathtt{a}}\hat{\mathtt{b}}\hat{\mathtt{c}}\hat{\mathtt{d}}} \in ({\bf{5},\bf{1}})_{(-8,-4,+4)}$ &  $\theta^{\mathtt{i}\mathtt{j}\hat{\mathtt{e}}8} $ \\
\hline
\end{tabular}
\caption{Type~IIA fluxes in the O$4$-plane duality frame of half-maximal supergravity and their identification with embedding tensor components. Note that the Romans mass is not present in this duality frame as it corresponds to the spinorial representation $F_{(0)} \in {(\bf{1,1})}_{(+10,-9,+3)}$.} 
\label{Table:O4_fluxes}
\end{center}
\end{table}
In this section we will consider the orientifold action induced by an O$4$-plane filling the external spacetime and a single two-cycle inside $\mathbb{T}^{7} = \mathbb{T}^{2}_{1} \times \mathbb{T}^{2}_{2} \times \mathbb{T}^{2}_{3} \times \mathbb{S}^{1}$ in (\ref{T7_factorisation}). We will proceed as before and, starting from the type~IIA models with an O$2$-plane, we will perform two T-dualities along the coordinates in an internal two-torus. Without loss of generality we select the internal two-torus $\mathbb{T}^{2}_{1} \subset \mathbb{T}^{7}$ in (\ref{T7_factorisation}). The presence of the O$4$-plane necessarily breaks the $\textrm{SL}(7)$ covariance down to a subgroup $\textrm{SL}(5) \times \textrm{SL}(2) \subset \textrm{SL}(7)$. The relevant branching rules for the type~IIA with O2 $\Leftrightarrow$ type~IIA with O4 correspondence are summarised in Table~\ref{Table:SL7-SL2xSL5_IIA}. The original type~IIA internal derivatives $\partial_{m} \in {\bf{7}'}_{(-1,+3)} \in \textrm{SL}(7) \times \mathbb{R}_{2} \times \mathbb{R}_{1}$ in (\ref{partial_SL(8)}) now branch under $\textrm{SL}(5) \times \textrm{SL}(2) \times \mathbb{R}_{3} \times \mathbb{R}_{2} \times \mathbb{R}_{1}$ as
\begin{equation}
\label{partial_SL(2)xSL(5)_IIA}
\begin{array}{rccccc}
{\bf{7}'}_{(-1,+3)} &\rightarrow& ({\bf{5'},\bf{1}})_{(-2,-1,+3)}  &\oplus& ({\bf{1},\bf{2}})_{(+5,-1,+3)} & , \\[2mm]
\partial_{m} &\rightarrow&  \partial_{\hat{\mathtt{a}}} &\oplus& \partial_{\mathtt{i}} & ,
\end{array}
\end{equation}
and the physical derivatives in this new type~IIA duality frame are identified as
\begin{equation}
\label{partial_IIA_O4}
\partial_{\hat{\mathtt{a}}} \in ({\bf{5'},\bf{1}})_{(-2,-1,+3)}
\hspace{10mm} \textrm{ and } \hspace{10mm}
\tilde{\partial}_{\mathtt{i}} \equiv \partial^{\mathtt{i}} \in ({\bf{1},\bf{2}})_{(-5,-6,+2)} \ .
\end{equation}
Note again the mixed nature of the physical derivatives in these type~IIA models: $\tilde{\partial}_{\mathtt{i}}$ are bosonic ($\mathbb{R}_{1}$ charge $+2$) and $\partial_{\hat{\mathtt{a}}}$ are spinorial ($\mathbb{R}_{1}$ charge $+3$).

We continue with the group-theoretical identification of the internal components of the various type~IIA fields. The bosonic scalars are identified as
\begin{equation}
\label{scalars_SL(5)xSL(2)_IIA_even}
\begin{array}{rclcrclcrclc}
e_{\mathtt{i}}{}^{\mathtt{j}} &\in& {\bf{(1,3+1)}}_{(0,0,0)} & , & B_{\mathtt{i}\hat{\mathtt{a}}} &\in& {\bf{(5',2)}}_{(-7,0,0)}  & , & C_{\mathtt{i}\mathtt{j}\hat{\mathtt{a}}} &\in& {\bf{(5',1)}}_{(-2,-8,0)}  & 
 \\[2mm]
e_{\hat{\mathtt{a}}}{}^{\hat{\mathtt{b}}} &\in& {\bf{(24+1,1)}}_{(0,0,0)} & , & B_{\mathtt{i}\mathtt{j}\hat{\mathtt{a}}\hat{\mathtt{b}}\hat{\mathtt{c}}\hat{\mathtt{d}}} &\in& {\bf{(5,1)}}_{(+2,-6,+2)}  & , & C_{\hat{\mathtt{a}}\hat{\mathtt{b}}\hat{\mathtt{c}}} &\in& {\bf{(10,1)}}_{(+4,+2,+2)} &   \\[2mm]
\Phi &\in& {\bf{(1,1)}}_{(0,0,0)}  & , & C_{\mathtt{i}} &\in& {\bf{(1,2)}}_{(+5,-8,0)} & , &  C_{\mathtt{i}\hat{\mathtt{a}}\hat{\mathtt{b}}\hat{\mathtt{c}}\hat{\mathtt{d}}} &\in& {\bf{(5,2)}}_{(-3,+2,+2)}   & 
\end{array}
\end{equation}
where $10+1$ compact scalars must be subtracted from $e_{\hat{\mathtt{a}}}{}^{\hat{\mathtt{b}}}$ and $e_{\mathtt{i}}{}^{\mathtt{j}}$ upon gauge-fixing of the internal $\textrm{SO}(5) \times \textrm{SO}(2)$ local symmetry, namely, $e_{\hat{\mathtt{a}}}{}^{\hat{\mathtt{b}}} \in \textrm{GL}(5)/\textrm{SO}(5)$ and $e_{\mathtt{i}}{}^{\mathtt{j}} \in \textrm{GL}(2)/\textrm{SO}(2)$. On the other hand, there are $64-2-1=61$ bosonic scalars in (\ref{scalars_SL(5)xSL(2)_IIA_even}) with the $2+1$ missing scalars being dual to the vectors $e_{\mu}{}^{\mathtt{i}}$ and $C_{\mu}$. The spinorial scalars are identified as
\begin{equation}
\label{scalars_SL(5)xSL(2)_IIA_odd}
\begin{array}{rclcrclcrclc}
e_{\mathtt{i}}{}^{\hat{\mathtt{a}}} &\in& {\bf{(5,2)}}_{(-3,-5,-1)} & , & B_{\hat{\mathtt{a}}\hat{\mathtt{b}}} &\in& {\bf{(10',1)}}_{(-4,+5,+1)}  & , & C_{\mathtt{i}\hat{\mathtt{a}}\hat{\mathtt{b}}} &\in& {\bf{(10',2)}}_{(+1,-3,+1)} & 
 \\[2mm]
e_{\hat{\mathtt{a}}}{}^{\mathtt{i}} &\in& {\bf{(5',2)}}_{(+3,+5,+1)} & , & B_{\mathtt{i}\hat{\mathtt{a}}\hat{\mathtt{b}}\hat{\mathtt{c}}\hat{\mathtt{d}}\hat{\mathtt{e}}} &\in& {\bf{(1,2)}}_{(+5,-1,+3)} &  & C_{\mathtt{i}\mathtt{j}\hat{\mathtt{a}}\hat{\mathtt{b}}\hat{\mathtt{c}}} &\in& {\bf{(10,1)}}_{(-6,-3,+1)} &  \\[2mm]
B_{\mathtt{i}\mathtt{j}} &\in& {\bf{(1,1)}}_{(-10,-5,-1)} & , & C_{\hat{\mathtt{a}}} &\in& {\bf{(5',1)}}_{(+8,-3,+1)}   & , & C_{\hat{\mathtt{a}}\hat{\mathtt{b}}\hat{\mathtt{c}}\hat{\mathtt{d}}\hat{\mathtt{e}}} &\in& {\bf{(1,1)}}_{(0,+7,+3)} & 
\end{array}
\end{equation}
They add up to $64-5=59$ with the $5$ missing scalars being dual to the vector fields $e_{\mu}{}^{\hat{\mathtt{a}}}$. Lastly, an explicit computation gives rise to the bosonic fluxes in Table~\ref{Table:O4_fluxes} together with additional spinorial fluxes which are projected out of the half-maximal theory.

\subsubsection*{Orientifold interpretation and O$4$-planes}

\begin{table}[t]
\begin{center}
\renewcommand{\arraystretch}{1.7}
\begin{tabular}{|c|c|c|c|c|}
\hline
Fields & $\Omega_{P}$ & $(-1)^{F_{L}}$   &  $\sigma_{\textrm{O}4}$ & $\mathcal{O}_{\mathbb{Z}_{2}}$  \\
\hline
\hline
$e_{\mathtt{i}}{}^{\mathtt{j}} \, , \, e_{\hat{\mathtt{a}}}{}^{\hat{\mathtt{b}}}$ & \multirow{2}{*}{$+$}  & \multirow{2}{*}{$+$}  & $+$  & $+$  \\
\cline{1-1}\cline{4-5}
$e_{\mathtt{i}}{}^{\hat{\mathtt{a}}} \, , \, e_{\hat{\mathtt{a}}}{}^{\mathtt{i}}$ &  & & $-$  & $-$ \\
\hline
\hline
$\Phi$ & $+$ & $+$ & $+$  & $+$ \\
\hline
\hline
$B_{\mathtt{i}\mathtt{j}}\, , \, B_{\hat{\mathtt{a}}\hat{\mathtt{b}}}$ & \multirow{2}{*}{$-$} & \multirow{2}{*}{$+$} & $+$  & $-$  \\
\cline{1-1}\cline{4-5}
$B_{\mathtt{i}\hat{\mathtt{a}}}$ &  & & $-$  & $+$ \\
\hline
\hline
$B_{\mathtt{i}\hat{\mathtt{a}}\hat{\mathtt{b}}\hat{\mathtt{c}}\hat{\mathtt{d}}\hat{\mathtt{e}}}$ & \multirow{2}{*}{$+$}  & \multirow{2}{*}{$+$} & $-$  & $-$  \\
\cline{1-1}\cline{4-5}
$B_{\mathtt{i}\mathtt{j}\hat{\mathtt{a}}\hat{\mathtt{b}}\hat{\mathtt{c}}\hat{\mathtt{d}}}$ & & & $+$  & $+$ \\
\hline
\end{tabular}
\hspace{5mm}
\begin{tabular}{|c|c|c|c|c|}
\hline
Fields & $\Omega_{P}$ & $(-1)^{F_{L}}$   &  $\sigma_{\textrm{O}4}$ & $\mathcal{O}_{\mathbb{Z}_{2}}$  \\
\hline
\hline
$C_{\hat{\mathtt{a}}}$ & \multirow{2}{*}{$-$} & \multirow{2}{*}{$-$}  & $-$  & $-$  \\
\cline{1-1}\cline{4-5}
$C_{\mathtt{i}}$ &  & &  $+$  & $+$ \\
\hline
\hline
$C_{\mathtt{i}\mathtt{j}\hat{\mathtt{a}}}\, , \, C_{\hat{\mathtt{a}}\hat{\mathtt{b}}\hat{\mathtt{c}}}$ & \multirow{2}{*}{$+$}  & \multirow{2}{*}{$-$}  & $-$  & $+$  \\
\cline{1-1}\cline{4-5}
$C_{\mathtt{i}\hat{\mathtt{a}}\hat{\mathtt{b}}}$ &  &  & $+$  & $-$ \\
\hline
\hline
$C_{\mathtt{i}\hat{\mathtt{a}}\hat{\mathtt{b}}\hat{\mathtt{c}}\hat{\mathtt{d}}}$ & \multirow{2}{*}{$-$}  &  \multirow{2}{*}{$-$}  &  $+$  & $+$  \\
\cline{1-1}\cline{4-5}
$C_{\mathtt{i}\mathtt{j}\hat{\mathtt{a}}\hat{\mathtt{b}}\hat{\mathtt{c}}} \, , \, C_{\hat{\mathtt{a}}\hat{\mathtt{b}}\hat{\mathtt{c}}\hat{\mathtt{d}}\hat{\mathtt{e}}}$ & & & $-$  & $-$ \\
\hline
\end{tabular}
\caption{Grading of the various type~IIA fields under the O$4$-plane orientifold action ${\mathcal{O}_{\mathbb{Z}_{2}} = \Omega_{P} \, (-1)^{F_{L}} \, \sigma_{\textrm{O}4}}$. The  $\mathcal{O}_{\mathbb{Z}_{2}}$-even fields match the ones in (\ref{scalars_SL(5)xSL(2)_IIA_even}), whereas the $\mathcal{O}_{\mathbb{Z}_{2}}$-odd fields match the ones in (\ref{scalars_SL(5)xSL(2)_IIA_odd}).} 
\label{Table:O4_fields}
\end{center}
\end{table}

The above class of type~IIA flux models can be string-theoretically interpreted as type~IIA orientifold reductions including O$4$-planes (and D$4$-branes) placed as
\begin{equation}
\label{O4_location}
\begin{array}{lll|lc|lc|lc|c}
x^{0} & x^{1} & x^{2} & \tilde{y}^{1}  & \tilde{y}^{2} & y^{3}  & y^{4}& y^{5} & y^{6} & y^{7}   \\
\hline
\times & \times & \times & \times & \times &  &  &  &  & 
\end{array}   
\end{equation}
The location of the O$4$-plane in (\ref{O4_location}) is this time compatible with the $\textrm{SL}(5) \times \textrm{SL}(2)$ covariance of the type~IIA models. The internal target space involution $\sigma_{\textrm{O}4}$ reflects the five coordinates (and derivatives) on the $\mathbb{T}_{2}^{2} \times \mathbb{T}_{3}^{2} \times \mathbb{S}^{1} \subset \mathbb{T}^{7}$ in (\ref{T7_factorisation}) transverse to the O$4$-plane. This is
\begin{equation}
\label{sigma_O4}
\begin{array}{cc}
\sigma_{\textrm{O}4} : & \hspace{5mm}  \tilde{y}^{\mathtt{i}} \,\, \rightarrow\,\, \tilde{y}^{\mathtt{i}} \hspace{5mm} , \hspace{5mm} y^{\hat{\mathtt{a}}} \,\, \rightarrow\,\,  - \, y^{\hat{\mathtt{a}}} \\[2mm]
 & \hspace{5mm}  \tilde{\partial}_{\mathtt{i}} \,\, \rightarrow\,\, \tilde{\partial}_{\mathtt{i}} \hspace{5mm} , \hspace{5mm}   \partial_{\hat{\mathtt{a}}} \,\, \rightarrow\,\,  - \, \partial_{\hat{\mathtt{a}}}
\end{array}
\hspace{6mm} \textrm{ with } \hspace{6mm} 
\mathtt{i} = 1,2
\hspace{4mm} , \hspace{4mm}
\hat{\mathtt{a}} = 3,\ldots, 7 \ .
\end{equation}
At the level of the type~IIA fields, the full orientifold action $\mathcal{O}_{\mathbb{Z}_{2}} = \Omega_{P} \,(-1)^{F_{L}}\, \sigma_{\textrm{O}4}$ acts as displayed in Table~\ref{Table:O4_fields}. As a check of consistency, the $\mathcal{O}_{\mathbb{Z}_{2}}$-even and $\mathcal{O}_{\mathbb{Z}_{2}}$-odd fields in Table~\ref{Table:O4_fields} are in one-to-one correspondence with the scalars in (\ref{scalars_SL(5)xSL(2)_IIA_even}) and (\ref{scalars_SL(5)xSL(2)_IIA_odd}), respectively. The same matching holds also at the level of the fluxes.

The set of QC's that descends from (\ref{QC_N=8}) upon plugging the embedding tensor/flux dictionary in Table~\ref{Table:O4_fluxes} is interpreted as follows. There is the nilpotency condition $(D^2=0)$ of the $D = d + \omega$ twisted exterior derivative on the internal space, as well as sourceless Bianchi identities of the form $DH_{(1)} = 0$ and $DH_{(3)} +\alpha H_{(1)}\wedge H_{(3)} = 0$ that follow from the total absence of NS$7$-branes (the first one) and NS$5$-branes (the second one) in the compactification scheme. In addition, there are Bianchi identities of the form $DF_{(2)} + \beta_{2}H_{(1)} \wedge F_{(2)} = 0$ and $DF_{(6)} = 0$ reflecting the absence of D$6$-branes and D$2$-branes, respectively. Lastly, there is a non-trivial Bianchi identity (\ref{Tadpole_p-form}) of the form $DF_{(4)} +\beta_{4} H_{(1)} \wedge F_{(4)} - H_{(3)} \wedge F_{(2)} = J_{\textrm{O}4/\textrm{D}4}$ involving O$4$/D$4$-sources. Since the orientifold action is generated by O$4$-planes that extend along the internal two-torus $\mathbb{T}_{1}^{2}$ in (\ref{T7_factorisation}), the QC's in (\ref{QC_N=8}) yield
\begin{equation}
\left. DF_{(4)}  - H_{(3)} \wedge F_{(2)} \, \right|_{dy^{\hat{\mathtt{a}}_1} \wedge \cdots \wedge \,dy^{\hat{\mathtt{a}}_5} } = \textrm{unrestricted} \ ,
\end{equation}
while any other component vanishes. Equivalently, O$4$/D$4$-sources threading the submanifold whose Poincar\'e dual is $dy^{\hat{\mathtt{a}}_1} \wedge \cdots \wedge \,dy^{\hat{\mathtt{a}}_5}$ are compatible with the half-maximal supersymmetry of the type~IIA flux models under inspection.

\subsection{Type~IIB with O$5$-planes}
\label{sec:O5-plane}

\begin{table}[t!]
\begin{center}
\scalebox{0.76}{
\renewcommand{\arraystretch}{1.7}
\begin{tabular}{|c|c|c|}
\hline 
& Half-Maximal &  ${\rm SL(4)} \times {\rm SL(3)} \times \mathbb{R}_3 \times \mathbb{R}_2 \times \mathbb{R}_1$ \\ 
\hline \hline
 & -- & $ [({\bf{4'}},{\bf{1}})_{(-3,-1,-1)} \oplus ({\bf{1}},{\bf{3'}})_{(+4,-1,-1)}  \oplus {\bf{1}}_{(0,+7,-1)}] \oplus [({\bf{4}},{\bf{1}})_{(+3,+1,+1)} \oplus ({\bf{1}},{\bf{3}})_{(-4,+1,+1)}  \oplus {\bf{1}}_{(0,-7,+1)}]$ \\
\hline \hline
\multirow{10}{*}{\rotatebox{90}{Scalars}} & \multirow{4}{*}{$L^{MN}$} & $[({\bf (6,1)}_{(-6,-2,-2)} \oplus {\bf (4',3')}_{(+1,-2,-2)} \oplus {\bf (1,3)}_{(+8,-2,-2)}) \oplus ({\bf (4',1)}_{(-3,+6,-2)} \oplus {\bf (1,3')}_{(+4,+6,-2)})]$ \\
 & & ${\color{RoyalBlue}{{\bf (1,1)}_{(0,0,0)}}} \oplus [( {\color{ForestGreen}{{\bf (4',3)}_{(-7,0,0)}}} \oplus {\color{RoyalBlue}{{\bf (15+1,1)}_{(0,0,0)}}}  \oplus {\color{RoyalBlue}{{\bf (1,8+1)}_{(0,0,0)}}} \oplus {\bf (4,3')}_{(+7,0,0)} )$ \\
& & $({\color{BrickRed}{{\bf (4',1)}_{(-3,-8,0)}}} \oplus {\color{BrickRed}{{\bf (1,3')}_{(+4,-8,0)}}}) \oplus ({\bf (4,1)}_{(+3,+8,0)} \oplus {\bf (1,3)}_{(-4,+8,0)})]$ \\ 
&  & $[({\color{ForestGreen}{{\bf (4,1)}_{(+3,-6,+2)}}} \oplus \boxed{{\bf (1,3)}_{(-4,-6,+2)})} \oplus ({\color{BrickRed}{{\bf (6,1)}_{(+6,+2,+2)}}} \oplus {\color{BrickRed}{{\bf (4,3)}_{(-1,+2,+2)}}} \oplus {\color{BrickRed}{{\bf (1,3')}_{(-8,+2,+2)}}})]$ \\[2mm]
\cline{2-3}
 & \multirow{6}{*}{--} & $[({\bf (4,1)}_{(+3,+1,-3)} \oplus {\bf (1,3)}_{(-4,+1,-3)}) \oplus {\bf (1,1)}_{(0,-7,-3)}]$ \\
& &  $[({\bf (4,3')}_{(-5,+3,-1)} \oplus {\bf (6,3)}_{(+2,+3,-1)} \oplus{\bf (4',1)}_{(+9,+3,-1)} \oplus{\bf (1,1)}_{(-12,+3,-1)})$ \\
& &  $( {\color{RoyalBlue}{{\bf (4,3)}_{(-1,-5,-1)}}} \oplus {\bf (6,1)}_{(+6,-5,-1)} \oplus {\color{ForestGreen}{{\bf (1,3')}_{(-8,-5,-1)}}} )]$ \\
& &  $( {\color{RoyalBlue}{{\bf (4',3')}_{(+1,+5,+1)}}} \oplus {\color{ForestGreen}{{\bf (6,1)}_{(-6,+5,+1)}}} \oplus {\bf (1,3)}_{(+8,+5,+1)} )]$ \\
& &  $[({\color{BrickRed}{{\bf (4',3)}_{(+5,-3,+1)}}} \oplus {\color{BrickRed}{{\bf (6,3')}_{(-2,-3,+1)}}} \oplus  {\color{BrickRed}{{\bf (4,1)}_{(-9,-3,+1)}}} \oplus   {\color{BrickRed}{{\bf (1,1)}_{(+12,-3,+1)})}}$ \\
& & $[(  \boxed{{\bf (4',1)}_{(-3,-1,+3)}} \oplus {\color{ForestGreen}{{\bf (1,3')}_{(+4,-1,+3)}}}) \oplus {\color{BrickRed}{{\bf (1,1)}_{(0,+7,+3)}}}]$ \\[2mm]
\hline
\end{tabular}}
\caption{Type~IIB with O$5$-plane branching rules for the embedding $\textrm{SL}(4) \times \textrm{SL}(3) \times \mathbb{R}_3 \times \mathbb{R}_2 \times \mathbb{R}_1 \subset {\rm SL(7)} \times \mathbb{R}_2 \times \mathbb{R}_1$. The subscripts in the third column indicate $(\mathbb{R}_3,\mathbb{R}_2,\mathbb{R}_1)$-charges. We have highlighted the scalars ${\color{RoyalBlue}{e_{m}{}^{n}}}$, ${\color{RoyalBlue}{\Phi}}$, ${\color{ForestGreen}{B_{(2)}}}$, ${\color{ForestGreen}{B_{(6)}}}$ and ${\color{BrickRed}{C_{(p)}}}$ listed in (\ref{scalars_SL(4)xSL(3)_IIB_even})-(\ref{scalars_SL(4)xSL(3)_IIB_odd}). The physical internal derivatives $\tilde{\partial}_{i} \equiv \partial^{i}$ and $\partial_{\hat{a}}$ have been put in a box for their quick identification too. This table should be understood as a continuation of Table~\ref{Table:SL8-SL7}.}
\label{Table:SL7-SL4xSL3}
\end{center}
\end{table}

Let us continue with the orientifold action induced by an O$5$-plane filling the external spacetime and a three-cycle inside $\mathbb{T}^{7}$. Taking again as a starting point the O$2$-plane of the type~IIA models but now applying three T-dualities along three type~IIA coordinates $y^{i}$, one is left with an O$5$-plane in the type~IIB theory with dual type~IIB coordinates $\tilde{y}^{i} \equiv y_{i}$ (notice the position change of the index $i$). The presence of the O$5$-plane breaks the $\textrm{SL}(7)$ covariance down to a subgroup $\textrm{SL}(4) \times \textrm{SL}(3) \subset \textrm{SL}(7)$. The relevant branching rules for the type~IIA with O2 $\Leftrightarrow$ type~IIB with O5 correspondence are summarised in Table~\ref{Table:SL7-SL4xSL3}. The original type~IIA internal derivatives $\partial_{m} \in {\bf{7}'}_{(-1,+3)} \in \textrm{SL}(7) \times \mathbb{R}_{2} \times \mathbb{R}_{1}$ in (\ref{partial_SL(8)}) branch under $\textrm{SL}(4) \times  \textrm{SL}(3) \times \mathbb{R}_{3} \times \mathbb{R}_{2} \times \mathbb{R}_{1}$ as
\begin{equation}
\label{partial_SL(4)xSL(3)}
\begin{array}{rccccc}
{\bf{7}'}_{(-1,+3)} &\rightarrow& {\bf (1,3')}_{(+4,-1,+3)}  &\oplus& {\bf (4',1)}_{(-3,-1,+3)} & , \\[2mm]
\partial_{m} &\rightarrow&  \partial_{i} &\oplus& \partial_{\hat{a}} & ,
\end{array}
\end{equation}
where, without loss of generality, we take $i=2,4,6$ and $\hat{a}=1,3,5,7$. Performing three T-dualities along the type~IIA coordinates $y^{i}$ replaces $\partial_{i} \rightarrow \tilde{\partial}_{i}$ and leaves $\partial_{\hat{a}}$ unaffected. The physical derivatives in the new type~IIB duality frame differ from the original type~IIA ones and are identified as
\begin{equation}
\label{partial_IIB}
\tilde{\partial}_{i} \equiv \partial^{i} \in {\bf (1,3)}_{(-4,-6,+2)} 
\hspace{10mm} \textrm{ and } \hspace{10mm}
\partial_{\hat{a}} \in {\bf (4',1)}_{(-3,-1,+3)} \ .
\end{equation}
Note the maximally-mixed nature of the physical derivatives in these type~IIB models:  $\tilde{\partial}_{i}$ are bosonic ($\mathbb{R}_{1}$ charge $+2$) whereas $\partial_{\hat{a}}$ are spinorial ($\mathbb{R}_{1}$ charge $+3$). This will translate into a more diverse set of type~IIB fluxes, as compared to the M-theory and previous type~IIA cases, and also a richer structure of flux vacua.

\begin{table}[t]
\begin{center}
\renewcommand{\arraystretch}{1.8}
\begin{tabular}{|c|c|c|}
\hline
Fluxes &  Flux components  &  Embedding tensor   \\
\hline
\hline
\multirow{3}{*}{$\omega$} & $\omega_{ij}{}^{k} \in {\bf{(1,6+3)}}_{(-4,-6,+2)}$ & $\theta^{ij8}{}_{k}$  \\
\cline{2-3}
 & $\omega_{\hat{a}\hat{b}}{}^{i} \in {\bf{(6,3')}}_{(-2,+4,+4)} $  &  $\theta^{\hat{c}\hat{d}jk}$   \\
\cline{2-3}
 & $ \omega_{\hat{a}i}{}^{\hat{b}} \in {\bf{(15+1,3)}}_{(-4,-6,+2)}$ &  $\theta^{i\hat{b}8}{}_{\hat{a}}$ \\
\hline
\hline
$H_{(1)}$ & $H_{i} \in {\bf{(1,3)}}_{(-4,-6,+2)}$  &  $\frac{1}{3\gamma+2\alpha-\beta}\theta^{i8},\frac{1}{\gamma+\alpha}\theta^{ij8}{}_{j},\frac{1}{\gamma}\theta^{i\hat{a}8}{}_{\hat{a}}$ \\
\hline
\hline
$F_{(1)}$ & $F_{\hat{a}} \in {\bf{(4',1)}}_{(+9,-4,+4)}$  &  $-\theta^{\hat{b}\hat{c}\hat{d}8}$ \\
\hline
\hline
\multirow{2}{*}{$H_{(3)}$} & $H_{\hat{a}\hat{b}\hat{c}} \in {\bf{(4,1)}}_{(-9,+4,+4)}$  & $-\theta^{\hat{d}ijk}$  \\
\cline{2-3}
 & $H_{ij\hat{c}} \in {\bf{(4',3')}}_{(-11,-6,+2)} $ & $\theta^{ij8}{}_{\hat{c}}$ \\
\hline
\hline
\multirow{2}{*}{$F_{(3)}$} & $F_{ijk} \in {\bf{(1,1)}}_{(0,-14,+2)}$ &  $-\theta^{88}$ \\
\cline{2-3}
 & $F_{\hat{a}\hat{b}k} \in {\bf{(6,3)}}_{(+2,-4,+4)}$ &  $\theta^{\hat{c}\hat{d}k8}$ \\
\hline
\hline
$F_{(5)}$ & $F_{\hat{a}\hat{b}\hat{c}ij} \in {\bf{(4,3')}}_{(-5,-4,+4)}$ &  $\theta^{\hat{d}ij8}$ \\
\hline
\hline
$F_{(7)}$ & $F_{\hat{a}\hat{b}\hat{c}\hat{d}ijk} \in {\bf{(1,1)}}_{(-12,-4,+4)}$ &  $-\theta^{ijk8}$ \\
\hline
\end{tabular}
\caption{Type~IIB fluxes in the O$5$-plane duality frame of half-maximal supergravity and their identification with embedding tensor components.} 
\label{Table:O5_fluxes}
\end{center}
\end{table}

\begin{table}[t]
\begin{center}
\renewcommand{\arraystretch}{1.7}
\begin{tabular}{|c|c|c|c|}
\hline
Fields & $\Omega_{P}$   &  $\sigma_{\textrm{O}5}$ & $\mathcal{O}_{\mathbb{Z}_{2}}$  \\
\hline
\hline
$e_{i}{}^{j} \, , \, e_{\hat{a}}{}^{\hat{b}}$ & \multirow{2}{*}{$+$}  & $+$  & $+$  \\
\cline{1-1}\cline{3-4}
$e_{i}{}^{\hat{a}} \, , \, e_{\hat{a}}{}^{i}$ &  & $-$  & $-$ \\
\hline
\hline
$\Phi$ & $+$ & $+$  & $+$ \\
\hline
\hline
$B_{ij}\, , \, B_{\hat{a}\hat{b}}$ & \multirow{2}{*}{$-$}  & $+$  & $-$  \\
\cline{1-1}\cline{3-4}
$B_{i\hat{a}}$ &  & $-$  & $+$ \\
\hline
\hline
$B_{ij\hat{a}\hat{b}\hat{c}\hat{d}}$ & \multirow{2}{*}{$-$}  & $+$  & $-$  \\
\cline{1-1}\cline{3-4}
$B_{ijk\hat{a}\hat{b}\hat{c}}$ &  & $-$  & $+$ \\
\hline
\end{tabular}
\hspace{10mm}
\begin{tabular}{|c|c|c|c|}
\hline
Fields & $\Omega_{P}$   &  $\sigma_{\textrm{O}5}$ & $\mathcal{O}_{\mathbb{Z}_{2}}$  \\
\hline
\hline
$C_{\hat{a}\hat{b}\hat{c}\hat{d}} \, , \, C_{ij\hat{a}\hat{b}}$ & \multirow{2}{*}{$-$}  & $+$  & $-$  \\
\cline{1-1}\cline{3-4}
$C_{i\hat{a}\hat{b}\hat{c}} \, , \, C_{ijk\hat{a} }$ &  & $-$  & $+$ \\
\hline
\hline
$C_{(0)}$ & $-$ & $+$  & $-$ \\
\hline
\hline
$C_{ij}\, , \, C_{\hat{a}\hat{b}}$ & \multirow{2}{*}{$+$}  & $+$  & $+$  \\
\cline{1-1}\cline{3-4}
$C_{i\hat{a}}$ &  & $-$  & $-$ \\
\hline
\hline
$C_{ij\hat{a}\hat{b}\hat{c}\hat{d}}$ & \multirow{2}{*}{$+$}  & $+$  & $+$  \\
\cline{1-1}\cline{3-4}
$C_{ijk\hat{a}\hat{b}\hat{c}}$ &  & $-$  & $-$ \\
\hline
\end{tabular}
\caption{Grading of the type~IIB fields under the O$5$-plane orientifold action $\mathcal{O}_{\mathbb{Z}_{2}} = \Omega_{P} \, \sigma_{\textrm{O}5}$. $\mathcal{O}_{\mathbb{Z}_{2}}$-even fields match the ones in (\ref{scalars_SL(4)xSL(3)_IIB_even}), whereas $\mathcal{O}_{\mathbb{Z}_{2}}$-odd fields match the ones in (\ref{scalars_SL(4)xSL(3)_IIB_odd}).} 
\label{Table:O5_fields_fluxes}
\end{center}
\end{table}
The next step is to group-theoretically identify the internal components of the various type~IIB fields. A careful analysis shows that they are given by
\begin{equation}
\label{scalars_SL(4)xSL(3)_IIB_even}
\begin{array}{rclcrclc}
e_{i}{}^{j} &\in& {\bf{(1,8+1)}}_{(0,0,0)} & \hspace{3mm} , & \hspace{3mm} C_{i\hat{a}\hat{b}\hat{c}} &\in& {\bf{(4,3)}}_{(-1,+2,+2)} & ,
 \\[2mm]
e_{\hat{a}}{}^{\hat{b}} &\in& {\bf{(15+1,1)}}_{(0,0,0)} & \hspace{3mm} , & \hspace{3mm} C_{ijk\hat{a}} &\in& {\bf{(4',1)}}_{(-3,-8,0)} & ,  \\[2mm]
\Phi &\in& {\bf{(1,1)}}_{(0,0,0)}  & \hspace{3mm} , & \hspace{3mm} C_{ij} &\in& {\bf{(1,3')}}_{(+4,-8,0)} & ,
 \\[2mm]
B_{i\hat{a}} &\in& {\bf{(4',3)}}_{(-7,0,0)}  & \hspace{3mm} , & \hspace{3mm} C_{\hat{a}\hat{b}} &\in& {\bf{(6,1)}}_{(+6,+2,+2)} & , \\[2mm]
B_{ijk\hat{a}\hat{b}\hat{c}} &\in& {\bf{(4,1)}}_{(+3,-6,+2)} & \hspace{3mm} , & \hspace{3mm} C_{ij\hat{a}\hat{b}\hat{c}\hat{d}} &\in& {\bf{(1,3')}}_{(-8,+2,+2)} & ,
\end{array}
\end{equation}
where $3+6$ compact scalars must be subtracted from $e_{i}{}^{j}$ and $e_{\hat{a}}{}^{\hat{b}}$ upon gauge-fixing of the internal $\textrm{SO}(3) \times \textrm{SO}(4)$ local symmetry, namely, $e_{i}{}^{j} \in \textrm{GL}(3)/\textrm{SO}(3)$ and $e_{\hat{a}}{}^{\hat{b}} \in \textrm{GL}(4)/\textrm{SO}(4)$. The spinorial scalars are identified as
\begin{equation}
\label{scalars_SL(4)xSL(3)_IIB_odd}
\begin{array}{rclcrclc}
e_{i}{}^{\hat{a}} &\in& {\bf{(4,3)}}_{(-1,-5,-1)} & \hspace{3mm} , & \hspace{3mm} C_{\hat{a}\hat{b}\hat{c}\hat{d}} &\in& {\bf{(1,1)}}_{(0,+7,+3)} & ,
 \\[2mm]
e_{\hat{a}}{}^{i} &\in& {\bf{(4',3')}}_{(+1,+5,+1)} & \hspace{3mm} , & \hspace{3mm} C_{ij\hat{a}\hat{b}} &\in& {\bf{(6,3')}}_{(-2,-3,+1)} & , \\[2mm]
B_{ij} &\in& {\bf{(1,3')}}_{(-8,-5,-1)}  & \hspace{3mm} , & \hspace{3mm} C_{(0)} &\in& {\bf{(1,1)}}_{(+12,-3,+1)} & ,
 \\[2mm]
B_{\hat{a}\hat{b}} &\in& {\bf{(6,1)}}_{(-6,+5,+1)}   & \hspace{3mm} , & \hspace{3mm} C_{i\hat{a}} &\in& {\bf{(4',3)}}_{(+5,-3,+1)} & , \\[2mm]
B_{ij\hat{a}\hat{b}\hat{c}\hat{d}} &\in& {\bf{(1,3')}}_{(+4,-1,+3)} & \hspace{3mm} , & \hspace{3mm} C_{ijk\hat{a}\hat{b}\hat{c}} &\in& {\bf{(4,1)}}_{(-9,-3,+1)}   & .
\end{array}
\end{equation}
In order to add up to $64$, the $64-3=61$ bosonic scalars (even $\mathbb{R}_{1}$-charges) in (\ref{scalars_SL(4)xSL(3)_IIB_even}) must be completed with $3$ additional scalars dual to the vectors $e_{\mu}{}^{i}$. Similarly, the $64-4=60$ spinorial scalars (odd $\mathbb{R}_{1}$-charges) in (\ref{scalars_SL(4)xSL(3)_IIB_odd}) must be completed with $4$ additional scalars dual to the vectors $e_{\mu}{}^{\hat{a}}$. Since there are both bosonic and spinorial physical derivatives in (\ref{partial_IIB}), the set of type~IIB fluxes that appear when acting upon the scalars (\ref{scalars_SL(4)xSL(3)_IIB_even})-(\ref{scalars_SL(4)xSL(3)_IIB_odd}) is very diverse. An explicit computation yields the bosonic fluxes in Table~\ref{Table:O5_fluxes} together with additional spinorial fluxes which are projected out of the half-maximal theory. The fluxes in Table~\ref{Table:O5_fluxes} therefore specify the embedding tensor (\ref{ET_N=8}) and gauge brackets (\ref{brackets_N=8}) of the type~IIB half-maximal supergravity models under consideration.

\subsubsection*{Orientifold interpretation and O$5$-planes}

These type~IIB flux models can be string-theoretically interpreted as type~IIB orientifold reductions including O$5$-planes (and D$5$-branes) located as
\begin{equation}
\label{O5_location}
\begin{array}{lll|lc|lc|lc|c}
x^{0} & x^{1} & x^{2} & y^{1}  & \tilde{y}^{2} & y^{3}  & \tilde{y}^{4}& y^{5} & \tilde{y}^{6} & y^{7}   \\
\hline
\times & \times &\times & & \times &  & \times &  & \times &  
\end{array}   
\end{equation}
The O$5$-plane is filling the external spacetime together with the three internal directions $\tilde{y}^{i}$ with $i=2,4,6$. The internal target space involution $\sigma_{\textrm{O}5}$ then reflects the four transverse coordinates (and their derivatives) on $\mathbb{T}^{7}$, namely,
\begin{equation}
\label{sigma_O5}
\begin{array}{cc}
\sigma_{\textrm{O}5} : & \hspace{5mm}  \tilde{y}^{i} \,\, \rightarrow\,\, \tilde{y}^{i} \hspace{5mm} , \hspace{5mm} y^{\hat{a}} \,\, \rightarrow\,\,  - \, y^{\hat{a}} \\[2mm]
 & \hspace{5mm}  \tilde{\partial}_{i} \,\, \rightarrow\,\, \tilde{\partial_{i}} \hspace{5mm} , \hspace{5mm}   \partial_{\hat{a}} \,\, \rightarrow\,\,  - \, \partial_{\hat{a}}
\end{array}
\hspace{6mm} \textrm{ with } \hspace{6mm} 
i = 2,4,6
\hspace{4mm} , \hspace{4mm}
\hat{a} = 1,3,5,7 \ .
\end{equation}
The corresponding orientifold action $\mathcal{O}_{\mathbb{Z}_{2}} = \Omega_{P} \, \sigma_{\textrm{O}5}$ acts on the type~IIB fields as displayed in Table~\ref{Table:O5_fields_fluxes}. As expected, the $\mathcal{O}_{\mathbb{Z}_{2}}$-even and $\mathcal{O}_{\mathbb{Z}_{2}}$-odd fields in Table~\ref{Table:O5_fields_fluxes} precisely match the ones in (\ref{scalars_SL(4)xSL(3)_IIB_even}) and (\ref{scalars_SL(4)xSL(3)_IIB_odd}), respectively. The same matching holds at the level of the fluxes.

The string-theoretic realisation of the half-maximal supergravity models again allows us to better understand the QC's restricting the type~IIB fluxes. These include the nilpotency condition $(D^2=0)$ of the $D = d + \omega$ twisted exterior derivative on the internal space, as well as sourceless Bianchi identities of the form $DH_{(1)} = 0$, $DF_{(1)} +\beta_{1} H_{(1)} \wedge F_{(1)} = 0$, $DH_{(3)} + \alpha H_{(1)}\wedge H_{(3)}= 0$ and $DF_{(5)} +\beta_{5} H_{(1)} \wedge F_{(5)} - H_{(3)} \wedge F_{(3)} = 0$. These follow from the total absence of $7$-branes (the first two), NS$5$-branes (the third one) and D$3$-branes (the fourth one) in the compactification scheme. More relevant are the Bianchi identities involving O$5$/D$5$-sources in (\ref{Tadpole_p-form}), namely, $DF_{(3)} +\beta_{3} {H_{(1)} \wedge F_{(3)}} - {H_{(3)} \wedge F_{(1)}} = J_{\textrm{O}5/\textrm{D}5}$. Since the orientifold action we are considering here is generated by O$5$-planes that extend along the three internal directions $\tilde{y}^{i}$, we don't expect any QC in the supergravity restricting the number of such sources. Indeed, an explicit computation of the QC's in (\ref{QC_N=8}) using the embedding tensor/flux correspondence in Table~\ref{Table:O5_fluxes} shows that 
\begin{equation}
\label{unrestrited_5_branes}
\left. DF_{(3)} - H_{(3)} \wedge F_{(1)} \, \right|_{dy^{\hat{a}} \wedge \,dy^{\hat{b}}  \wedge \,dy^{\hat{c}} \wedge \,dy^{\hat{d}} } = \textrm{unrestricted} \ ,
\end{equation}
while any other component must vanish. In other words, O$5$/D$5$-sources threading the submanifold whose Poincar\'e dual is $dy^{\hat{a}} \wedge \,dy^{\hat{b}}  \wedge \,dy^{\hat{c}} \wedge \,dy^{\hat{d}}$ are compatible with the half-maximal supersymmetry of the type~IIB flux models under consideration.

\subsection{Type~IIA with O$6$-planes}
\label{sec:O6-plane}

The next orientifold action to be considered is the one induced by an O$6$-plane filling the external spacetime and a four-cycle inside $\mathbb{T}^{7}$. This case can be obtained from the O$2$-plane setup by performing four T-dualities along four coordinates $y^{\hat{a}}$. Without loss of generality, we will take such four coordinates to span the four-cycle $\mathbb{T}_{2}^{2} \times \mathbb{T}_{3}^{2} \subset \mathbb{T}^{7}$ in (\ref{T7_factorisation}). The O$6$-plane then breaks the $\textrm{SL}(7)$ covariance down to a subgroup $\textrm{SL}(3) \times \textrm{SL}(4) \subset \textrm{SL}(7)$. The relevant branching rules for the type~IIA with O2 $\Leftrightarrow$ type~IIA with O6 correspondence are summarised in Table~\ref{Table:SL7-SL4xSL3_IIA}. The original type~IIA internal derivatives $\partial_{m} \in {\bf{7}'}_{(-1,+3)} \in \textrm{SL}(7) \times \mathbb{R}_{2} \times \mathbb{R}_{1}$ in (\ref{partial_SL(8)}) branch under $\textrm{SL}(4) \times \textrm{SL}(3) \times \mathbb{R}_{3} \times \mathbb{R}_{2} \times \mathbb{R}_{1}$ as
\begin{equation}
\label{partial_SL(4)xSL(3)_IIA}
\begin{array}{rccccc}
{\bf{7}'}_{(-1,+3)} &\rightarrow& ({\bf{1},\bf{3'}})_{(+4,-1,+3)}  &\oplus& ({\bf{4'},\bf{1}})_{(-3,-1,+3)} & , \\[2mm]
\partial_{m} &\rightarrow&  \partial_{i} &\oplus& \partial_{\hat{a}} & .
\end{array}
\end{equation}
Then, the physical derivatives in the type~IIA with O$6$-planes duality frame are identified as
\begin{equation}
\label{partial_IIA_O6}
\partial_i \in ({\bf{1},\bf{3'}})_{(+4,-1,+3)}
\hspace{10mm} \textrm{ and } \hspace{10mm}
\tilde{\partial}_{\hat{a}} \equiv \partial^{\hat{a}} \in ({\bf{4},\bf{1}})_{(+3,-6,+2)} \ .
\end{equation}
Note again the maximally-mixed nature of the physical derivatives in this type~IIA duality frame: $\tilde{\partial}_{\hat{a}}$ are bosonic ($\mathbb{R}_{1}$ charge $+2$) and $\partial_{i}$ are spinorial ($\mathbb{R}_{1}$ charge $+3$).

\begin{table}[t!]
\begin{center}
\scalebox{0.74}{
\renewcommand{\arraystretch}{1.7}
\begin{tabular}{|c|c|c|}
\hline 
& Half-Maximal &  ${\rm SL(4)} \times {\rm SL(3)} \times \mathbb{R}_3 \times \mathbb{R}_2 \times \mathbb{R}_1$ \\ 
\hline \hline
 & -- & $ [({\bf{4'}},{\bf{1}})_{(-3,-1,-1)} \oplus ({\bf{1}},{\bf{3'}})_{(+4,-1,-1)}  \oplus {\bf{1}}_{(0,+7,-1)}] \oplus [({\bf{4}},{\bf{1}})_{(+3,+1,+1)} \oplus ({\bf{1}},{\bf{3}})_{(-4,+1,+1)}  \oplus {\bf{1}}_{(0,-7,+1)}]$ \\
\hline \hline
\multirow{10}{*}{\rotatebox{90}{Scalars}} & \multirow{4}{*}{$L^{MN}$} & $[({\bf (6,1)}_{(-6,-2,-2)} \oplus {\bf (4',3')}_{(+1,-2,-2)} \oplus {\color{ForestGreen}{{\bf (1,3)}_{(+8,-2,-2)}}}) \oplus ({\bf (4',1)}_{(-3,+6,-2)} \oplus {\bf (1,3')}_{(+4,+6,-2)})]$ \\
 & & ${\color{RoyalBlue}{{\bf (1,1)}_{(0,0,0)}}} \oplus [( {\bf (4',3)}_{(-7,0,0)} \oplus {\color{RoyalBlue}{{\bf (15+1,1)}_{(0,0,0)}}}  \oplus {\color{RoyalBlue}{{\bf (1,8+1)}_{(0,0,0)}}} \oplus {\color{ForestGreen}{{\bf (4,3')}_{(+7,0,0)}}} )$ \\
& & $({\color{BrickRed}{{\bf (4',1)}_{(-3,-8,0)}}} \oplus {\color{BrickRed}{{\bf (1,3')}_{(+4,-8,0)}}}) \oplus ({\bf (4,1)}_{(+3,+8,0)} \oplus {\bf (1,3)}_{(-4,+8,0)})]$ \\ 
&  & $[(\boxed{{\bf (4,1)}_{(+3,-6,+2)}} \oplus {\bf (1,3)}_{(-4,-6,+2)}) \oplus ({\color{BrickRed}{{\bf (6,1)}_{(+6,+2,+2)}}} \oplus {\color{BrickRed}{{\bf (4,3)}_{(-1,+2,+2)}}} \oplus {\color{BrickRed}{{\bf (1,3')}_{(-8,+2,+2)}}})]$ \\[2mm]
\cline{2-3}
 & \multirow{6}{*}{--} & $[({\bf (4,1)}_{(+3,+1,-3)} \oplus {\bf (1,3)}_{(-4,+1,-3)}) \oplus {\bf (1,1)}_{(0,-7,-3)}]$ \\
& &  $[({\bf (4,3')}_{(-5,+3,-1)} \oplus {\bf (6,3)}_{(+2,+3,-1)} \oplus {\color{ForestGreen}{{\bf (4',1)}_{(+9,+3,-1)}}} \oplus{\bf (1,1)}_{(-12,+3,-1)})$ \\
& &  $( {\color{RoyalBlue}{{\bf (4,3)}_{(-1,-5,-1)}}} \oplus {\color{ForestGreen}{{\bf (6,1)}_{(+6,-5,-1)}}} \oplus {\bf (1,3')}_{(-8,-5,-1)} )]$ \\
& &  $( {\color{RoyalBlue}{{\bf (4',3')}_{(+1,+5,+1)}}} \oplus {\bf (6,1)}_{(-6,+5,+1)} \oplus {\color{ForestGreen}{{\bf (1,3)}_{(+8,+5,+1)}}} )]$ \\
& &  $[({\color{BrickRed}{{\bf (4',3)}_{(+5,-3,+1)}}} \oplus {\color{BrickRed}{{\bf (6,3')}_{(-2,-3,+1)}}} \oplus  {\color{BrickRed}{{\bf (4,1)}_{(-9,-3,+1)}}} \oplus   {\bf (1,1)}_{(+12,-3,+1)})$ \\
& & $[( {\bf (4',1)}_{(-3,-1,+3)} \oplus \boxed{ {\bf (1,3')}_{(+4,-1,+3)}}) \oplus {\color{BrickRed}{{\bf (1,1)}_{(0,+7,+3)}}}]$ \\[2mm]
\hline
\end{tabular}}
\caption{Type~IIA with O$6$-plane branching rules for the embedding $\textrm{SL}(4) \times \textrm{SL}(3) \times \mathbb{R}_3 \times \mathbb{R}_2 \times \mathbb{R}_1 \subset {\rm SL(7)} \times \mathbb{R}_2 \times \mathbb{R}_1$. The subscripts in the third column indicate $(\mathbb{R}_3,\mathbb{R}_2,\mathbb{R}_1)$-charges. We have highlighted the scalars ${\color{RoyalBlue}{e_{m}{}^{n}}}$, ${\color{RoyalBlue}{\Phi}}$, ${\color{ForestGreen}{B_{(2)}}}$, ${\color{ForestGreen}{B_{(6)}}}$ and ${\color{BrickRed}{C_{(p)}}}$ listed in (\ref{scalars_SL(4)xSL(3)_IIA_even})-(\ref{scalars_SL(4)xSL(3)_IIA_odd}). The physical internal derivatives $\tilde{\partial}_{\hat{a}} \equiv \partial^{\hat{a}}$ and $\partial_{i}$ have been put in a box for their quick identification too. This table should be understood as a continuation of Table~\ref{Table:SL8-SL7}.}
\label{Table:SL7-SL4xSL3_IIA}
\end{center}
\end{table}

\begin{table}[t]
\begin{center}
\renewcommand{\arraystretch}{1.8}
\begin{tabular}{|c|c|c|}
\hline
Fluxes &  Flux components  &  Embedding tensor   \\
\hline
\hline
\multirow{3}{*}{$\omega$} & $\omega_{\hat{a}\hat{b}}{}^{\hat{c}} \in ({\bf{20}+\bf{4},\bf{1}})_{(+3,-6,+2)}$ & $-\theta^{\hat{a}\hat{b}8}{}_{\hat{c}}$  \\
\cline{2-3}
 & $\omega_{\hat{a}i}{}^{j} \in ({\bf{4},\bf{8}+\bf{1}})_{(+3,-6,+2)} $  &  $\theta^{\hat{a}j8}{}_{i}$  \\
 \cline{2-3}
 & $\omega_{ij}{}^{\hat{a}} \in ({\bf{4'},\bf{3}})_{(+5,+4,+4)} $  &  $\theta^{k\hat{b}\hat{c}\hat{d}}$   \\
\hline
\hline
$H_{(1)}$ & $H_{\hat{a}} \in ({\bf{4},\bf{1}})_{(+3,-6,+2)}$  &  $ \frac{1}{3\gamma+\alpha+\beta}\theta^{\hat{a}8},\,\frac{1}{\gamma}\theta^{\hat{a}\hat{b}8}{}_{\hat{b}},\,\frac{1}{\gamma+\alpha}\theta^{\hat{a}i8}{}_{i}$ \\
\hline
\hline
$F_{(0)}$ & $F_{(0)} \in ({{\bf{1},\bf{1}}})_{(-12,-4,+4)}$  &  $-\theta^{ijk8} $ \\
\hline
\hline
\multirow{2}{*}{$H_{(3)}$} & $H_{i\hat{a}\hat{b}} \in ({\bf{6},\bf{3'}})_{(+10,-6,+2)}$  & $\theta^{\hat{a}\hat{b}8}{}_{i} $  \\
\cline{2-3}
& $H_{ijk} \in ({\bf{1},\bf{1}})_{(+12,+4,+4)}$  & $-\theta^{\hat{a}\hat{b}\hat{c}\hat{d}} $\\
\hline
\hline
$F_{(2)}$ & $F_{i\hat{a}} \in ({\bf{4},\bf{3'}})_{(-5,-4,+4)}$ &  $ \theta^{jk\hat{a}8}$ \\
\hline
\multirow{2}{*}{$F_{(4)}$} & $F_{\hat{a}\hat{b}\hat{c}\hat{d}} \in ({\bf{1},\bf{1}})_{(0,-14,+2)}$ &  $\theta^{88} $ \\
\cline{2-3}
& $F_{ij\hat{a}\hat{b}} \in ({\bf{6},\bf{3}})_{(+2,-4,+4)}$  & $\theta^{k\hat{a}\hat{b}8} $\\
\hline
$F_{(6)}$ & $F_{ijk \hat{a}\hat{b}\hat{c}} \in ({\bf{4'},\bf{1}})_{(+9,-4,+4)}$ &  $-\theta^{\hat{a}\hat{b}\hat{c}8} $ \\
\hline
\end{tabular}
\caption{Type~IIA fluxes in the O$6$-plane duality frame of half-maximal supergravity and their identification with embedding tensor components.} 
\label{Table:O6_fluxes}
\end{center}
\end{table}

The group-theoretical identification of the internal components of the various type~IIA fields results as follows. The bosonic scalars are identified as
\begin{equation}
\label{scalars_SL(4)xSL(3)_IIA_even}
\begin{array}{rclcrclc}
e_{i}{}^{j} &\in& {\bf{(1,8+1)}}_{(0,0,0)} & \hspace{3mm} , & \hspace{3mm} C_{i\hat{a}\hat{b}\hat{c}\hat{d}} &\in& {\bf{(1,3')}}_{(+4,-8,0)} & ,
 \\[2mm]
e_{\hat{a}}{}^{\hat{b}} &\in& {\bf{(15+1,1)}}_{(0,0,0)} & \hspace{3mm} , & \hspace{3mm} C_{ijk\hat{a}\hat{b}} &\in& {\bf{(6,1)}}_{(+6,+2,+2)} & ,  \\[2mm]
\Phi &\in& {\bf{(1,1)}}_{(0,0,0)}  & \hspace{3mm} , & \hspace{3mm} C_{\hat{a}\hat{b}\hat{c}} &\in& {\bf{(4',1)}}_{(-3,-8,0)} & ,
 \\[2mm]
B_{i\hat{a}} &\in& {\bf{(4,3')}}_{(+7,0,0)}  & \hspace{3mm} , & \hspace{3mm} C_{ij\hat{a}} &\in& {\bf{(4,3)}}_{(-1,+2,+2)} & , \\[2mm]
B_{ij\hat{a}\hat{b}\hat{c}\hat{d}} &\in& {\bf{(1,3)}}_{(+8,-2,-2)} & \hspace{3mm} , & \hspace{3mm} C_{i} &\in& {\bf{(1,3')}}_{(-8,+2,+2)} & ,
\end{array}
\end{equation}
where $3+6$ compact scalars must be subtracted from $e_{i}{}^{j}$ and $e_{\hat{a}}{}^{\hat{b}}$ upon gauge-fixing of the internal $\textrm{SO}(3) \times \textrm{SO}(4)$ local symmetry, namely, $e_{i}{}^{j} \in \textrm{GL}(3)/\textrm{SO}(3)$ and $e_{\hat{a}}{}^{\hat{b}} \in \textrm{GL}(4)/\textrm{SO}(4)$. There are $64-4=60$ bosonic scalars in (\ref{scalars_SL(4)xSL(3)_IIA_even}), the four missing scalars being dual to the vectors $e_{\mu}{}^{\hat{a}}$. In addition, the spinorial scalars are given by
\begin{equation}
\label{scalars_SL(4)xSL(3)_IIA_odd}
\begin{array}{rclcrclcrclc}
e_{i}{}^{\hat{a}} &\in& {\bf{(4',3')}}_{(+1,+5,+1)} &  , & B_{\hat{a}\hat{b}} &\in& {\bf{(6,1)}}_{(+6,-5,-1)}   & , &  C_{ijk} &\in& {\bf{(1,1)}}_{(0,+7,+3)}  & ,
 \\[2mm]
e_{\hat{a}}{}^{i} &\in& {\bf{(4,3)}}_{(-1,-5,-1)} & , & B_{ijk\hat{a}\hat{b}\hat{c}} &\in& {\bf{(4',1)}}_{(+9,+3,-1)} & , &  C_{i\hat{a}\hat{b}} &\in& {\bf{(6,3')}}_{(-2,-3,+1)}  & , \\[2mm]
B_{ij} &\in& {\bf{(1,3)}}_{(+8,+5,+1)}  & , &  C_{ij\hat{a}\hat{b}\hat{c}} &\in& {\bf{(4',3)}}_{(+5,-3,+1)}  & , & C_{\hat{a}} &\in& {\bf{(4,1)}}_{(-9,-3,+1)} & ,
\end{array}
\end{equation}
adding up to $64-3-1=60$. The $3+1$ missing scalars are dual to the vectors $e_{\mu}{}^{i}$ and $C_{\mu}$. Lastly, an explicit computation yields the bosonic fluxes in Table~\ref{Table:O6_fluxes} together with additional spinorial fluxes which are projected out of the half-maximal theory.

\subsubsection*{Orientifold interpretation and O$6$-planes}

\begin{table}[t]
\begin{center}
\renewcommand{\arraystretch}{1.7}
\begin{tabular}{|c|c|c|c|c|}
\hline
Fields & $\Omega_{P}$ &  $\sigma_{\textrm{O}6}$ & $\mathcal{O}_{\mathbb{Z}_{2}}$  \\
\hline
\hline
$e_{i}{}^{j} \, , \, e_{\hat{a}}{}^{\hat{b}}$ & \multirow{2}{*}{$+$}  & $+$  & $+$  \\
\cline{1-1}\cline{3-4}
$e_{i}{}^{\hat{a}} \, , \, e_{\hat{a}}{}^{i}$ &  & $-$  & $-$ \\
\hline
\hline
$\Phi$ & $+$  & $+$  & $+$ \\
\hline
\hline
$B_{ij}\, , \, B_{\hat{a}\hat{b}}$ & \multirow{2}{*}{$-$} & $+$  & $-$  \\
\cline{1-1}\cline{3-4}
$B_{i\hat{a}}$ & & $-$  & $+$ \\
\hline
\hline
$B_{ijk\hat{a}\hat{b}\hat{c}}$ & \multirow{2}{*}{$+$}  & $-$  & $-$  \\
\cline{1-1}\cline{3-4}
$B_{ij\hat{a}\hat{b}\hat{c}\hat{d}}$ & & $+$  & $+$ \\
\hline
\end{tabular}
\hspace{10mm}
\begin{tabular}{|c|c|c|c|c|}
\hline
Fields & $\Omega_{P}$ &   $\sigma_{\textrm{O}6}$ & $\mathcal{O}_{\mathbb{Z}_{2}}$  \\
\hline
\hline
$C_{\hat{a}}$ & \multirow{2}{*}{$-$}   & $+$  & $-$  \\
\cline{1-1}\cline{3-4}
$C_{i}$ &  &  $-$  & $+$ \\
\hline
\hline
$C_{ij\hat{a}}\, , \, C_{\hat{a}\hat{b}\hat{c}}$ & \multirow{2}{*}{$+$}  &  $+$  & $+$  \\
\cline{1-1}\cline{3-4}
$C_{ijk} \, , \, C_{i\hat{a}\hat{b}}$ &  & $-$  & $-$ \\
\hline
\hline
$C_{ij\hat{a}\hat{b}\hat{c}}$ & \multirow{2}{*}{$-$}  &  $+$  & $-$  \\
\cline{1-1}\cline{3-4}
$C_{i\hat{a}\hat{b}\hat{c}\hat{d}} \, , \, C_{ijk\hat{a}\hat{b}}$ & & $-$  & $+$ \\
\hline
\end{tabular}
\caption{Grading of the various type~IIA fields under the O$6$-plane orientifold action ${\mathcal{O}_{\mathbb{Z}_{2}} = \Omega_{P} \sigma_{\textrm{O}6}}$. The  $\mathcal{O}_{\mathbb{Z}_{2}}$-even fields match the ones in (\ref{scalars_SL(4)xSL(3)_IIA_even}), whereas the $\mathcal{O}_{\mathbb{Z}_{2}}$-odd fields match the ones in (\ref{scalars_SL(4)xSL(3)_IIA_odd}).} 
\label{Table:O6_fields}
\end{center}
\end{table}

These type~IIA flux models can be string-theoretically interpreted as type~IIA orientifold reductions including O$6$-planes (and D$6$-branes) placed as
\begin{equation}
\label{O6_location}
\begin{array}{lll|lc|lc|lc|c}
x^{0} & x^{1} & x^{2} & \tilde{y}^{1}  & y^{2} & \tilde{y}^{3}  & y^{4}& \tilde{y}^{5} & y^{6} & \tilde{y}^{7}   \\
\hline
\times & \times & \times & \times &  & \times &  & \times &  & \times
\end{array}   
\end{equation}
The O$6$-plane is filling the external spacetime together with the four internal directions $\tilde{y}^{\hat{a}}$ with $\hat{a}=1,3,5,7$. The internal target space involution $\sigma_{\textrm{O}6}$ then reflects the three transverse coordinates (and their derivatives) on $\mathbb{T}^{7}$, namely,
\begin{equation}
\label{sigma_O6}
\begin{array}{cc}
\sigma_{\textrm{O}6} : & \hspace{5mm}  \tilde{y}^{\hat{a}} \,\, \rightarrow\,\, \tilde{y}^{\hat{a}} \hspace{5mm} , \hspace{5mm} y^{i} \,\, \rightarrow\,\,  - \, y^{i} \\[2mm]
 & \hspace{5mm}  \tilde{\partial}_{\hat{a}} \,\, \rightarrow\,\, \tilde{\partial}_{\hat{a}} \hspace{5mm} , \hspace{5mm}   \partial_{i} \,\, \rightarrow\,\,  - \, \partial_{i}
\end{array}
\hspace{6mm} \textrm{ with } \hspace{6mm} 
\hat{a} = 1,3,5,7
\hspace{4mm} , \hspace{4mm}
i = 2,4,6 \ .
\end{equation}
The corresponding orientifold action $\mathcal{O}_{\mathbb{Z}_{2}} = \Omega_{P} \, \sigma_{\textrm{O}6}$ acts on the type~IIB fields as summarised in Table~\ref{Table:O6_fields}. The $\mathcal{O}_{\mathbb{Z}_{2}}$-even and $\mathcal{O}_{\mathbb{Z}_{2}}$-odd fields in Table~\ref{Table:O6_fields} match the ones in (\ref{scalars_SL(4)xSL(3)_IIA_even}) and (\ref{scalars_SL(4)xSL(3)_IIA_odd}), respectively. The same matching holds at the level of the fluxes.

Employing the embedding tensor/flux dictionary in Table~\ref{Table:O6_fluxes}, the computation of the QC's in (\ref{QC_N=8}) produces the following constraints on the type~IIA fluxes. There is the by now standard nilpotency condition $(D^2=0)$ of the $D = d + \omega$ twisted exterior derivative on the internal space, as well as sourceless Bianchi identities of the form $DH_{(1)} = 0$ and $DH_{(3)} +\alpha H_{(1)}\wedge H_{(3)}= 0$ forbidding the presence of NS$7$-branes and NS$5$-branes in the compactification scheme, respectively. There are also Bianchi identities of the form $DF_{(4)} +\beta_{4} H_{(1)} \wedge F_{(4)} - H_{(3)} \wedge F_{(2)} = 0$ and $DF_{(6)} +\beta_{6} H_{(1)} \wedge F_{(6)} - H_{(3)} \wedge F_{(4)} = 0$ reflecting the absence of D$4$-branes and D$2$-branes, respectively. Finally there is a non-trivial Bianchi identity (\ref{Tadpole_p-form}) of the form $DF_{(2)} +\beta_{2} H_{(1)} \wedge F_{(2)} - H_{(3)} \wedge F_{(0)} = J_{\textrm{O}6/\textrm{D}6}$ involving O$6$/D$6$-sources. Recalling that the orientifold action is generated by O$6$-planes extending along the $y^{\hat{a}}$ coordinates, the QC's in (\ref{QC_N=8}) yield
\begin{equation}
\left. DF_{(2)} - H_{(3)} \wedge F_{(0)} \, \right|_{dy^{i} \wedge dy^{j} \wedge \,dy^{k} } = \textrm{unrestricted} \ ,
\end{equation}
while any other component vanishes. Therefore, O$6$/D$6$-sources threading the submanifold whose Poincar\'e dual is $dy^{i} \wedge dy^{j} \wedge \,dy^{k}$ are compatible with the half-maximal supersymmetry of the type~IIA flux models.

\subsection{Type~IIB with O$7$-planes}
\label{sec:O7-plane}

\begin{table}[t!]
\begin{center}
\scalebox{0.70}{
\renewcommand{\arraystretch}{1.7}
\begin{tabular}{|c|c|c|}
\hline 
& Half-Maximal &  ${\rm SL(5)} \times {\rm SL(2)} \times \mathbb{R}_3 \times \mathbb{R}_2 \times \mathbb{R}_1$ \\ 
\hline \hline
 & -- & $ [({\bf{5'}},{\bf{1}})_{(-2,-1,-1)} \oplus ({\bf{1}},{\bf{2}})_{(+5,-1,-1)}  \oplus ({\bf{1},\bf{1}})_{(0,+7,-1)}] \oplus [({\bf{5}},{\bf{1}})_{(+2,+1,+1)} \oplus ({\bf{1}},{\bf{2}})_{(-5,+1,+1)}  \oplus ({\bf{1},\bf{1}})_{(0,-7,+1)}]$ \\
\hline \hline
\multirow{10}{*}{\rotatebox{90}{Scalars}} & \multirow{4}{*}{$L^{MN}$} & $[({\bf (10',1)}_{(-4,-2,-2)} \oplus {\bf (5',2)}_{(+3,-2,-2)} \oplus {\bf (1,1)}_{(+10,-2,-2)}) \oplus ({\bf (5',1)}_{(-2,+6,-2)} \oplus {\bf (1,2)}_{(+5,+6,-2)})]$ \\
 & & ${\color{RoyalBlue}{\bf (1,1)}_{(0,0,0)}} \oplus [( {\bf (5',2)}_{(-7,0,0)} \oplus {\color{RoyalBlue}{\bf (24,1)}_{(0,0,0)}}  \oplus {\color{RoyalBlue}{\bf (1,1)}_{(0,0,0)}} \oplus {\color{RoyalBlue}{\bf (1,3)}_{(0,0,0)}}$  \\
& & ${\color{ForestGreen}{\bf (5,2)}_{(+7,0,0)}} ) \oplus {\color{RoyalBlue}{\bf (1,1)}_{(0,0,0)}}\oplus({\color{BrickRed}{{\bf (5',1)}_{(-2,-8,0)}}} \oplus {\color{BrickRed}{{\bf (1,2)}_{(+5,-8,0)}}}) \oplus ({\bf (5,1)}_{(+2,+8,0)} \oplus {\bf (1,2)}_{(-5,+8,0)})]$ \\ 
&  & $[(\boxed{{\bf (5,1)}_{(+2,-6,+2)}} \oplus {\color{ForestGreen}{\bf (1,2)}_{(-5,-6,+2)}}) \oplus ({\color{BrickRed}{{\bf (10,1)}_{(+4,+2,+2)}}} \oplus {\color{BrickRed}{{\bf (5,2)}_{(-3,+2,+2)}}} \oplus {\color{BrickRed}{{\bf (1,1)}_{(-10,+2,+2)}}})]$ \\[2mm]
\cline{2-3}
 & \multirow{6}{*}{--} & $[({\bf (5,1)}_{(+2,+1,-3)} \oplus {\bf (1,2)}_{(-5,+1,-3)}) \oplus {\bf (1,1)}_{(0,-7,-3)}]$ \\
& &  $[({\bf (5,1)}_{(-8,+3,-1)} \oplus {\bf (10,2)}_{(-1,+3,-1)} \oplus{\bf (10',1)}_{(+6,+3,-1)})$ \\
& &  $(  {\color{RoyalBlue}{\bf (5,2)}_{(-3,-5,-1)}} \oplus {\color{ForestGreen}{\bf (10,1)}_{(+4,-5,-1)}} \oplus {\bf (1,1)}_{(-10,-5,-1)} )]$ \\
& &  $[(  {\color{RoyalBlue}{\bf (5',2)}_{(+3,+5,+1)}} \oplus {\bf (10',1)}_{(-4,+5,+1)} \oplus {\color{ForestGreen}{\bf (1,1)}_{(+10,+5,+1)}} )$ \\
& &  $({\color{BrickRed}{{\bf (5',1)}_{(+8,-3,+1)}}} \oplus {\color{BrickRed}{{\bf (10',2)}_{(+1,-3,+1)}}} \oplus  {\color{BrickRed}{{\bf (10,1)}_{(-6,-3,+1)}}} )]$ \\
& & $[( {\color{ForestGreen}{\bf (5',1)}_{(-2,-1,+3)}} \oplus \boxed{{\bf (1,2)}_{(+5,-1,+3)}}) \oplus {\color{BrickRed}{{\bf (1,1)}_{(0,+7,+3)}}}]$ \\[2mm]
\hline
\end{tabular}}
\caption{Type~IIB with O$7$-plane branching rules for the embedding $\textrm{SL}(5) \times \textrm{SL}(2) \times \mathbb{R}_3 \times \mathbb{R}_2 \times \mathbb{R}_1 \subset {\rm SL(7)} \times \mathbb{R}_2 \times \mathbb{R}_1$. The subscripts in the third column indicate $(\mathbb{R}_3,\mathbb{R}_2,\mathbb{R}_1)$-charges. We have highlighted the scalars ${\color{RoyalBlue}{e_{m}{}^{n}}}$, ${\color{RoyalBlue}{\Phi}}$, ${\color{ForestGreen}{B_{(2)}}}$, ${\color{ForestGreen}{B_{(6)}}}$ and ${\color{BrickRed}{C_{(p)}}}$ listed in (\ref{scalars_SL(5)xSL(2)_IIB_even})-(\ref{scalars_SL(5)xSL(2)_IIB_odd}). The physical internal derivatives $\tilde{\partial}_{\hat{\mathtt{a}}} \equiv \partial^{\hat{\mathtt{a}}}$ and $\partial_{\mathtt{i}}$ have been put in a box for their quick identification too. This table should be understood as a continuation of Table~\ref{Table:SL8-SL7}.}
\label{Table:SL7-SL2xSL5_IIB}
\end{center}
\end{table}

The next duality frame we are considering is type~IIB with an O$7$-plane filling the external spacetime and an internal five-cycle. We will choose it to be $\mathbb{T}_{2}^{2} \times \mathbb{T}_{3}^{2} \times \mathbb{S}^{1} \subset \mathbb{T}^{7}$ in (\ref{T7_factorisation}) without loss of generality. This type~IIB with an O$7$-plane duality frame can be generated from the O$2$-plane case by applying this time five T-dualities along the aforementioned five-cycle. The presence of the O$7$-plane breaks $\textrm{SL}(7)$ covariance down to a subgroup $\textrm{SL}(2) \times \textrm{SL}(5) \subset \textrm{SL}(7)$. The relevant branching rules for the type~IIA with O2 $\Leftrightarrow$ type~IIB with O7 correspondence are summarised in Table~\ref{Table:SL7-SL2xSL5_IIB}. The original type~IIA internal derivatives $\partial_{m} \in {\bf{7}'}_{(-1,+3)} \in \textrm{SL}(7) \times \mathbb{R}_{2} \times \mathbb{R}_{1}$ in (\ref{partial_SL(8)}) branch under $\textrm{SL}(5) \times  \textrm{SL}(2) \times \mathbb{R}_{3} \times \mathbb{R}_{2} \times \mathbb{R}_{1}$ as
\begin{equation}
\label{partial_SL(5)xSL(2)_IIB}
\begin{array}{rccccc}
{\bf{7}'}_{(-1,+3)} &\rightarrow& {\bf (5',1)}_{(-2,-1,+3)}  &\oplus& {\bf (1,2)}_{(+5,-1,+3)} & , \\[2mm]
\partial_{m} &\rightarrow&  \partial_{\hat{\mathtt{a}}} &\oplus& \partial_{\mathtt{i}} & ,
\end{array}
\end{equation}
where, without loss of generality, we take $\mathtt{i}=1,2$ and $\hat{\mathtt{a}}=3,4,5,6,7$. The physical derivatives in this type~IIB duality frame are given by
\begin{equation}
\label{partial_O7_IIB}
\tilde{\partial}_{\hat{\mathtt{a}}} \equiv \partial^{\hat{\mathtt{a}}} \in {\bf (5,1)}_{(+2,-6,+2)} 
\hspace{10mm} \textrm{ and } \hspace{10mm}
\partial_{\mathtt{i}} \in {\bf (1,2)}_{(+5,-1,+3)} \ .
\end{equation}
Note that the derivatives $\tilde{\partial}_{\hat{\mathtt{a}}}$ are bosonic (with $\mathbb{R}_1$ charge $+2$) whereas $\partial_{\mathtt{i}}$ are spinorial (with $\mathbb{R}_1$ charge $+3$).

\begin{table}[t]
\begin{center}
\renewcommand{\arraystretch}{1.8}
\begin{tabular}{|c|c|c|}
\hline
Fluxes &  Flux components  &  Embedding tensor   \\
\hline
\hline
\multirow{3}{*}{$\omega$} & $\omega_{\hat{\mathtt{a}}\hat{\mathtt{b}}}{}^{\hat{\mathtt{c}}} \in ({{\bf{45}}+{\bf{5}},\bf{1}})_{(+2,-6,+2)}$ & $\theta^{\hat{\mathtt{a}}\hat{\mathtt{b}}8}{}_{\hat{\mathtt{c}}}$  \\
\cline{2-3}
 & $\omega_{\hat{\mathtt{a}}\mathtt{i}}{}^{\mathtt{j}} \in ({\bf{5},\bf{3}+{\bf{1}}})_{(+2,-6,+2)} $  &  $-\theta^{\hat{\mathtt{a}}\mathtt{j}8}{}_{\mathtt{i}}$  \\
 \cline{2-3}
 & $\omega_{\mathtt{i}\mathtt{j}}{}^{\hat{\mathtt{a}}} \in ({\bf{5'},\bf{1}})_{(+8,+4,+4)} $  &  $\theta^{\hat{\mathtt{b}}\hat{\mathtt{c}}\hat{\mathtt{d}}\hat{\mathtt{e}}}$   \\
\hline
\hline
$H_{(1)}$ & $H_{\hat{\mathtt{a}}} \in ({\bf{5},\bf{1}})_{(+2,-6,+2)}$  &  $\frac{1}{3\gamma-\beta}\theta^{\hat{\mathtt{a}}8},\,\frac{1}{\gamma-\alpha}\theta^{\hat{\mathtt{a}}\mathtt{i}8}{}_{\mathtt{i}},\,\frac{1}{\gamma}\theta^{\hat{\mathtt{a}}\hat{\mathtt{b}}8}{}_{\hat{\mathtt{b}}}$ \\
\hline
\hline
$H_{(3)}$ & $H_{\mathtt{i}\hat{\mathtt{a}}\hat{\mathtt{b}}} \in ({\bf{10},\bf{2}})_{(+9,-6,+2)}$  & $\theta^{\hat{\mathtt{a}}\hat{\mathtt{b}}8}{}_{\mathtt{i}} $ \\
\hline
\hline
$F_{(1)}$ & $F_{\hat{\mathtt{a}}}  \in ({\bf{5},\bf{1}})_{(-8,-4,+4)}$  &  $\theta^{\hat{\mathtt{a}}\mathtt{i}\mathtt{j}8}$ \\
\hline
$F_{(3)}$ & $F_{\mathtt{i}\hat{\mathtt{a}}\hat{\mathtt{b}}} \in ({\bf{10},\bf{2}})_{(-1,-4,+4)}$ &  $ \theta^{\mathtt{j}\hat{\mathtt{a}}\hat{\mathtt{b}}8}$ \\
\hline
\multirow{2}{*}{$F_{(5)}$} & $F_{\hat{\mathtt{a}}\hat{\mathtt{b}}\hat{\mathtt{c}}\hat{\mathtt{d}}\hat{\mathtt{e}}} \in ({\bf{1},\bf{1}})_{(0,-14,+2)}$ &  $\theta^{88} $ \\
\cline{2-3}
& $F_{\mathtt{i}\mathtt{j}\hat{\mathtt{a}}\hat{\mathtt{b}}\hat{\mathtt{c}}} \in ({\bf{10'},\bf{1}})_{(+6,-4,+4)}$ &  $-\theta^{\hat{\mathtt{a}}\hat{\mathtt{b}}\hat{\mathtt{c}}8} $ \\
\hline
\end{tabular}
\caption{Type~IIB fluxes in the O$7$-plane duality frame of half-maximal supergravity and their identification with embedding tensor components.} 
\label{Table:O7_fluxes}
\end{center}
\end{table}

The group-theoretical identification of the internal components of the various type~IIB fields yields the following bosonic scalars
\begin{equation}
\label{scalars_SL(5)xSL(2)_IIB_even}
\begin{array}{rclcrclc}
e_{\mathtt{i}}{}^{\mathtt{j}} &\in& {\bf{(1,3+1)}}_{(0,0,0)} & \hspace{3mm} , & \hspace{3mm} C_{\mathtt{i}\hat{\mathtt{a}}\hat{\mathtt{b}}\hat{\mathtt{c}}\hat{\mathtt{d}}\hat{\mathtt{e}}} &\in& {\bf{(1,2)}}_{(+5,-8,0)} & ,
 \\[2mm]
e_{\hat{\mathtt{a}}}{}^{\hat{\mathtt{b}}} &\in& {\bf{(24+1,1)}}_{(0,0,0)} & \hspace{3mm} , & \hspace{3mm} C_{\hat{\mathtt{a}}\hat{\mathtt{b}}\hat{\mathtt{c}}\hat{\mathtt{d}}} &\in& {\bf{(5',1)}}_{(-2,-8,0)} & ,  \\[2mm]
\Phi &\in& {\bf{(1,1)}}_{(0,0,0)}  & \hspace{3mm} , & \hspace{3mm} C_{\mathtt{i}\mathtt{j}\hat{\mathtt{a}}\hat{\mathtt{b}}} &\in& {\bf{(10,1)}}_{(+4,+2,+2)} & ,
 \\[2mm]
B_{\mathtt{i}\hat{\mathtt{a}}} &\in& {\bf{(5,2)}}_{(+7,0,0)}  & \hspace{3mm} , & \hspace{3mm} C_{\mathtt{i}\hat{\mathtt{a}}} &\in& {\bf{(5,2)}}_{(-3,+2,+2)} & , \\[2mm]
B_{\mathtt{i}\hat{\mathtt{a}}\hat{\mathtt{b}}\hat{\mathtt{c}}\hat{\mathtt{d}}\hat{\mathtt{e}}} &\in& {\bf{(1,2)}}_{(-5,-6,+2)}  & \hspace{3mm} , & \hspace{3mm} C_{(0)}  &\in& {\bf{(1,1)}}_{(-10,+2,+2)} & ,
\end{array}
\end{equation}
where $10+1$ compact scalars must be subtracted from $e_{\hat{\mathtt{a}}}{}^{\hat{\mathtt{b}}}$ and $e_{\mathtt{i}}{}^{\mathtt{j}}$ upon gauge-fixing of the internal $\textrm{SO}(5) \times \textrm{SO}(2)$ local symmetry, namely, $e_{\hat{\mathtt{a}}}{}^{\hat{\mathtt{b}}} \in \textrm{GL}(5)/\textrm{SO}(5)$ and $e_{\mathtt{i}}{}^{\mathtt{j}} \in \textrm{GL}(2)/\textrm{SO}(2)$. The bosonic scalars add up to $64-5=59$ with the $5$ missing scalars being dual to the five vectors $e_{\mu}{}^{\hat{\mathtt{a}}}$. The fermionic scalars are given by
\begin{equation}
\label{scalars_SL(5)xSL(2)_IIB_odd}
\begin{array}{rclcrclcrclc}
e_{\mathtt{i}}{}^{\hat{\mathtt{a}}} &\in& {\bf{(5',2)}}_{(+3,+5,+1)} & , & B_{\mathtt{i}\mathtt{j}} &\in& {\bf{(1,1)}}_{(+10,+5,+1)}  & , &  C_{\mathtt{i}\mathtt{j}\hat{\mathtt{a}}\hat{\mathtt{b}}\hat{\mathtt{c}}\hat{\mathtt{d}}} &\in& {\bf{(5',1)}}_{(+8,-3,+1)} & 
 \\[2mm]
e_{\hat{\mathtt{a}}}{}^{\mathtt{i}} &\in& {\bf{(5,2)}}_{(-3,-5,-1)} & , & B_{\mathtt{i}\mathtt{j}\hat{\mathtt{a}}\hat{\mathtt{b}}\hat{\mathtt{c}}\hat{\mathtt{d}}} &\in& {\bf{(5',1)}}_{(-2,-1,+3)}  & , &  C_{\mathtt{i}\mathtt{j}} &\in& {\bf{(1,1)}}_{(0,+7,+3)}  &   \\[2mm]
B_{\hat{\mathtt{a}}\hat{\mathtt{b}}} &\in& {\bf{(10,1)}}_{(+4,-5,-1)}  & , &    C_{\hat{\mathtt{a}}\hat{\mathtt{b}}} &\in& {\bf{(10,1)}}_{(-6,-3,+1)} & , & C_{\mathtt{i}\hat{\mathtt{a}}\hat{\mathtt{b}}\hat{\mathtt{c}}} &\in& {\bf{(10',2)}}_{(+1,-3,+1)} & 
\end{array}
\end{equation}
adding up to $64-2=62$. The $2$ missing spinorial scalars are dual to the vectors $e_{\mu}{}^{\mathtt{i}}$. Lastly, an explicit computation yields the bosonic fluxes in Table~\ref{Table:O7_fluxes} together with additional spinorial fluxes which are projected out of the half-maximal theory.

\subsubsection*{Orientifold interpretation and O$7$-planes}

This class of type~IIB flux models has a string-theoretic description in terms of type~IIB orientifold reductions including O$7$-planes (and D$7$-branes) placed at
\begin{equation}
\label{O7_location}
\begin{array}{lll|lc|lc|lc|c}
x^{0} & x^{1} & x^{2} & y^{1}  & y^{2} & \tilde{y}^{3}  & \tilde{y}^{4}& \tilde{y}^{5} & \tilde{y}^{6} & \tilde{y}^{7}   \\
\hline
\times & \times & \times &  &  & \times & \times & \times & \times  & \times
\end{array}   
\end{equation}
The O$7$-plane fills the external spacetime together with the five internal directions $\tilde{y}^{\hat{\mathtt{a}}}$ with $\hat{\mathtt{a}}=3,\ldots,7$. The internal target space involution $\sigma_{\textrm{O}7}$ reflects the two transverse coordinates (and their derivatives) on $\mathbb{T}^{7}$, namely,
\begin{equation}
\label{sigma_O7}
\begin{array}{cc}
\sigma_{\textrm{O}7} : & \hspace{5mm}  \tilde{y}^{\hat{\mathtt{a}}} \,\, \rightarrow\,\, \tilde{y}^{\hat{\mathtt{a}}} \hspace{5mm} , \hspace{5mm} y^{\mathtt{i}} \,\, \rightarrow\,\,  - \, y^{\mathtt{i}} \\[2mm]
 & \hspace{5mm}  \tilde{\partial}_{\hat{\mathtt{a}}} \,\, \rightarrow\,\, \tilde{\partial}_{\hat{\mathtt{a}}} \hspace{5mm} , \hspace{5mm}   \partial_{\mathtt{i}} \,\, \rightarrow\,\,  - \, \partial_{\mathtt{i}}
\end{array}
\hspace{6mm} \textrm{ with } \hspace{6mm} 
\hat{\mathtt{a}} = 3,\ldots,7
\hspace{4mm} , \hspace{4mm}
\mathtt{i} = 1,2
 \ .
\end{equation}
The orientifold action $\mathcal{O}_{\mathbb{Z}_{2}} = \Omega_{P} \, (-1)^{F_{L}}\, \sigma_{\textrm{O}7}$ acts on the type~IIB fields as summarised in Table~\ref{Table:O7_fields}. Note again the perfect matching between the $\mathcal{O}_{\mathbb{Z}_{2}}$-even and $\mathcal{O}_{\mathbb{Z}_{2}}$-odd fields in Table~\ref{Table:O7_fields} and those in (\ref{scalars_SL(5)xSL(2)_IIB_even}) and (\ref{scalars_SL(5)xSL(2)_IIB_odd}), respectively. The same matching holds at the level of the fluxes.

\begin{table}[t]
\begin{center}
\renewcommand{\arraystretch}{1.7}
\begin{tabular}{|c|c|c|c|c|}
\hline
Fields & $\Omega_{P}$ & $(-1)^{F_{L}}$   &  $\sigma_{\textrm{O}7}$ & $\mathcal{O}_{\mathbb{Z}_{2}}$  \\
\hline
\hline
$e_{\mathtt{i}}{}^{\mathtt{j}} \, , \, e_{\hat{\mathtt{a}}}{}^{\hat{\mathtt{b}}}$ & \multirow{2}{*}{$+$}  & \multirow{2}{*}{$+$}  & $+$  & $+$  \\
\cline{1-1}\cline{4-5}
$e_{\mathtt{i}}{}^{\hat{\mathtt{a}}} \, , \, e_{\hat{\mathtt{a}}}{}^{\mathtt{i}}$ &  & & $-$  & $-$ \\
\hline
\hline
$\Phi$ & $+$ & $+$ & $+$  & $+$ \\
\hline
\hline
$B_{\mathtt{i}\mathtt{j}}\, , \, B_{\hat{\mathtt{a}}\hat{\mathtt{b}}}$ & \multirow{2}{*}{$-$} & \multirow{2}{*}{$+$} & $+$  & $-$  \\
\cline{1-1}\cline{4-5}
$B_{\mathtt{i}\hat{\mathtt{a}}}$ &  & & $-$  & $+$ \\
\hline
\hline
$B_{\mathtt{i}\hat{\mathtt{a}}\hat{\mathtt{b}}\hat{\mathtt{c}}\hat{\mathtt{d}}\hat{\mathtt{e}}}$ & \multirow{2}{*}{$-$}  & \multirow{2}{*}{$+$} & $-$  & $+$  \\
\cline{1-1}\cline{4-5}
$B_{\mathtt{i}\mathtt{j}\hat{\mathtt{a}}\hat{\mathtt{b}}\hat{\mathtt{c}}\hat{\mathtt{d}}}$ & & & $+$  & $-$ \\
\hline
\end{tabular}
\hspace{5mm}
\begin{tabular}{|c|c|c|c|c|}
\hline
Fields & $\Omega_{P}$ & $(-1)^{F_{L}}$   &  $\sigma_{\textrm{O}7}$ & $\mathcal{O}_{\mathbb{Z}_{2}}$  \\
\hline
\hline
$C_{\hat{\mathtt{a}}\hat{\mathtt{b}}\hat{\mathtt{c}}\hat{\mathtt{d}}} \,, \, C_{\mathtt{i}\mathtt{j}\hat{\mathtt{a}}\hat{\mathtt{b}}}$ & \multirow{2}{*}{$-$} & \multirow{2}{*}{$-$}  & $+$  & $+$  \\
\cline{1-1}\cline{4-5}
$C_{\mathtt{i}\hat{\mathtt{a}}\hat{\mathtt{b}}\hat{\mathtt{c}}}$ &  & &  $-$  & $-$ \\
\hline
\hline
$C_{(0)}$ & $-$ & $-$ & $+$ & $+$ \\
\hline
\hline
$C_{\mathtt{i}\mathtt{j}}\, , \, C_{\hat{\mathtt{a}}\hat{\mathtt{b}}}$ & \multirow{2}{*}{$+$}  & \multirow{2}{*}{$-$}  & $+$  & $-$  \\
\cline{1-1}\cline{4-5}
$C_{\mathtt{i}\hat{\mathtt{a}}}$ &  &  & $-$  & $+$ \\
\hline
\hline
$C_{\mathtt{i}\hat{\mathtt{a}}\hat{\mathtt{b}}\hat{\mathtt{c}}\hat{\mathtt{d}}\hat{\mathtt{e}}}$ & \multirow{2}{*}{$+$}  &  \multirow{2}{*}{$-$}  &  $-$  & $+$  \\
\cline{1-1}\cline{4-5}
$C_{\mathtt{i}\mathtt{j}\hat{\mathtt{a}}\hat{\mathtt{b}}\hat{\mathtt{c}}\hat{\mathtt{d}}}$ & & & $+$  & $-$ \\
\hline
\end{tabular}
\caption{Grading of the various type~IIB fields under the O$7$-plane orientifold action ${\mathcal{O}_{\mathbb{Z}_{2}} = \Omega_{P} \, (-1)^{F_{L}} \, \sigma_{\textrm{O}7}}$. The  $\mathcal{O}_{\mathbb{Z}_{2}}$-even fields match the ones in (\ref{scalars_SL(5)xSL(2)_IIB_even}), whereas the $\mathcal{O}_{\mathbb{Z}_{2}}$-odd fields match the ones in (\ref{scalars_SL(5)xSL(2)_IIB_odd}).} 
\label{Table:O7_fields}
\end{center}
\end{table}

The QC's in (\ref{QC_N=8}) reduce to the following string-theoretic conditions on the type~IIB fluxes. First of all, there is the nilpotency condition $(D^2=0)$ of the $D = d + \omega$ twisted exterior derivative on the internal space. There are also the sourceless Bianchi identities $DH_{(1)} = 0$ and $DH_{(3)} +\alpha H_{(1)}\wedge H_{(3)} = 0$ that follow from the absence of NS$7$-branes and NS$5$-branes in the compactification scheme, respectively. Moreover, sourceless Bianchi identities $DF_{(5)} +\beta_{5} H_{(1)} \wedge F_{(5)} - H_{(3)} \wedge F_{(3)} = 0$ and $DF_{(3)} +\beta_{3} H_{(1)} \wedge F_{(3)} - H_{(3)} \wedge F_{(1)} = 0$ appear associated with the absence of D$3$-branes and D$5$-branes. Lastly, one finds Bianchi identities involving O$7$/D$7$-sources in (\ref{Tadpole_p-form}), namely, $DF_{(1)} + \beta_{1} H_{(1)} \wedge F_{(1)} = J_{\textrm{O}7/\textrm{D}7}$. Since the orientifold action is generated by O$7$-planes that extend along the five internal directions $\tilde{y}^{\hat{\mathtt{a}}}$, there are no QC's restricting the number of such sources, namely,
\begin{equation}
\left. DF_{(1)}  \, \right|_{dy^{\mathtt{i}} \wedge \, dy^{\mathtt{j}} } = \textrm{unrestricted} \ ,
\end{equation}
while any other component must vanish. Therefore, O$7$/D$7$-sources threading the submanifold whose Poincar\'e dual is $dy^{\mathtt{i}} \wedge \, dy^{\mathtt{j}}$ are compatible with the half-maximal supersymmetry of the type~IIB flux models.

\subsection{Type~IIA with O$8$-planes}
\label{sec:O8-plane}

In this section we will consider the orientifold action induced by an O$8$-plane filling the external spacetime and a single six-cycle inside $\mathbb{T}^{7}$. Without loss of generality we will select $\mathbb{T}_{1}^{2} \times \mathbb{T}_{2}^{2} \times \mathbb{T}_{3}^{2} \subset \mathbb{T}^{7}$ in (\ref{T7_factorisation}). This O$8$-plane case is generated by performing six T-dualities on the O$2$-plane setup along such a particular six-cycle. The presence of the O$8$-plane breaks $\textrm{SL}(7)$ covariance down to a subgroup $\textrm{SL}(6) \subset \textrm{SL}(7)$. The relevant branching rules for the type~IIA with O2 $\Leftrightarrow$ type~IIA with O8 correspondence are summarised in Table~\ref{Table:SL7-SL6_IIA}. The original type~IIA internal derivatives $\partial_{m} \in {\bf{7}'}_{(-1,+3)} \in \textrm{SL}(7) \times \mathbb{R}_{2} \times \mathbb{R}_{1}$ in (\ref{partial_SL(8)}) branch under $\textrm{SL}(6) \times \mathbb{R}_{3} \times \mathbb{R}_{2} \times \mathbb{R}_{1}$ as
\begin{equation}
\label{partial_SL(6)_IIA}
\begin{array}{rccccc}
{\bf{7}'}_{(-1,+3)} &\rightarrow& {\bf 6'}_{(-1,-1,+3)}  &\oplus& {\bf 1}_{(+6,-1,+3)} & , \\[2mm]
\partial_{m} &\rightarrow&  \partial_{\mathsf{m}} &\oplus& \partial_{7} & ,
\end{array}
\end{equation}
with $\mathsf{m}=1,\ldots,6$. As a result, the physical derivatives in this type~IIA duality frame are identified as
\begin{equation}
\label{partial_IIA_O8}
\tilde{\partial}_{\mathsf{m}} \equiv \partial^{\mathsf{m}} \in {\bf 6}_{(+1,-6,+2)} 
\hspace{10mm} \textrm{ and } \hspace{10mm}
\partial_{7} \in {\bf 1}_{(+6,-1,+3)} \ .
\end{equation}
As in previous cases, the physical derivatives in this type~IIA frame are of mixed nature:  $\tilde{\partial}_{\mathsf{m}}$ are bosonic ($\mathbb{R}_{1}$ charge $+2$) and $\partial_{7}$ is spinorial ($\mathbb{R}_{1}$ charge $+3$).

\begin{table}[t!]
\begin{center}
\scalebox{0.84}{
\renewcommand{\arraystretch}{1.7}
\begin{tabular}{|c|c|c|}
\hline 
& Half-Maximal &  ${\rm SL(6)} \times \mathbb{R}_3 \times \mathbb{R}_2 \times \mathbb{R}_1$ \\ 
\hline \hline
 & -- & $ [({\bf{6'}}_{(-1,-1,-1)} \oplus {\bf{1}}_{(+6,-1,-1)})  \oplus {\bf{1}}_{(0,+7,-1)}] \oplus [({\bf{6}}_{(+1,+1,+1)} \oplus {\bf{1}}_{(-6,+1,+1)})  \oplus {\bf{1}}_{(0,-7,+1)}]$ \\
\hline \hline
\multirow{10}{*}{\rotatebox{90}{Scalars}} & \multirow{4}{*}{$L^{MN}$} & $[({\bf 15'}_{(-2,-2,-2)} \oplus {\bf 6'}_{(+5,-2,-2)}) \oplus ({\bf 6'}_{(-1,+6,-2)} \oplus {\bf 1}_{(+6,+6,-2)})]$ \\
 & & ${\color{RoyalBlue}{{\bf 1}_{(0,0,0)}}} \oplus [({\bf 6'}_{(-7,0,0)} \oplus {\color{RoyalBlue}{{\bf 35}_{(0,0,0)}  \oplus {\bf 1}_{(0,0,0)}}}
 \oplus {\color{ForestGreen}{{\bf 6}_{(+7,0,0)}}}) \oplus {\color{RoyalBlue}{{\bf 1}_{(0,0,0)}}}$ \\
& & $({\color{BrickRed}{{\bf 6'}_{(-1,-8,0)}}} \oplus {\bf 1}_{(+6,-8,0)}) \oplus ({\bf 6}_{(+1,+8,0)} \oplus {\bf 1}_{(-6,+8,0)})]$ \\ 
&  & $[(\boxed{{\bf 6}_{(+1,-6,+2)}} \oplus {\color{ForestGreen}{{\bf 1}_{(-6,-6,+2)}}}) \oplus ({\color{BrickRed}{{\bf 15}_{(+2,+2,+2)}}} \oplus {\color{BrickRed}{{\bf 6}_{(-5,+2,+2)}}} )]$ \\[2mm]
\cline{2-3}
 & \multirow{6}{*}{--} & $[({\bf 6}_{(+1,+1,-3)} \oplus {\bf 1}_{(-6,+1,-3)}) \oplus {\bf 1}_{(0,-7,-3)}]$ \\
& &  $[({\bf 20}_{(+3,+3,-1)} \oplus {\bf 15}_{(-4,+3,-1)}) \oplus ( {\color{ForestGreen}{{\bf 15}_{(+2,-5,-1)}}} \oplus {\color{RoyalBlue}{{\bf 6}_{(-5,-5,-1)}}}]$ \\
& &  $[ ({\bf 15'}_{(-2,+5,+1)} \oplus {\color{RoyalBlue}{{\bf 6'}_{(+5,+5,+1)}}}) \oplus ({\color{BrickRed}{{\bf 20}_{(-3,-3,+1)}}} \oplus {\color{BrickRed}{{\bf 15'}_{(+4,-3,+1)}}}) ]$ \\
& & $[( {\color{ForestGreen}{{\bf 6'}_{(-1,-1,+3)}}} \oplus \boxed{ {\bf 1}_{(+6,-1,+3)} } ) \oplus {\color{BrickRed}{{\bf 1}_{(0,+7,+3)}}}]$ \\[2mm]
\hline
\end{tabular}}
\caption{Type~IIA with O$8$-plane branching rules for the embedding $\textrm{SL}(6) \times \mathbb{R}_3 \times \mathbb{R}_2 \times \mathbb{R}_1 \subset {\rm SL(7)} \times \mathbb{R}_2 \times \mathbb{R}_1$. The subscripts in the third column indicate $(\mathbb{R}_3,\mathbb{R}_2,\mathbb{R}_1)$-charges. We have highlighted the scalars ${\color{RoyalBlue}{e_{m}{}^{n}}}$, ${\color{RoyalBlue}{\Phi}}$, ${\color{ForestGreen}{B_{(2)}}}$, ${\color{ForestGreen}{B_{(6)}}}$ and ${\color{BrickRed}{C_{(p)}}}$ listed in (\ref{scalars_SL(6)_IIA_even})-(\ref{scalars_SL(6)_IIA_odd}). The physical internal derivatives $\tilde{\partial}_{\mathsf{m}} \equiv \partial^{\mathsf{m}}$ and $\partial_{7}$ have been put in a box for their quick identification too. This table should be understood as a continuation of Table~\ref{Table:SL8-SL7}.}
\label{Table:SL7-SL6_IIA}
\end{center}
\end{table}

The group-theoretical identification of the internal components of the various type~IIA fields gives
\begin{equation}
\label{scalars_SL(6)_IIA_even}
\begin{array}{rclcrclc}
e_{\mathsf m}{}^{\mathsf n} &\in& {(\bf{35}+\bf{1})}_{(0,0,0)} & \hspace{3mm} , & \hspace{3mm} B_{\mathsf{m}\mathsf{n}\mathsf{p}\mathsf{q}\mathsf{r}\mathsf{s}} &\in& {\bf{1}}_{(-6,-6,+2)} & ,
 \\[2mm]
e_{7}{}^{7} &\in& {\bf{1}}_{(0,0,0)} & \hspace{3mm} , & \hspace{3mm}  C_{\mathsf{m}\mathsf{n}\mathsf{p}\mathsf{q}\mathsf{r}} &\in& {\bf{6'}}_{(-1,-8,0)} &, \\[2mm]
\Phi &\in& {\bf{1}}_{(0,0,0)}  & \hspace{3mm} , & \hspace{3mm} C_{\mathsf{m}\mathsf{n}7} &\in& {\bf{15}}_{(+2,+2,+2)}  & ,
 \\[2mm]
B_{\mathsf{m}7} &\in& {\bf{6}}_{(+7,0,0)}  & \hspace{3mm} , & \hspace{3mm} C_{\mathsf{m}} &\in& {\bf{6}}_{(-5,+2,+2)} & ,
\end{array}
\end{equation}
where $15$ compact scalars must be subtracted from $e_{\mathsf{m}}{}^{\mathsf{n}}$ upon gauge-fixing of the internal $\textrm{SO}(6)$ local symmetry, namely, $e_{\mathsf{m}}{}^{\mathsf{n}} \in \textrm{GL}(6)/\textrm{SO}(6)$. The number of bosonic scalars in (\ref{scalars_SL(6)_IIA_even}) is $64-6-1=57$, the $6+1$ missing scalars being dual to the vectors $e_{\mu}{}^{\mathsf{m}}$ and $C_{\mu}$. The spinorial scalars are given by
\begin{equation}
\label{scalars_SL(6)_IIA_odd}
\begin{array}{rclcrclcrclc}
e_{\mathsf{m}}{}^{7} &\in& {\bf{6}}_{(-5,-5,-1)} & , & B_{\mathsf{m}\mathsf{n}\mathsf{p}\mathsf{q}\mathsf{r}7} &\in& {\bf{6'}}_{(-1,-1,+3)} & , &  C_{7} &\in& {\bf{1}}_{(0,+7,+3)} & ,
 \\[2mm]
e_{7}{}^{\mathsf{m}} &\in& {\bf{6'}}_{(+5,+5,+1)}  & , &  C_{\mathsf{m}\mathsf{n}\mathsf{p}\mathsf{q}7} &\in& {\bf{15'}}_{(+4,-3,+1)} & , \\[2mm]
B_{\mathsf{m}\mathsf{n}} &\in& {\bf{15}}_{(+2,-5,-1)}  & , & C_{\mathsf{m}\mathsf{n}\mathsf{p}} &\in& {\bf{20}}_{(-3,-3,+1)}  & .
\end{array}
\end{equation}
They add up to $64-1=63$ with the missing spinorial scalar being dual to the vector $e_{\mu}{}^{7}$. Finally, an explicit computation yields the bosonic fluxes displayed in Table~\ref{Table:O8_fluxes} together with additional spinorial fluxes which are projected out of the half-maximal theory.

\begin{table}[t]
\begin{center}
\renewcommand{\arraystretch}{1.8}
\begin{tabular}{|c|c|c|}
\hline
Fluxes &  Flux components  &  Embedding tensor   \\
\hline
\hline
\multirow{2}{*}{$\omega$} & $\omega_{\mathsf{m}\mathsf{n}}{}^{\mathsf{p}} \in ({\bf{84}+\bf{6}})_{(+1,-6,+2)}$ & $\theta^{\mathsf{m}\mathsf{n}8}{}_{\mathsf{p}}$  \\
\cline{2-3}
 & $\omega_{\mathsf{m}7}{}^{7} \in {\bf{6}}_{(+1,-6,+2)} $  &   $\theta^{\mathsf{m}78}{}_{7}$  \\
\hline
\hline
$H_{(1)}$ & $H_{\mathsf{m}} \in {\bf{6}}_{(+1,-6,+2)}$  &  $\frac{1}{\gamma}\theta^{\mathsf{m}8},\,\frac{3}{\gamma+\beta}\theta^{\mathsf{m}\mathsf{n}8}{}_{\mathsf{n}},\,\frac{1}{\delta}\theta^{\mathsf{m}78}{}_{7}$\\
\hline
$H_{(3)}$ & $H_{\mathsf{m}\mathsf{n}7} \in {\bf{15}}_{(+8,-6,+2)}$  & $-\theta^{\mathsf{m}\mathsf{n}8}{}_{7} $  \\
\hline
\hline
$F_{(2)}$ & $F_{\mathsf{m}\mathsf{n}} \in {\bf{15}}_{(-4,-4,+4)}$ &  $ \theta^{\mathsf{m}\mathsf{n}78}$ \\
\hline
$F_{(4)}$ & $F_{\mathsf{m}\mathsf{n}\mathsf{p}7} \in {\bf{20}}_{(+3,-4,+4)}$ &  $\theta^{\mathsf{m}\mathsf{n}\mathsf{p}8} $ \\
\hline
$F_{(6)}$ & $F_{\mathsf{m}\mathsf{n}\mathsf{p}\mathsf{q}\mathsf{r}\mathsf{s}} \in {\bf{1}}_{(0,-14,+2)}$ &  $\theta^{88}$ \\
\hline
\end{tabular}
\caption{Type~IIA fluxes in the O$8$-plane duality frame of half-maximal supergravity and their identification with embedding tensor components. Note that the Romans mass is not present in this duality frame as it corresponds to the spinorial representation $F_{(0)} \in {\bf{1}}_{(-6,+1,+5)}$.} 
\label{Table:O8_fluxes}
\end{center}
\end{table}

\subsubsection*{Orientifold interpretation and O$8$-planes}

The type~IIA flux models here can be string-theoretically described as type~IIA orientifold reductions including O$8$-planes (and D$8$-branes) located at
\begin{equation}
\label{O8_location}
\begin{array}{lll|lc|lc|lc|c}
x^{0} & x^{1} & x^{2} & \tilde{y}^{1}  & \tilde{y}^{2} & \tilde{y}^{3}  & \tilde{y}^{4}& \tilde{y}^{5} & \tilde{y}^{6} & y^{7}   \\
\hline
\times & \times & \times & \times & \times & \times & \times & \times & \times &
\end{array}  
\end{equation}
The O$8$-plane fills the external spacetime and the six internal directions $\tilde{y}^{\mathsf{m}}$. The internal target space involution $\sigma_{\textrm{O}8}$ reflects the unique transverse coordinate (and its derivative) on $\mathbb{T}^{7}$, namely,
\begin{equation}
\label{sigma_O8}
\begin{array}{cc}
\sigma_{\textrm{O}8} : & \hspace{5mm}  \tilde{y}^{\textsf{m}} \,\, \rightarrow\,\, \tilde{y}^{\textsf{m}} \hspace{5mm} , \hspace{5mm} y^{7} \,\, \rightarrow\,\,  - \, y^{7} \\[2mm]
 & \hspace{5mm}  \tilde{\partial}_{\textsf{m}} \,\, \rightarrow\,\, \tilde{\partial}_{\textsf{m}} \hspace{5mm} , \hspace{5mm}   \partial_{7} \,\, \rightarrow\,\,  - \, \partial_{7} 
\end{array}
\hspace{6mm} \textrm{ with } \hspace{6mm} 
\textsf{m} = 1,\ldots,6 \ .
\end{equation}
The corresponding orientifold action $\mathcal{O}_{\mathbb{Z}_{2}} = \Omega_P \, (-1)^{F_{L}} \, \sigma_{\textrm{O}8}$ acts on the type~IIA fields as summarised in Table~\ref{Table:O8_fields}. The $\mathcal{O}_{\mathbb{Z}_{2}}$-even and $\mathcal{O}_{\mathbb{Z}_{2}}$-odd fields in Table~\ref{Table:O8_fields} match the ones in (\ref{scalars_SL(6)_IIA_even}) and (\ref{scalars_SL(6)_IIA_odd}), respectively. The same matching holds at the level of the fluxes.

Upon use of the embedding tensor/flux dictionary in Table~\ref{Table:O8_fluxes}, the computation of the QC's in (\ref{QC_N=8}) gives rise to the following constraints on the type~IIA fluxes. There is the nilpotency condition $(D^2=0)$ of the $D = d + \omega$ twisted exterior derivative on the internal space. In addition, there are sourceless Bianchi identities of the form $DH_{(1)} = 0$, $DH_{(3)} + \alpha H_{(1)}\wedge H_{(3)}= 0$, $DF_{(2)} +\beta_{2} H_{(1)} \wedge F_{(2)} = 0$ and $DF_{(4)} +\beta_{4} H_{(1)} \wedge F_{(4)} - H_{(3)} \wedge F_{(2)} = 0$ reflecting the absence of NS$7$-branes, NS$5$-branes, D$6$-branes and D$4$-branes in the compactification scheme, respectively. Note also that the constraint $DF_{(6)} = 0$, reflecting the absence of D$2$-branes, trivialises with the permitted fluxes in Table~\ref{Table:O8_fluxes}. One is then left with no non-trivial constraints on the gauge fluxes. In other words, no flux-induced tadpole can be generated for the O$8$/D$8$ sources. Therefore, these type~IIA flux models satisfy the extra constraints in \eqref{QC_N=16_extra} and are embeddable into maximal three-dimensional supergravity.

\begin{table}[t]
\begin{center}
\renewcommand{\arraystretch}{1.7}
\begin{tabular}{|c|c|c|c|c|}
\hline
Fields & $\Omega_{P}$ & $(-1)^{F_{L}}$   &  $\sigma_{\textrm{O}8}$ & $\mathcal{O}_{\mathbb{Z}_{2}}$  \\
\hline
\hline
$e_{7}{}^{7} \, , \, e_{\mathsf{m}}{}^{\mathsf{n}}$ & \multirow{2}{*}{$+$}  & \multirow{2}{*}{$+$}  & $+$  & $+$  \\
\cline{1-1}\cline{4-5}
$e_{7}{}^{\mathsf{m}} \, , \, e_{\mathsf{m}}{}^{7}$ &  & & $-$  & $-$ \\
\hline
\hline
$\Phi$ & $+$ & $+$ & $+$  & $+$ \\
\hline
\hline
$B_{\mathsf{m}\mathsf{n}}$ & \multirow{2}{*}{$-$} & \multirow{2}{*}{$+$} & $+$  & $-$  \\
\cline{1-1}\cline{4-5}
$B_{\mathsf{m}7}$ &  & & $-$  & $+$ \\
\hline
\hline
$B_{\mathsf{m}\mathsf{n}\mathsf{p}\mathsf{q}\mathsf{r}7}$ & \multirow{2}{*}{$+$}  & \multirow{2}{*}{$+$} & $-$  & $-$  \\
\cline{1-1}\cline{4-5}
$B_{\mathsf{m}\mathsf{n}\mathsf{p}\mathsf{q}\mathsf{r}\mathsf{s}}$ & & & $+$  & $+$ \\
\hline
\end{tabular}
\hspace{5mm}
\begin{tabular}{|c|c|c|c|c|}
\hline
Fields & $\Omega_{P}$ & $(-1)^{F_{L}}$   &  $\sigma_{\textrm{O}8}$ & $\mathcal{O}_{\mathbb{Z}_{2}}$  \\
\hline
\hline
$C_{\mathsf{m}}$ & \multirow{2}{*}{$-$} & \multirow{2}{*}{$-$}  & $+$  & $+$  \\
\cline{1-1}\cline{4-5}
$C_{7}$ &  & &  $-$  & $-$ \\
\hline
\hline
$C_{\mathsf{m}\mathsf{n}\mathsf{p}}$ & \multirow{2}{*}{$+$}  & \multirow{2}{*}{$-$}  & $+$  & $-$  \\
\cline{1-1}\cline{4-5}
$C_{\mathsf{m}\mathsf{n}7}$ &  &  & $-$  & $+$ \\
\hline
\hline
$C_{\mathsf{m}\mathsf{n}\mathsf{p}\mathsf{q}7}$ & \multirow{2}{*}{$-$}  &  \multirow{2}{*}{$-$}  &  $-$  & $-$  \\
\cline{1-1}\cline{4-5}
$C_{\mathsf{m}\mathsf{n}\mathsf{p}\mathsf{q}\mathsf{r}} $ & & & $+$  & $+$ \\
\hline
\end{tabular}
\caption{Grading of the various type~IIA fields under the O$8$-plane orientifold action ${\mathcal{O}_{\mathbb{Z}_{2}} = \Omega_{P} \, (-1)^{F_{L}} \, \sigma_{\textrm{O}8}}$. The  $\mathcal{O}_{\mathbb{Z}_{2}}$-even fields match the ones in (\ref{scalars_SL(6)_IIA_even}), whereas the $\mathcal{O}_{\mathbb{Z}_{2}}$-odd fields match the ones in (\ref{scalars_SL(6)_IIA_odd}).} 
\label{Table:O8_fields}
\end{center}
\end{table}

\subsection{Type~IIB with O$9$-planes: type~I/Heterotic}
\label{sec:O9-plane}

\begin{table}[t!]
\begin{center}
\scalebox{0.78}{
\renewcommand{\arraystretch}{1.7}
\begin{tabular}{|c|c||c|}
     \hline 
     & Half-Maximal &  ${\rm SL(7)} \times \mathbb{R}_2 \times \mathbb{R}_1$ \\ 
     \hline \hline
     & -- &   $[{\bf{7'}}_{(-1,-1)} \oplus {\bf{1}}_{(+7,-1)}] \oplus [{\bf{7}}_{(+1,+1)} \oplus {\bf{1}}_{(-7,+1)}]$ \\
     \hline\hline  
     \multirow{4}{*}{\rotatebox{90}{Scalars\,\,}} &  \multirow{2}{*}{$L^{MN}$}   & $[{\bf 21'}_{(-2,-2)} \oplus {\bf 7'}_{(+6,-2)}] \oplus [{\color{BrickRed}{{\bf 7'}_{(-8,0)}}} \oplus {\color{RoyalBlue}{{\bf (48+1)}_{(0,0)}}} \oplus {\bf 7}_{(+8,0)}] $ 
     \\
     &   & ${\color{RoyalBlue}{{\bf 1}_{(0,0)}}}  \oplus [\boxed{  {\bf 7}_{(-6,+2)} } \oplus {\color{BrickRed}{{\bf 21}_{(+2,+2)}}}]$
     \\[2mm] \cline{2-3}
     & \multirow{2}{*}{--} &   $ [{\bf 7}_{(+1,-3)} \oplus {\color{BrickRed}{{\bf 1}_{(-7,-3)}}}] \oplus [{\bf 35}_{(+3,-1)} \oplus {\color{ForestGreen}{{\bf 21}_{(-5,-1)}}}] $
     \\
     &  & $[  {\color{BrickRed}{{\bf 35'}_{(-3,+1)}}} \oplus {\bf 21'}_{(+5,+1)}] \oplus [{\color{ForestGreen}{{\bf 7'}_{(-1,+3)}}} \oplus {\bf 1}_{(+7,+3)}]$ 
     \\
     \hline \hline
      \multirow{20}{*}{\rotatebox{90}{Embedding Tensor / Fluxes}} & $\theta$ &   ${\bf 1}_{(0,0)}$  
     \\\cline{2-3}
     & \multirow{3}{*}{$\theta_{MN}$} &   $[{\bf 28'}_{(-2,-2)} \oplus {\bf 7'}_{(+6,-2)} \oplus {\bf 1}_{(+14,-2)}]$
     \\
     &  & $[{\bf 7'}_{(-8,0)} \oplus {\bf 48}_{(0,0)} \oplus {\bf 1}_{(0,0)} \oplus {\bf 7}_{(+8,0)}]$
     \\
     &  & $[{\bf 28}_{(+2,+2)} \oplus {\color{RoyalBlue}{{\bf 7}_{(-6,+2)}}} \oplus {\color{BrickRed}{{\bf 1}_{(-14,+2)}}}]$
     \\\cline{2-3}
     & \multirow{6}{*}{$\theta_{MNPQ}$}   & $[{\bf 35}_{(-4,-4)} \oplus {\bf 35'}_{(+4,-4)}] \oplus [{\bf 21'}_{(-2,-2)} \oplus {\bf 7'}_{(+6,-2)}]$
     \\
     &  & $[{\bf 35'}_{(-10,-2)} \oplus {\bf 21'}_{(-2,-2)} \oplus {\bf 224'}_{(-2,-2)} \oplus {\bf 140'}_{(+6,-2)}]$
     \\ \cline{3-3}
     & &   $[{\bf 140'}_{(-8,0)} \oplus {\bf 392}_{(0,0)} \oplus {\bf 48}_{(0,0)} \oplus {\bf 140}_{(+8,0)}]$
     \\
     &  & $[{\bf 7'}_{(-8,0)} \oplus {\bf 48}_{(0,0)} \oplus {\bf 1}_{(0,0)} \oplus {\bf 7}_{(+8,0)}] \oplus {\bf 1}_{(0,0)}$
     \\ \cline{3-3}
     & &  $[{\color{BrickRed}{{\bf 35}_{(-4,+4)}}} \oplus{\bf 35'}_{(+4,+4)}] \oplus [{\color{RoyalBlue}{{\bf 7}_{(-6,+2)}}} \oplus {\bf 21}_{(+2,+2)}]$
     \\
     &  & $[{\bf 35}_{(+10,+2)} \oplus {\bf 21}_{(+2,+2)} \oplus {\bf 224}_{(+2,+2)} \oplus {\color{RoyalBlue}{{\bf 140}_{(-6,+2)}}}]$
     \\\cline{2-3}
     & \multirow{10}{*}{--}  &   $[{\bf 7'}_{(-1,-5)} \oplus {\bf 1}_{(+7,-5)}] \oplus [{\bf 1}_{(-7,-3)} \oplus {\bf 7}_{(+1,-3)}]$
     \\
     &  & $[{\bf 48}_{(-7,-3)} \oplus {\bf 140}_{(+1,-3)} \oplus {\bf 7}_{(+1,-3)} \oplus {\bf 21}_{(+9,-3)}]$
     \\ \cline{3-3}
     & &  $[{\bf 21}_{(-5,-1)} \oplus {\bf 35}_{(+3,-1)}]$
     \\
     &  & $[{\bf 224}_{(-5,-1)} \oplus {\bf 210}_{(+3,-1)} \oplus {\bf 35}_{(+3,-1)} \oplus {\bf 35'}_{(+11,-1)}]$
     \\
     &  & $[{\color{BrickRed}{{\bf 7}_{(-13,-1)}}} \oplus {\bf 21}_{(-5,-1)} \oplus {\bf 28}_{(-5,-1)} \oplus {\bf 112}_{(+3,-1)}]$
     \\ \cline{3-3}
     & &   $[{\bf 21'}_{(+5,+1)} \oplus {\bf 35'}_{(-3,+1)}]$
     \\
     &  & $[{\bf 224'}_{(+5,+1)} \oplus {\bf 210'}_{(-3,+1)} \oplus {\bf 35'}_{(-3,+1)} \oplus {\color{ForestGreen}{{\bf 35}_{(-11,+1)}}}]$
     \\
     &  & $[{\bf 7'}_{(+13,+1)} \oplus {\bf 21'}_{(+5,+1)} \oplus {\bf 28'}_{(+5,+1)} \oplus {\bf 112'}_{(-3,+1)}]$
     \\ \cline{3-3}
     & &   $[{\bf 7}_{(+1,+5)} \oplus {\color{ForestGreen}{{\bf 1}_{(-7,+5)}}}] \oplus [{\bf 7'}_{(-1,+3)} \oplus {\bf 1}_{(+7,+3)}]$
     \\
     &  & $[{\bf 48}_{(+7,+3)} \oplus {\bf 140'}_{(-1,+3)} \oplus {\bf 7'}_{(-1,+3)} \oplus {\color{BrickRed}{{\bf 21'}_{(-9,+3)}}}]$
     \\
     \hline
\end{tabular}
}
\caption{Type~I branching rules for the embedding ${\textrm{SO(8,8)} \supset {\rm SL(7)} \times \mathbb{R}_2 \times \mathbb{R}_1}$. The subscripts in the third column indicate $(\mathbb{R}_2,\mathbb{R}_1)$-charges. We have highlighted the scalars ${\color{RoyalBlue}{e_{m}{}^{n}}}$, ${\color{RoyalBlue}{\Phi}}$, ${\color{ForestGreen}{B_{(2)}}}$, ${\color{ForestGreen}{B_{(6)}}}$ and ${\color{BrickRed}{C_{(p)}}}$ listed in (\ref{scalars_type_I_bosonic})-(\ref{scalars_type_I_fermionic}) as well as their associated fluxes in (\ref{fluxes_type_I_bosonic})-(\ref{fluxes_type_I_fermionic}). The physical internal derivatives $\tilde{\partial}_{m}$ have been put in a box for their quick identification.} 
\label{Table:SO(8,8)-SL7}
\end{center}
\end{table}

Let us now consider the duality frame of the type~I theory, \textit{i.e.} type~IIB with O$9$-planes. This duality frame is obtained upon further T-dualisation of the $y^{\hat{a}}$ coordinates in (\ref{sigma_O5}). As a result of this procedure, the physical coordinates and their associated derivatives turn out to be
\begin{equation}
\label{partial_SL(7)_typeI}
\tilde{y}^{m} \in {\bf{7}'}_{(+6,-2)} 
\hspace{10mm} \textrm{ and } \hspace{10mm}
\tilde{\partial}_{m} \in {\bf{7}}_{(-6,+2)} \ ,
\end{equation}
so that they all become bosonic as far as their SO$(8,8)$ origin is concerned (see Table~\ref{Table:SO(8,8)-SL7}). A group-theoretical identification of the internal components of the type~IIB fields yields
\begin{equation}
\label{scalars_type_I_bosonic}
\begin{array}{rclcrclc}
e_{m}{}^{n} &\in& ({\bf{48+1}})_{(0,0)} & \hspace{3mm}  , & \hspace{3mm} C_{(2)} &\in& {\bf{21}}_{(+2,+2)}  & ,
 \\[2mm]
\Phi &\in& {\bf{1}}_{(0,0)}  & \hspace{3mm}  , & \hspace{3mm} C_{(6)} &\in&  {\bf{7'}}_{(-8,0)} & ,
\end{array}
\end{equation}
together with
\begin{equation}
\label{scalars_type_I_fermionic}
\begin{array}{rclcrclc}
B_{(2)} &\in& {\bf{21}}_{(-5,-1)} & \hspace{3mm}  , & \hspace{3mm}  C_{(0)} &\in& {\bf{1}}_{(+7,+3)}  & ,
 \\[2mm]
B_{(6)} &\in&  {\bf{7'}}_{(-1,+3)} & \hspace{3mm}  , & \hspace{3mm}  C_{(4)} &\in&  {\bf{35'}}_{(-3,+1)}  & ,
\end{array}
\end{equation}
where, as before, $21$ compact scalars must be subtracted from $e_{m}{}^{n}$ upon gauge-fixing of the internal $\textrm{SO}(7)$ local symmetry, namely, $e_{m}{}^{n} \in \textrm{GL}(7)/\textrm{SO}(7)$. The $64-7=57$ bosonic scalars in (\ref{scalars_type_I_bosonic}) must be completed with $7$ additional scalars dual to the vectors $e_{\mu}{}^{m}$ in order to match the $64$ scalars of half-maximal supergravity. The spinorial scalars in (\ref{scalars_type_I_fermionic}) correctly add up to $64$. Acting on the scalars (\ref{scalars_type_I_bosonic}) and (\ref{scalars_type_I_fermionic}) with the physical derivatives in (\ref{partial_SL(7)_typeI}) produces fluxes of the form
\begin{equation}
\label{fluxes_type_I_bosonic}
\begin{array}{rclcrclc}
\omega_{m n}{}^{p} &\in& ({\bf{140}}+{\bf{7}})_{(-6,+2)} & \hspace{3mm}  , & \hspace{3mm}  F_{(3)} &\in& {\bf{35}}_{(-4,+4)}  & ,
 \\[2mm]
H_{(1)} &\in& {\bf{7}}_{(-6,+2)}  & \hspace{3mm}  , & \hspace{3mm}  F_{(7)} &\in&  {\bf{1}}_{(-14,+2)} & ,
\end{array}
\end{equation}
together with
\begin{equation}
\label{fluxes_type_I_fermionic}
\begin{array}{rclcrclc}
H_{(3)} &\in& {\bf{35}}_{(-11,+1)} & \hspace{3mm}  , & \hspace{3mm}  F_{(1)} &\in& {\bf{7}}_{(+1,+5)}  & ,
 \\[2mm]
H_{(7)} &\in&  {\bf{1}}_{(-7,+5)} & \hspace{3mm}  , & \hspace{3mm}  F_{(5)} &\in&  {\bf{21'}}_{(-9,+3)}  & .
\end{array}
\end{equation}

Truncating away the scalars in (\ref{scalars_type_I_fermionic}) and the fluxes in (\ref{fluxes_type_I_fermionic}), which descend from spinorial representations of $\textrm{SO}(8,8)$, the field content of the type~I models is given by
\begin{equation}
\label{Fields&Fluxes_bosonic_Type_I}
\begin{array}{lcl}
\textrm{Scalars} & : & e_{m}{}^{n} \in \frac{\textrm{GL}(7)}{\textrm{SO}(7)}
\hspace{3mm} , \hspace{3mm}
\Phi
\hspace{3mm} , \hspace{3mm}
C_{(2)}
\hspace{3mm} , \hspace{3mm}
C_{(6)} \ , \\[2mm]
\textrm{Fluxes} & : & 
\omega_{mn}{}^{p}
\hspace{3mm} , \hspace{3mm}
H_{(1)}
\hspace{3mm} , \hspace{3mm}
F_{(3)} 
\hspace{3mm} , \hspace{3mm}
F_{(7)} \ .
\end{array}
\end{equation}
Note that the content of fields and fluxes in (\ref{Fields&Fluxes_bosonic_Type_I}) becomes the universal NS--NS sector of the Heterotic string upon a S-duality: $\Phi \rightarrow - \Phi$, $C_{(2)} \rightarrow B_{(2)}$ and $C_{(6)} \rightarrow B_{(6)}$, with the corresponding fluxes also being exchanged. The fluxes in (\ref{Fields&Fluxes_bosonic_Type_I}) induce a gauging with embedding tensor components
\begin{equation}
\begin{array}{rclcrclc}
\label{ET_type_I-theory}
\theta^{mn8}{}_{p} &=& \omega_{mn}{}^{p}   \, +\,\tfrac{\gamma-\alpha+\beta}{3} \,\delta^{p}_{[m} \, H_{n]} &
\hspace{4mm} , & \hspace{4mm}
\theta^{mnp8} &=&  F_{mnp}
\hspace{4mm}  & , \\[2mm]
\theta^{m8} &=& \,\gamma H_{m} &
\hspace{4mm} , & \hspace{4mm}
\theta^{88} &=& \tfrac{1}{7!} \, \varepsilon^{mnpqrst} F_{mnpqrst}  & .
\end{array}
\end{equation}
An explicit computation of the QC's in (\ref{QC_N=8}) yields
\begin{equation}
\label{QC_type_I}
\omega_{[mn}{}^{r} \, \omega_{p]r}{}^{s} = 0
\hspace{5mm} , \hspace{5mm}
\omega_{[mn}{}^{r} \, F_{pq]r} +(\beta-\alpha) H_{[m} \, F_{npq]}  = 0 
\hspace{5mm} , \hspace{5mm}
\omega_{mn}{}^{r} \, H_{r} = 0 \ .
\end{equation}

\subsubsection*{Orientifold interpretation and O$9$-planes}

The O$9$-plane is somehow special. Being placed as
\begin{equation}
\label{O9_location}
\begin{array}{lll|lc|lc|lc|c}
x^{0} & x^{1} & x^{2} & \tilde{y}^{1}  & \tilde{y}^{2} & \tilde{y}^{3}  & \tilde{y}^{4}& \tilde{y}^{5} & \tilde{y}^{6} & \tilde{y}^{7}   \\
\hline
\times & \times &\times & \times & \times &  \times & \times &  \times & \times &  \times
\end{array}   
\end{equation}
it does fill the entire ten-dimensional spacetime. As a consequence, the internal target space involution $\sigma_{\textrm{O}9}$ does not reflect any of the internal coordinates (and derivatives) on $\mathbb{T}^{7}$, namely,
\begin{equation}
\label{sigma_O9}
\begin{array}{cc}
\sigma_{\textrm{O}9} : & \hspace{5mm}  \tilde{y}^{m} \,\, \rightarrow\,\, \tilde{y}^{m} \\[2mm]
 & \hspace{5mm}  \tilde{\partial}_{m} \,\, \rightarrow\,\,  \tilde{\partial}_{m}
\end{array}
\hspace{10mm} \textrm{ with } \hspace{10mm} m = 1 ,\ldots, 7 \ .
\end{equation}
In the string-theoretic language of orientifold reductions, the QC's in (\ref{QC_type_I}) are reinterpreted as follows. The first condition is again the nilpotency condition $(D^2=0)$ of the twisted exterior derivative $D = d + \omega$ on the internal space. The second and third conditions are interpreted as the sourceless Bianchi identities $DF_{(3)} +\beta_{3} H_{(1)} \wedge F_{(3)} = 0$ and $DH_{(1)} = 0$ forbidding D$5$-branes and NS$7$-branes, respectively. In the S-dual Heterotic picture, these map into the absence of NS$5$-branes and (again) NS$7$-branes. Finally, although in principle permitted by half-maximal supersymmetry, no flux-induced tadpole can be defined for the O$9$/D$9$ sources. Therefore, these type I/Heterotic flux models are embeddable into maximal three-dimensional supergravity.

\begin{table}[t]
\begin{center}
\renewcommand{\arraystretch}{1.7}
\begin{tabular}{|c|c|c|c|c|c|c|c|c|c|}
\hline
Fields &    $e_{n}{}^{p}$ &  $\Phi$ &  $B_{(2)}$ & $B_{(6)}$ &  $C_{(0)}$ & $C_{(2)}$ & $C_{(4)}$ & $C_{(6)}$ \\
\hline
$\Omega_{P}$ & $+$  & $+$  & $-$ & $-$ & $-$ & $+$ &  $-$   &  $+$   \\
\hline
$\sigma_{\textrm{O}9}$  & $+$  & $+$ & $+$ & $+$ & $+$ & $+$ & $+$ & $+$    \\
\hline
$\mathcal{O}_{\mathbb{Z}_{2}}$  & $+$  & $+$ & $-$ & $-$ &  $-$ & $+$ & $-$  & $+$  \\
\hline\hline
Fluxes &    $\omega_{mn}{}^{p}$ &  $H_{(1)}$ &  $H_{(3)}$ & $H_{(7)}$ &   $F_{(1)}$ & $F_{(3)}$ & $F_{(5)}$ & $F_{(7)}$ \\
\hline  
$\mathcal{O}_{\mathbb{Z}_{2}}$  & $+$  & $+$ & $-$ & $-$ & $-$ & $+$ & $-$ & $+$  \\
\hline
\end{tabular}
\caption{Grading of the various type~IIB fields and fluxes under the O$9$-plane orientifold action $\mathcal{O}_{\mathbb{Z}_{2}} = \Omega_{P} \, \sigma_{\textrm{O}9}$.} 
\label{Table:O9_fields_fluxes}
\end{center}
\end{table}

\section{Flux models and vacua}
\label{sec:flux_models&vacua}

In the previous section we have established a precise correspondence between single (spacetime filling) O$p$-plane orientifold reductions of type~II strings in the presence of background fluxes and half-maximal gauged supergravities in three dimensions. Equipped with this correspondence we can start charting the landscape of half-maximal flux compactifications in three dimensions.

\subsection{Invariant sectors}
\label{sec:invariant_sectors}

Half-maximal supergravity coupled to eight matter multiplets contains $64$ scalars. In the light-cone basis of $\textrm{SO}(8,8)$ they parameterise the scalar geometry (\ref{scalar_geometry_N=8}) in terms of a coset representative of the form
\begin{equation}
\label{coset_parameterisation_N=8}
\mathcal{V} \,\, = \,\, \left( 
\begin{matrix}  \mathbb{I} & 0 \\ 
\boldsymbol{b} & \mathbb{I}
\end{matrix} 
\right) 
\left( 
\begin{matrix}  \boldsymbol{e} & 0 \\ 
0 & \boldsymbol{e}^{-T}
\end{matrix} 
\right) 
\,\, \in \,\, \frac{\textrm{SO}(8,8)}{\textrm{SO}(8) \times \textrm{SO}(8)} \ ,
\end{equation}
where $\boldsymbol{e} \in \textrm{GL}(8)/\textrm{SO}(8)$ and $\boldsymbol{b}=-\boldsymbol{b}^{T}$ are $8 \times 8$ matrices containing $36$ and $28$ scalars, respectively. The scalar-dependent and symmetric matrix $M_{MN}$ entering (\ref{L_bosonic}) takes the form
\begin{equation}
\label{M_parameterisation_N=8}
M \,\, = \,\, \mathcal{V} \, \mathcal{V}^{T}  \,\, = \,\, \left( 
\begin{matrix}  \boldsymbol{g} & -\boldsymbol{g} \, \boldsymbol{b} \\ 
\boldsymbol{b} \, \boldsymbol{g} & \boldsymbol{g}^{-1} - \boldsymbol{b} \, \boldsymbol{g} \, \boldsymbol{b}
\end{matrix} 
\right) \,\, \in \,\, \textrm{SO}(8,8) \ ,
\end{equation}
with $\boldsymbol{g} = \boldsymbol{e} \, \boldsymbol{e}^{T}$. In addition, there is a (worrying) large number of both metric and gauge flux components which complicates the analysis of the vacuum structure of the models. In order to derive more tractable models, still able to capture relevant physics, we will require invariance of the fields and fluxes under an additional $\,\mathbb{Z}_{2}^{2}\,$ or $\,\textrm{SO}(3)\,$ symmetry.

A few comments are in order at this point. Firstly, and even though we are looking at group-theoretically consistent sectors of half-maximal supergravity that are invariant under a $\,\mathbb{Z}_{2}^{2}\,$ or $\,\textrm{SO}(3)\,$ symmetry, we are still finding extrema of the scalar potential of the full theory, which  contains additional fields. Therefore, in order to assess the perturbative stability of a given extremum of the scalar potential, the complete set of $64$ scalar masses must be computed (as we will do). Secondly, it will turn out that the three-dimensional flux vacua generically come along with a number of massless (or unstabilised) scalars. However, in some particular cases (see Section~\ref{Delta's&scale-separation}), the eight dilaton-like moduli inside $\,\boldsymbol{e}\,$ happen to be stabilised. These are the moduli codifying the string coupling constant as well as various phenomenologically relevant scales in the compactification scheme. Lastly, both the $\,\mathbb{Z}_{2}^{2}\,$ and $\,\textrm{SO}(3)\,$ invariant sectors should be describable as $\mathcal{N}=2$ supersymmetric models by themselves, although we leave this reformulation for future work.

\subsubsection{The $\mathbb{Z}_{2}^{2}$ invariant sector: the eight-chiral model}
\label{sec:eight-chiral_model}

The first symmetry we will mod out the theory by is a $\mathbb{Z}_{2}^{2}$ discrete symmetry. It consists of two independent elements acting on the vector index $\,A=1,\ldots,8\,$ of $\,\textrm{SL}(8) \subset \textrm{SO}(8,8)\,$ as
\begin{equation}
\label{Z2xZ2_generators}
\begin{array}{cccccc}
\mathbb{Z}_{2}^{(\alpha)} : &  \left( \, y^{1} \, , \, y^{2} \, , \, y^{3} \, , \, y^{4} \, , \, y^{5} \, , \, y^{6} \, , \, y^{7} \, , \, y^{8}  \,  \right) & \rightarrow &  \left( \, -y^{1} \, , \, -y^{2} \, , \, -y^{3} \, , \, -y^{4} \, , \, y^{5} \, , \, y^{6} \, , \, y^{7} \, , \, y^{8}  \,  \right)     \\[2mm]
\mathbb{Z}_{2}^{(\beta)} : &  \left( \, y^{1} \, , \, y^{2} \, , \, y^{3} \, , \, y^{4} \, , \, y^{5} \, , \, y^{6} \, , \, y^{7} \, , \, y^{8}  \,  \right) & \rightarrow &  \left( \, -y^{1} \, , \, -y^{2} \, , \, y^{3} \, , \, y^{4} \, , \, -y^{5} \, , \, -y^{6} \, , \, y^{7} \, , \, y^{8}  \,  \right) 
\end{array}
\end{equation}
together with the additional elements $\mathbb{I}$ and $\mathbb{Z}_{2}^{(\alpha)}\mathbb{Z}_{2}^{(\beta)}$. This $\mathbb{Z}_{2}^{2}$ discrete symmetry leaves invariant the gravitini $(\psi_{\mu}{}^{7},\psi_{\mu}{}^{8})$, hence the $\mathcal{N}=2$ supersymmetry of this sector, and $8+8$ non-compact scalars within the coset space (\ref{scalar_coset_N=8}). In the Lorentzian basis, these are associated to the generators $L^{\mathcal{I}r}$ in (\ref{coset_rep_Ltz}) with $(\mathcal{I},r)=(1,1), (2,2), \ldots,(8,8)$ as well as $(\mathcal{I},r)=(1,2),(2,1),(3,4),(4,3),(5,6),(6,5),(7,8),(8,7)$. The scalar geometry describing the $\mathbb{Z}_{2}^{2}$ invariant sector is identified with
\begin{equation}
\label{scalar_geometry_8_chiral}
{\mathcal{M}}_{\textrm{scal}} = \left[ \frac{\textrm{SL}(2)}{\textrm{SO}(2)}\right]^{8} \subset \frac{{\rm SO}(8,8)}{{\rm SO}(8) \times {\rm SO}(8)}  \ ,
\end{equation}
and comprises eight copies of the Poincar\'e disk $\textrm{SL}(2)/\textrm{SO}(2)$ parameterised by eight complex scalars 
\begin{equation}
T_{I}=B_{I} + i A_{I}
\hspace{8mm} \textrm{ and } \hspace{8mm}
U_{I}=u_{I}+i \mu_{I} 
\hspace{8mm} \textrm{ with } \hspace{8mm}
I=1,\ldots,4 \ .
\end{equation}
Using the upper-half plane parameterisation of the Poincar\'e disk, \textit{i.e.} $A_{I}>0$, $\mu_{I}>0$, the coset representative in (\ref{coset_parameterisation_N=8}) adopts the block-diagonal structure
\begin{equation}
\label{coset_blocks_Z2^3}
\boldsymbol{e} \,\, = \,\, \left( 
\begin{matrix}  
\boldsymbol{e}_{1} & 0 & 0 & 0 \\ 
0 & \boldsymbol{e}_{2} & 0 & 0 \\ 
0  & 0 & \boldsymbol{e}_{3} & 0 \\ 
0 & 0 & 0 & \boldsymbol{e}_{4}
\end{matrix} 
\right)
\hspace{8mm} , \hspace{8mm}
\boldsymbol{b} \,\, = \,\,
\left( 
\begin{matrix}  
\boldsymbol{b}_{1} & 0 & 0 & 0 \\ 
0 & \boldsymbol{b}_{2} & 0 & 0 \\ 
0  & 0 & \boldsymbol{b}_{3} & 0 \\ 
0 & 0 & 0 & \boldsymbol{b}_{4}
\end{matrix}
\right)  \ ,
\end{equation}
with
\begin{equation}
\label{coset_blocks_2x2}
\boldsymbol{e}_{I} \,\, = \,\, (A_{I} \, \mu_{I})^{-\frac{1}{2}} \left( 
\begin{matrix}  
1 & 0 \\ 
u_{I} & \mu_{I}  
\end{matrix} 
\right)
\hspace{8mm} , \hspace{8mm}
\boldsymbol{b}_{I} \,\, = \,\,
\left( 
\begin{matrix}  
0 & B_{I}  \\ 
-B_{I} & 0 
\end{matrix}
\right)  \ .
\end{equation}

Constructing the scalar-dependent matrix in (\ref{M_parameterisation_N=8}) from (\ref{coset_blocks_Z2^3})-(\ref{coset_blocks_2x2}), and truncating away the vectors in the theory, the Lagrangian (\ref{L_bosonic}) reduces to an Einstein-scalar model of the form
\begin{equation}
\label{L_8_chiral}
\mathcal{L} = - \frac{1}{4} \, e \,  R  + \frac{1}{8} \, e \, \sum_{I=1}^{4} \frac{(\partial B_{I})^{2}+(\partial A_{I})^{2}}{A_{I}^{2}} + \frac{1}{8} \, e \, \sum_{I=1}^{4} \frac{(\partial u_{I})^{2}+(\partial \mu_{I})^{2}}{\mu_{I}^{2}}  - e \, V  \ ,
\end{equation}
where the scalar potential $V$ depends on the M-theory/string fluxes compatible with the specific duality frame under consideration. From now on, we will refer to the $\mathbb{Z}_{2}^{2}$ invariant sector of half-maximal supergravity as the eight-chiral model.\footnote{Although there are no chiral spinors in three dimensions, we have coined the term from its four-dimensional counterpart, \textit{i.e.} the seven-chiral model (see \textit{e.g.} \cite{Guarino:2019snw,Chamorro-Burgos:2023vzw}).} This model contains the eight dilatons $(A_{I},\mu_{I})$ associated with the Cartan subalgebra of $\textrm{SO}(8,8)$ and their partners $(B_{I},u_{I})$. From (\ref{coset_blocks_Z2^3})-(\ref{coset_blocks_2x2}) one sees that $(A_{I},\mu_{I} \,; u_{I})$ belong to $\boldsymbol{e} \in \textrm{GL}(8)/\textrm{SO}(8)$ whereas the axions $B_{I}$ belong to the completion into $\textrm{SO}(8,8)/(\textrm{SO}(8) \times \textrm{SO}(8))$. The eight-chiral model presented here provides a natural extension of the models investigated in \cite{Farakos:2020phe,Emelin:2021gzx} which only included the eight dilatons $(A_{I},\mu_{I})$ (see Section~\ref{Flux_vacua_IIA_O2}).

In the M-theory duality frame of Section~\ref{sec:M-theory}, the maximal subgroup $\textrm{GL}(8) \subset \textrm{SO}(8,8)$ is realised purely geometrically. The parameterisation in (\ref{coset_blocks_Z2^3})-(\ref{coset_blocks_2x2}) then gives rise to a matrix $\boldsymbol{g}$ in (\ref{M_parameterisation_N=8}) that is interpreted as the metric on the internal $\mathbb{T}^{8}$. It is
\begin{equation}
\label{T8_metric}
ds^2_{8} = \sum_{I=1}^{4} \frac{A^{-1}_{I}}{\textrm{Im}U_{I}} \left[ \left(\eta^{2I-1}\right)^{2} + |U_{I}|^{2} \left(\eta^{2I}\right)^{2} + 2 \, \left(\textrm{Re}U_{I}\right) \, \eta^{2I-1} \, \eta^{2I}\right] \ ,
\end{equation}
where $\eta^{A}=\bigoplus_{I=1}^{4}(\eta^{2I-1},\eta^{2I})$ denotes a basis of one-forms in the internal ${\mathbb{T}^{8}=\bigotimes_{I=1}^{4} \mathbb{T}_{I}^{2}}$. Looking at the metric (\ref{T8_metric}), the scalars $A_{I}$ play the role of K\"ahler moduli parameterising the size of the four $\mathbb{T}_{I}^{2} \subset \mathbb{T}^{8}$, whereas $U_{I}$ are complex structure moduli parameterising the shape. Note that $\sqrt{|\boldsymbol{g}|}=(A_{1}A_{2}A_{3}A_{4})^{-1}$ in agreement with the previous identification of moduli fields in the M-theory duality frame. A similar identification holds in the type~IIA with O$2$-planes duality frame of Section~\ref{sec:O2-plane} that appears after dimensional reduction, except that the radius of the $8^{\textrm{th}}$ circle on $\mathbb{T}^{8}$ is no longer interpreted geometrically but becomes the type~IIA dilaton $\Phi$. Further moving to other O$p$-plane duality frames involves the action of T-dualities which generically invert the radius of the internal circles upon which they act.

\subsubsection{The $\textrm{SO}(3)$ invariant sector: the RSTU-model}

The second symmetry we will mod out the theory by is a $\textrm{SO}(3)$ symmetry embedded in the $\textrm{SO}(8,8)$ duality group of half-maximal supergravity as
\begin{equation}
\label{SO(8,8)_to_SO(3)_embedding}
\begin{array}{cccccc}
\textrm{SO}(8,8) & \supset & \textrm{SO}(2,2) \times \textrm{SO}(6,6) & \supset & \textrm{SO}(2,2) \times \textrm{SO}(2,2) \times \textrm{SO}(3) \ . 
\end{array}
\end{equation}
It then follows from (\ref{SO(8,8)_to_SO(3)_embedding}) that the commutant of $\textrm{SO}(3)$ within $\textrm{SO}(8,8)$ is $\textrm{SO}(2,2) \times \textrm{SO}(2,2)$. Using the fact that $\textrm{SO}(2,2) \sim \textrm{SL}(2) \times \textrm{SL}(2)$, the scalar geometry describing the $\textrm{SO}(3)$ invariant sector of half-maximal supergravity is given by
\begin{equation}
\label{scalar_geometry_RSTU}
{\mathcal{M}}_{\textrm{scal}} = \left[ \frac{\textrm{SL}(2)}{\textrm{SO}(2)}\right]^{4} \subset \frac{{\rm SO}(8,8)}{{\rm SO}(8) \times {\rm SO}(8)}  \ ,
\end{equation}
and consists of four copies of the Poincar\'e disk parameterised by four complex scalars which we will denote as $R$, $S$, $T$ and $U$. It is worth mentioning here that the $\textrm{SO}(3)$ invariant sector is contained within the $\mathbb{Z}_{2}^{2}$ invariant sector previously discussed. More concretely, it is obtained upon the identifications
\begin{equation}
\label{RSTU_identifications}
T_{4} \equiv R
\hspace{5mm} , \hspace{5mm}
U_{4} \equiv S
\hspace{5mm} , \hspace{5mm}
T_{1}=T_{2}=T_{3} \equiv T
\hspace{5mm} , \hspace{5mm}
U_{1}=U_{2}=U_{3} \equiv U \ .
\end{equation}
This $\mathcal{N}=2$ sector of the theory -- we will refer to it as the RSTU-model -- is the three-dimensional analogue of the extensively studied $\mathcal{N}=1$ STU-model in four dimensions \cite{Kachru:2002he,Derendinger:2004jn,DeWolfe:2004ns,Camara:2005dc,Villadoro:2005cu,Derendinger:2005ph,DeWolfe:2005uu,Aldazabal:2006up,Dibitetto:2011gm}. The Einstein-scalar Lagrangian describing this sector is given by (\ref{L_8_chiral}) subject to the identifications in (\ref{RSTU_identifications}).

\subsection{Flux vacua in the eight-chiral model: type~IIA with O$2$-planes}
\label{Flux_vacua_IIA_O2}

The eight-chiral model has previously been considered in  \cite{Farakos:2020phe,Farakos:2023nms} within the context of type~IIA orientifold reductions with O$2$-planes discussed in Section~\ref{sec:O2-plane}. However, refs.~\cite{Farakos:2020phe,Farakos:2023nms} set the axions to zero, \textit{i.e.} $u_{I}=B_{I}=0$ in (\ref{coset_blocks_2x2}), so that $\boldsymbol{b}=0$ in (\ref{coset_blocks_Z2^3}). This follows from imposing an extra $\mathbb{Z}_{2}^{*}$ symmetry on the eight-chiral model acting as
\begin{equation}
\label{Z*_involution}
\begin{array}{cccccc}
\mathbb{Z}_{2}^{*} : &  \left( \, y^{1} \, , \, y^{2} \, , \, y^{3} \, , \, y^{4} \, , \, y^{5} \, , \, y^{6} \, , \, y^{7} \, , \, y^{8}  \,  \right) & \rightarrow &  \left( \, -y^{1} \, , \, y^{2} \, , \, -y^{3} \, , \, y^{4} \, , \, -y^{5} \, , \, y^{6} \, , \, -y^{7} \, , \, y^{8}  \,  \right)  \ .
\end{array}
\end{equation}
This additional $\mathbb{Z}_{2}^{*}$ projects out the gravitino $\psi_{\mu}{}^{7}$ and reduces the $\mathcal{N}=2$ supersymmetry of the $\mathbb{Z}_{2}^{2}$ invariant sector down to $\mathcal{N}=1$. In what follows it will prove convenient to redefine the eight dilatons $(A_{I},\mu_{I})$ entering the $\boldsymbol{e}_{I}$ matrices in (\ref{coset_blocks_2x2}) so that $\boldsymbol{e}$ in (\ref{coset_blocks_Z2^3}) becomes
\begin{equation}
\label{coset_IIA_O2}
\begin{array}{rcl}
\boldsymbol{e} &=& \textrm{diag}\left( 
\dfrac{1}{\sqrt{A_{1} \mu_{1}}} \, , \, 
\sqrt{\dfrac{\mu_{1}}{A_{1}}} \, , \, 
\dfrac{1}{\sqrt{A_{2} \mu_{2}}} \, , \, 
\sqrt{\dfrac{\mu_{2}}{A_{2}}} \, , \, 
\dfrac{1}{\sqrt{A_{3} \mu_{3}}} \, , \, 
\sqrt{\dfrac{\mu_{3}}{A_{3}}} \, , \, 
\dfrac{1}{\sqrt{A_{4} \mu_{4}}} \, , \, 
\sqrt{\dfrac{\mu_{4}}{A_{4}}} \, \right) \\[5mm]
 & \equiv & \sqrt{\sigma} \, \textrm{diag} \left( 
\dfrac{1}{\sqrt{\rho_{1}}} \, , \,   
\dfrac{1}{\sqrt{\rho_{2}}} \, , \,   
\dfrac{1}{\sqrt{\rho_{3}}} \, , \,   
\dfrac{1}{\sqrt{\rho_{4}}} \, , \,   
\dfrac{1}{\sqrt{\rho_{5}}} \, , \,   
\dfrac{1}{\sqrt{\rho_{6}}} \, , \,   
\dfrac{1}{\sqrt{\rho_{7}}} \, , \,   
\sqrt{\rho_{1} \, \rho_{2} \, \rho_{3} \, \rho_{4} \, \rho_{5} \, \rho_{6} \, \rho_{7}}
\right) \ , 
\end{array}
\end{equation}
in terms of eight new dilatons $(\sigma,\rho_{n})$ with $n=1,\ldots,7$. This redefinition will make some features of the type~IIA flux vacua more manifest.

\subsubsection{Fluxes and tadpoles}

The $\mathbb{Z}_{2}^{*} \times \mathbb{Z}_{2}^{2}$ invariance of the supergravity models we have discussed is string-theoretically compatible with demanding $\textrm{G}_{2}$ holonomy on the internal space of the corresponding type~IIA reductions (see \cite{Farakos:2020phe,Farakos:2023nms} for an explicit realisation). As a consequence, the $\mathbb{Z}_{2}^{*} \times \mathbb{Z}_{2}^{2}$ invariant gauge fluxes $H_{(3)}$ and $F_{(4)}$ have an expansion
\begin{equation}
\label{H3&F4_G2_inv}
H_{(3)} = \sum_{n=1}^{7} h_{n} \, \varphi^{n}_{(3)}
\hspace{10mm} , \hspace{10mm}
F_{(4)} = \sum_{n=1}^{7} f_{n} \, \varphi^{n}_{(4)}
\end{equation}
in terms of flux parameters $(h_{n},f_{n})$ and basis elements $\varphi^{n}_{(3)}$ and $\varphi^{n}_{(4)}$. These $7+7$ basis elements specify the $\textrm{G}_{2}$ invariant three-form and four-form as
\begin{equation}
\begin{array}{rcll}
\label{G2-inv_forms}
\varphi_{(3)} &=& dy^{127}  +  dy^{347} + dy^{567}  - dy^{136} - dy^{235} -  dy^{145}   +  dy^{246} & , \\[2mm]
\varphi_{(4)} &=& dy^{3456}  +  dy^{1256} + dy^{1234} -  dy^{2457}  -  dy^{1467} -  dy^{2367}  +  dy^{1357} & , \\[2mm]
\end{array}
\end{equation}
where we have introduced the notation $dy^{127}=dy^{1} \wedge dy^{2} \wedge dy^{7}$, etc.\footnote{Note that $\varphi_{(4)} = \star{_{7d}} \, \varphi_{(3)}$.} Together with the gauge fluxes in (\ref{H3&F4_G2_inv}), the type~IIA with O$2$-planes duality frame also allowed for the Romans mass parameter $F_{(0)}$. Importantly, the set of QC's of half-maximal supergravity gave in this case the tadpole cancellation condition in (\ref{QC_IIA_O2}), namely,
\begin{equation}
\label{QC_IIA_O2_G2_sec}
F_{(0)} \, H_{(3)} = 0 \ ,    
\end{equation}
related to the absence of O$6$/D$6$ sources in the compactification scheme. This condition is solved by setting
\begin{equation}
\label{QC_IIA_O2_solutions}
F_{(0)} =0  
\hspace{10mm} \textrm{ or } \hspace{10mm}
h_{n} = 0 \ .
\end{equation}
In other words, $F_{(0)}$ and $H_{(3)}$ cannot be simultaneously activated in a type~IIA reduction with O$2$-planes while preserving half-maximal supersymmetry.

\subsubsection{Mkw$_{3}$ flux vacua}

An explicit computation of the scalar potential using the gauge fluxes in (\ref{H3&F4_G2_inv}), the Romans mass $F_{(0)}$ and the $\boldsymbol{e}$ matrix in (\ref{coset_IIA_O2}) yields
\begin{equation}
\label{Scalar_potential_IIA_O2}
g^{-2} \, V_{\mathcal{N}=8} = \frac{\sigma^4}{32} \left[ \, \sum_{n=1}^{7} \left( \frac{h_n}{\sqrt{\rho^{n}_{(4)}}} - f_n \sqrt{\rho^{n}_{(4)}} \right)^2 \, \right] + \frac{\sigma^2}{32} \, F_{(0)}^2 \,  \prod_{n=1}^7 \rho_{n}^2  \ ,
\end{equation}
where we have introduced the short-hand notation $\rho^{n}_{(4)}$ ($n=1,\ldots,7$) for the seven quartic terms ${\rho^{1}_{(4)} \equiv \rho^{3}\rho^{4}\rho^{5}\rho^{6}}$, $\rho^{2}_{(4)} \equiv \rho^{1}\rho^{2}\rho^{5}\rho^{6}$, $\ldots$ , $\rho^{7}_{(4)} \equiv \rho^{1}\rho^{3}\rho^{5}\rho^{7}$ in one-to-one correspondence with the basis elements for $\varphi_{(4)}$ in (\ref{G2-inv_forms}). As a result, the scalar potential in (\ref{Scalar_potential_IIA_O2}) is non-negative definite.\footnote{Up to a term accounting for O$6$/D$6$ sources (to be discussed in a moment), the scalar potential (\ref{Scalar_potential_IIA_O2}) maps into the one in \cite{Farakos:2020phe,Farakos:2023nms} via a simple redefinition of the eight dilatons:
\begin{equation}
\label{Connection:Scalar_potential_IIA_O2-Farakos}
    \left( \rho_1, \rho_2, \rho_3, \rho_4, \rho_5, \rho_6, \rho_7 \right) = e^{-\frac{1}{2\sqrt{7}} x} \left( \frac{1}{\tilde{s}_1\tilde{s}_4\tilde{s}_6}, \tilde{s}_2\tilde{s}_3\tilde{s}_4\tilde{s}_6, \frac{1}{\tilde{s}_2\tilde{s}_4\tilde{s}_5}, \tilde{s}_1\tilde{s}_3\tilde{s}_4\tilde{s}_5, \frac{1}{\tilde{s}_3\tilde{s}_5\tilde{s}_6}, \tilde{s}_1\tilde{s}_2\tilde{s}_5\tilde{s}_6, \frac{1}{\tilde{s}_1\tilde{s}_2\tilde{s}_3} \right) \,,
\end{equation}
together with $\sigma^2 = e^{y}$.}

Solving the constraint (\ref{QC_IIA_O2_G2_sec}) by choosing $F_{(0)}=0$ in (\ref{QC_IIA_O2_solutions}) kills the second term in (\ref{Scalar_potential_IIA_O2}) and leaves an overall $\sigma^{4}$ dependence in the scalar potential. In this case, the extremisation of $V$ with respect to $\sigma$ forces the bracket in (\ref{Scalar_potential_IIA_O2}) to vanish. This gives rise to a Minkowski vacuum ($V_{0} \equiv \left\langle V \right\rangle = 0$) at
\begin{equation}
\label{rho_stabilisation}
\rho^{n}_{(4)} = \frac{h_{n}}{f_{n}} \ ,
\end{equation}
with the overall modulus $\sigma$ being the (unstabilised) massless ``no-scale'' direction of \cite{Farakos:2020phe}. Note also that, having $F_{(0)}=0$, these Minkowski vacua can be embedded in the M-theory setup of Section~\ref{sec:M-theory}. On the contrary, solving the constraint (\ref{QC_IIA_O2_G2_sec}) by choosing $H_{(3)}=0$ in (\ref{QC_IIA_O2_solutions}) retains both the $\sigma^{4}$ and $\sigma^{2}$ terms in the scalar potential (\ref{Scalar_potential_IIA_O2}). In this case, since both terms are positive definite, the extremisation of $V$ requires them to vanish separately. However, this cannot be achieved unless all the fluxes are zero (trivialising $V$) or the dilatons vanish (singular behaviour). Therefore, there are no vacua whenever $H_{(3)}=0$.

An explicit evaluation of the gravitini and scalar masses can be performed at the above class of no-scale Mkw$_{3}$ flux vacua. In order to present the results, it proves convenient to introduce the seven flux combinations
\begin{equation}
\label{lambda_parameters_def}
\lambda_{n} = \sqrt{h_{n} \, f_{n}}  \ ,
\end{equation}
and to define the vector $\vec{\lambda} \equiv (\lambda_{1},\ldots, \lambda_{7})$. Then, the masses of the eight gravitini are given by
\begin{equation}
\label{O2_spectrum_gravitini}
m_{3/2, \, \mathcal{A}}^2 = \dfrac{g^2}{16} \, \sigma^{4} \left( {\vec a}_\mathcal{A} \cdot {\vec \lambda}  \right)^2   
\hspace{10mm} \text{with} \hspace{10mm} 
\mathcal{A} = (0,\, \mathcal{I})
\hspace{6mm} \text{and} \hspace{6mm}
\mathcal{I}=1,\ldots,7 \ ,
\end{equation}
and are determined by eight flux-independent constant vectors $\,\vec{a}_{\mathcal{A}}=(\vec{a}_{0},\vec{a}_{\mathcal{I}})$. The first of these vectors takes the simple form
\begin{equation}
{\vec a}_0=\left( +1, +1, +1, +1, +1, +1, +1 \right) \ .
\end{equation}
The seven remaining vectors $\,{\vec a}_\mathcal{I}\,$ have entries $\,\pm 1\,$ with the positions of the $-1$'s specified by the Steiner triple system $\,({\cal I}, \cal{T})\,$:  $\,{\cal I}\,$ is the set $\,{\cal I}=\{ 1,2,3,4,5,6,7\}\,$ and $\,{\cal T}\,$ is the specific set of triplets $\,{{\cal T} = \{123,146,157,245,267,347,356\}}$ in one-to-one correspondence with the elements of $\,\mathcal{I}$. Then, for $\,\mathcal{I}=1,2,\ldots\,$, one has
\begin{equation}
{\vec a}_1=\left(-1, -1, -1, +1, +1, +1, +1\right)
\hspace{3mm} , \hspace{3mm}
{\vec a}_2=\left(-1, +1, +1, -1, +1, -1, +1\right)
\hspace{3mm} , \hspace{3mm} \ldots \ .
\end{equation}
The scalar mass spectrum consists of $\,29\,$ massless ($m^2=0$) and $\,35=7 + (7 \times 4)\,$ massive scalars. Amongst the massive ones, there are $\,7\,$ masses of the form
\begin{equation}
m_{n}^2 = \dfrac{g^2}{4} \, \sigma^{4} \,  \lambda_n^2   \ .
\end{equation}
The remaining $\,28=7 \times 4\,$ masses are organised in seven groups of four masses, each of the groups being specified by four $\lambda$-parameters in (\ref{lambda_parameters_def}). In order to establish which $\lambda$'s specify each group, it proves convenient to define the complement of $\,\mathcal{T}\,$ in the Steiner triple system as $\,{{\cal T}^{\perp} = \{4567,2357,2346,1367,1345,1256,1247\}}$. The four masses in the $\,\mathcal{I}^{\textrm{th}}\,$ group are then given by
\begin{equation}
m_{\mathcal{I},\alpha}^2 = \dfrac{g^2}{16} \, \sigma^{4} \left[(-1)^{\delta_{\alpha 1}} \lambda_{n_1} + (-1)^{\delta_{\alpha 2}} \lambda_{n_2} + (-1)^{\delta_{\alpha 3}} \lambda_{n_3} + (-1)^{\delta_{\alpha 4}} \lambda_{n_4}\right]^2 \ , 
\end{equation}
with $\,\alpha=1,\ldots,4\,$, and where the specific $4$-plet $(\lambda_{n_{1}},\lambda_{n_{2}},\lambda_{n_{3}},\lambda_{n_{4}})$ is given by the corresponding element of $\,\mathcal{T}^{\perp}$. For example, the four masses in the $\,\mathcal{I}=1\,$ group are given by
\begin{equation}
\begin{array}{cccc}
    m_{1,1}^2 = \dfrac{g^2}{16} \,  \sigma^{4} \left( -\lambda_{4} + \lambda_{5}+\lambda_{6}+\lambda_{7} \right)^2  \hspace{5mm} ,  \hspace{5mm} 
    m_{1,2}^2 = \dfrac{g^2}{16} \, \sigma^{4} \left(+\lambda_{4} - \lambda_{5}+\lambda_{6}+\lambda_{7}\right)^2 & , \\[4mm] 
   m_{1,3}^2 = \dfrac{g^2}{16} \, \sigma^{4} \left(+\lambda_{4} + \lambda_{5}-\lambda_{6}+\lambda_{7}\right)^2 \hspace{5mm} , \hspace{5mm}
   m_{1,4}^2 = \dfrac{g^2}{16} \, \sigma^{4} \left(+\lambda_{4} + \lambda_{5}+\lambda_{6}-\lambda_{7} \right)^2 & .
   \end{array} 
\end{equation}
By inspection of the gravitini masses in (\ref{O2_spectrum_gravitini}), we conclude that a generic no-scale Mkw$_{3}$ vacuum is non-supersymmetric. However, upon adjustment of the flux parameters in (\ref{lambda_parameters_def}), supersymmetry can be restored within a range $\,\mathcal{N}=1,\ldots, 6\,$ as counted by the number of massless gravitini. Finally, the additional quadratic constraints in (\ref{QC_N=16_extra}) give rise to a unique condition $\,\sum_{n} \lambda_{n}^{2}=0\,$ which trivialises the fluxes. Therefore, the no-scale Mkw$_{3}$ vacua cannot be embedded into maximal supergravity.

\subsubsection{Adding O$6$-planes to type~IIA with O$2$-planes}

Let us consider the possibility of adding O$6$-planes to the type~IIA with O$2$ duality frame. The first type of O$6$-planes we will consider is the O$6$-plane discussed in Section~\ref{sec:O6-plane}. The location of this O$6$-plane was displayed in (\ref{O6_location}) and its associated orientifold action given in (\ref{sigma_O6}). In order to distinguish this O$6$-plane from other types (to be introduced in a moment) we will denote it as O$6_{*}$ and its orientifold action as $\sigma_{\textrm{O}6_{*}}$. Note that $\sigma_{\textrm{O}6_{*}}$ descends from $\mathbb{Z}_{2}^{*}$ in (\ref{Z*_involution}). Together with the O$6_{*}$-plane, we will also introduce other types of O$6$-planes placed differently in the internal (twisted) $\mathbb{T}^{7}$. Following the construction in \cite{Farakos:2020phe}, let us consider two more types of O$6$-planes -- we will denote them as O$6_{\alpha}$ and O$6_{\beta}$ -- whose orientifold actions $\sigma_{\textrm{O}6_{\alpha}}$ and $\sigma_{\textrm{O}6_{\beta}}$ are inherited from $\mathbb{Z}_{2}^{(\alpha)}$ and $\mathbb{Z}_{2}^{(\beta)}$ in (\ref{Z2xZ2_generators}). Finally, we will also consider four more types of O$6$-planes that appear upon composition of O$6_{*}$, O$6_{\alpha}$ and O$6_{\beta}$. The resulting O$2$/O$6$-plane system is summarised in Table~\ref{Table:O2/O6_planes}.

As it was shown in Section~\ref{sec:O2-plane}, the O$2$-plane in the type~IIA with O$2$ duality frame was responsible for halving maximal into half-maximal supergravity, \textit{i.e.} $\mathcal{N}=16 \underset{\textrm{O}2}{\longrightarrow} \mathcal{N}=8$. Adding now the independent O$6_{*}$, O$6_{\alpha}$ and O$6_{\beta}$ further triggers three additional halvings, namely
\begin{equation}
\begin{array}{cccccccccc}
\mathcal{N}=16 &\underset{\textrm{O}2}{\longrightarrow}& \mathcal{N}=8  & \underset{\textrm{O}6_{*}}{\longrightarrow} &
\mathcal{N}=4  & \underset{\textrm{O}6_{\alpha}}{\longrightarrow} &
\mathcal{N}=2  & \underset{\textrm{O}6_{\beta}}{\longrightarrow} &
\mathcal{N}=1 \ .
\end{array}
\end{equation}
This breaking of supersymmetry from half-maximal to minimal due to the presence of O$6$-planes translates into the violation of the QC's of half-maximal supergravity in (\ref{QC_IIA_O2_G2_sec}). From (\ref{Tadpole_p-form}) one finds
\begin{equation}
\label{QC_IIA_O2_G2_sec_violation}
-F_{(0)} \, H_{(3)} = J_{\textrm{O}6/\textrm{D}6} \neq 0 \ ,    
\end{equation}
so that both the Romans mass $F_{(0)}$ and the gauge flux $H_{(3)}$ can be activated simultaneously. Having O$6$-planes in the compactification scheme produces a new contribution to the scalar potential of the form
\begin{equation}
\label{VO6_IIA}
g^{-2} \, V_{{\rm O}6} = \frac{\sigma^3}{16} \, F_{(0)}  \, \sum_{n=1}^7 \left(h_n \, \rho_{(3)}^n \sqrt{\rho_{(4)}^n}  \right) \ ,
\end{equation}
where we have again introduced a short-hand notation $\rho^{n}_{(3)}$ ($n=1,\ldots,7$) for the seven cubic terms $\rho^{1}_{(3)} \equiv \rho^{1}\rho^{2}\rho^{7}$, $\rho^{2}_{(3)} \equiv \rho^{3}\rho^{4}\rho^{7}$, $\ldots$ , $\rho^{7}_{(3)} \equiv \rho^{2}\rho^{4}\rho^{6}$ in one-to-one correspondence with the basis elements of $\varphi_{(3)}$ in (\ref{G2-inv_forms}). The contribution in (\ref{VO6_IIA}) was shown to descend from the ten-dimensional DBI action of the O$6$/D$6$ sources in \cite{Farakos:2020phe}. Being proportional to $-F_{(0)} \, H_{(3)}$, the contribution $V_{{\rm O}6}$ is not part of the scalar potential of half-maximal supergravity due to the QC's in (\ref{QC_IIA_O2_G2_sec}).

The scalar potential of the minimal supergravity models, \textit{i.e.} $V_{\mathcal{N}=1} = V_{\mathcal{N}=8} + V_{\textrm{O}6}\,$,\footnote{This potential matches the corresponding expression in \cite{Farakos:2020phe,Farakos:2023nms} upon the field redefinition (\ref{Connection:Scalar_potential_IIA_O2-Farakos}).} has been shown to accommodate AdS$_{3}$ vacua ($V_{0} \equiv \left\langle V \right\rangle < 0$) with interesting phenomenological properties, like scale separation between the AdS$_{3}$ scale $L^2=-2/V_{0}$ and the scale of the internal space \cite{Farakos:2020phe,Farakos:2023nms} or between different internal cycles (anisotropy) \cite{Farakos:2023wps}. 

Our next goal will be to show that the simpler half-maximal supergravity models arising from a single stack of O$p$-planes already suffice to accommodate both Mkw$_{3}$ and AdS$_{3}$ flux vacua with interesting phenomenology. We will illustrate this in the context of type~IIB reductions with O$5$-planes.

\begin{table}[t]
\begin{center}
\renewcommand{\arraystretch}{1.5}
\begin{tabular}{|c|ccc|cc|cc|cc|c|}
\cline{2-11}
\multicolumn{1}{c|}{} & \multicolumn{3}{c|}{External s-t} & \multicolumn{2}{c|}{$\mathbb{T}_{1}^{2}$}  & \multicolumn{2}{c|}{$\mathbb{T}_{2}^{2}$} & \multicolumn{2}{c|}{$\mathbb{T}_{3}^{2}$} & \multicolumn{1}{c|}{$\mathbb{S}^{1}$} \\ 
\cline{2-11}
\multicolumn{1}{c|}{} & $x^{0}$ & $x^{1}$ & $x^{2}$ & $y^{1}$  & $y^{2}$ & $y^{3}$  & $y^{4}$ & $y^{5}$ & $y^{6}$ & $y^{7}$  \\
\hline
\textrm{O$6_{*}$-plane} & $\times$ & $\times$ & $\times$ & $\times$ & $ $  & $\times$ &  $ $  & $\times$ &  $ $ & $\times$ \\
\hline
\textrm{O$6_{\alpha}$-plane} & $\times$ & $\times$ & $\times$ & $\times$ & $\times$ & $\times$ & $\times$ & $ $ & $ $ & $ $  \\
\hline
\textrm{O$6_{\beta}$-plane} & $\times$ & $\times$ & $\times$ & $\times$ & $\times$ & $ $ & $ $ & $\times$ & $\times$ & $ $  \\
\hline
\textrm{O$6_{\alpha\beta}$-plane} & $\times$ & $\times$ & $\times$ & $ $ & $ $ & $\times$ & $\times$ & $\times$ & $\times$ & $ $  \\
\hline
\textrm{O$6_{*\alpha}$-plane} & $\times$ & $\times$ & $\times$ & $ $ & $\times$ & $ $ & $\times$ & $\times$ & $ $ & $\times$  \\
\hline
\textrm{O$6_{*\beta}$-plane} & $\times$ & $\times$ & $\times$ & $ $ & $\times$ & $\times$ & $ $ & $ $ & $\times$ & $\times$  \\
\hline
\textrm{O$6_{*\alpha\beta}$-plane} & $\times$ & $\times$ & $\times$ & $\times$ & $ $ & $ $ & $\times$ & $ $ & $\times$ & $\times$  \\
\hline
\hline
\textrm{O$2$-planes} & $\times$ & $\times$ & $\times$ &  $ $  &  $ $ &  $ $  & $ $ &  $ $  & $ $ &  $ $ \\
\hline
\end{tabular}
\caption{Multiple O$2$/O$6$-planes configuration. The combined action of the three independent O$6_{\alpha}$, O$6_{\beta}$ and O$6_{*}$ orientifold actions, together with the O$2$-plane orientifold action, breaks half-maximal ${\mathcal{N}=8}$ supergravity into minimal $\mathcal{N}=1$ supergravity.}
\label{Table:O2/O6_planes}
\end{center}
\end{table}

\subsection{Flux vacua in the RSTU-model: a type~IIB with O$5$-planes appetizer}
\label{Section:RSTU_model}

\begin{table}[t]
\begin{center}
\renewcommand{\arraystretch}{1.5}
\begin{tabular}{|c|ccc|cc|cc|cc|c|c|}
\cline{2-11}
\multicolumn{1}{c|}{} & \multicolumn{3}{c|}{External s-t} & \multicolumn{2}{c|}{$\mathbb{T}_{1}^{2}$}  & \multicolumn{2}{c|}{$\mathbb{T}_{2}^{2}$} & \multicolumn{2}{c|}{$\mathbb{T}_{3}^{2}$} & \multicolumn{1}{c|}{$\mathbb{S}^{1}$} \\ 
\cline{2-12}
\multicolumn{1}{c|}{} & $x^{0}$ & $x^{1}$ & $x^{2}$ & $y^{1}$  & $y^{2}$ & $y^{3}$  & $y^{4}$ & $y^{5}$ & $y^{6}$ & $y^{7}$ & Internal Diff \\
\hline
\textrm{O2-plane} & $\times$ & $\times$ & $\times$ & &  &  &  &  &  &  & $\textrm{SL}(7)$ \\
\hline
\textrm{O3-plane} & $\times$ & $\times$ & $\times$ & &  &  &  &  &  & $\times$ &  $\textrm{SL}(6)$ \\
\hline
\textrm{O5-plane} & $\times$ & $\times$ & $\times$ & &   $\times$  &  &  $\times$  &  &  $\times$  & & $\textrm{SL}(3) \times \textrm{SL}(4)$ \\
\hline
\textrm{O6-plane} & $\times$ & $\times$ & $\times$ &  $\times$  &  &  $\times$  &  &  $\times$  &  &  $\times$ & $\textrm{SL}(3) \times \textrm{SL}(4)$ \\
\hline
\textrm{O8-plane} & $\times$ & $\times$ &$\times$ & $\times$ & $\times$  & $\times$ & $\times$ & $\times$ & $\times$ & & $\textrm{SL}(6)$ \\
\hline
\textrm{O9-plane} & $\times$ & $\times$ &$\times$ & $\times$ & $\times$  & $\times$ & $\times$ & $\times$ & $\times$ & $\times$ & $\textrm{SL}(7)$ \\
\hline
\end{tabular}
\caption{Single O$p$-plane duality frames compatible with the $\textrm{SO}(3)$ invariance of the RSTU-models. In all the cases the orientifold action $\mathcal{O}_{\mathbb{Z}_{2}}$ amounts to a discrete $\mathbb{Z}_{2}$ symmetry that halves maximal into half-maximal supergravity.}
\label{Table:Op_planes}
\end{center}
\end{table}

Let us now investigate half-maximal RSTU-models with a single class of O$p$-planes. The group-theoretical embedding (\ref{SO(8,8)_to_SO(3)_embedding}) of the $\textrm{SO}(3)$ symmetry of the RSTU-models implies the set of branching rules summarised in Table~\ref{Table:SO(3)_branchings}. In particular, using the light-cone basis for $\textrm{SO}(8,8)$, the vector index $M$ of the $\,{\bf 16}\,$ splits as
\begin{equation}
\label{decomposition_16_SO(3)}
\begin{array}{ccc}
T_{M} \,=\, \left( \, T_{A} \,;\, T^{A} \,  \right)  & \rightarrow & \left( \, T_{i} \,,\, T_{a} \,,\, T_{7} \,,\, T_{8} \,;\, T^{i} \,,\, T^{a} \,,\, T^{7} \,,\, T^{8} \,  \right) \ ,
\end{array}
\end{equation}
where $\,a=1,3,5\,$ and $\,i=2,4,6\,$ transform simultaneously under the action of the diagonal $\,\textrm{SO}(3) \subset \textrm{SO}(3)_{a} \times \textrm{SO}(3)_{i}$, whereas $7$ and $8$ are singlets. As a result, there exist $\textrm{SO}(3)$ invariant tensors of the form\footnote{Tensors with a mixed type of indices are invariant under the diagonal $\,\textrm{SO}(3) \subset \textrm{SO}(3)_{a} \times \textrm{SO}(3)_{i}$. For example, $\,\delta_{ai}\,$ has non-zero components $\,\delta_{12}=\delta_{34}=\delta_{56}=1$.}
\begin{equation}
\label{SO(3)_inv_tensors}
\delta_{ab} \,\, , \,\, \delta_{ij} \,\, , \,\, \delta_{ai} \, \, , \,\, \epsilon_{abc}  \,\, , \,\,  \epsilon_{abk}  \,\, , \,\, \epsilon_{ajk} \,\, , \,\, \epsilon_{ijk} \ ,
\end{equation}
which can be used to construct the set of $\textrm{SO}(3)$ invariant fluxes in a given O$p$-plane duality frame (see Table~\ref{Table:Op_planes}). As an example, we will focus the rest of the discussion on the type~IIB with O$5$-plane duality frame presented in Section~\ref{sec:O5-plane}.

\begin{table}[t]
\begin{center}
\scalebox{0.9}{
\renewcommand{\arraystretch}{1.7}
\begin{tabular}{|c||c||c|}
     \hline 
     ${\rm SO}(8,8)$ &  ${\rm SL}(2) \times {\rm SL}(2) \times {\rm SO}(6,6)$ & ${\rm SL}(2) \times {\rm SL}(2) \times {\rm SL}(2) \times {\rm SL}(2) \times {\rm SO}(3)$  \\ 
     \hline \hline
     ${\bf 16}$  &   $({\bf 1},{\bf 1} ; {\bf 12}) \oplus ({\bf 2}, { \bf 2}; {\bf 1})$ & $({\bf 1},{ \bf 1};{\bf 2},{\bf 2};{\bf 3}) \oplus ({\bf 2},{\bf 2};{\bf 1},{\bf 1};{\bf 1})$  \\
     \hline \hline
     ${\bf 1}$ & $({\bf 1},{\bf 1};{\bf 1})$ & $ ({\bf 1},{\bf 1};{\bf 1},{\bf 1};{\bf 1}) $  
     \\
     \hline
     ${\bf 135}$ & $({\bf 1},{\bf 1};{\bf 1}) \oplus ({\bf 3},{\bf 3};{\bf 1})$ & $ ({\bf 1},{\bf 1};{\bf 1},{\bf 1};{\bf 1}) \oplus ({\bf 3},{\bf 3};{\bf 1},{\bf 1};{\bf 1})$
     \\
     $ $ & $({\bf 1},{\bf 1};{\bf 77}) \oplus ({\bf 2},{\bf 2};{\bf 12})$ & $[({\bf 1},{\bf 1};{\bf 3},{\bf 3};{\bf 1})\oplus ({\bf 1},{\bf 1};{\bf 1},{\bf 3};{\bf 3}) \oplus ({\bf 1},{\bf 1};{\bf 3},{\bf 1};{\bf 3})$
     \\
     $ $ & $ $ & $({\bf 1},{\bf 1};{\bf 1},{\bf 1};{\bf 5}) \oplus ({\bf 1},{\bf 1};{\bf 3},{\bf 3};{\bf 5})] \oplus ({\bf 2},{\bf 2};{\bf 2},{\bf 2};{\bf 3})$
     \\
     \hline
     ${\bf 1820}$ & $({\bf 1},{\bf 1};{\bf 1})$ & $({\bf 1},{\bf 1};{\bf 1},{\bf 1};{\bf 1})$
     \\
     $ $ & $({\bf 1},{\bf 3};{\bf 66}) \oplus ({\bf 3},{\bf 1};{\bf 66})$ & $[({\bf 1},{\bf 3};{\bf 1},{\bf 3};{\bf 1}) \oplus ({\bf 1},{\bf 3};{\bf 3},{\bf 1};{\bf 1}) \oplus \ldots]$
     \\
     $ $ & $ $ & $[({\bf 3},{\bf 1};{\bf 1},{\bf 3};{\bf 1}) \oplus ({\bf 3},{\bf 1};{\bf 3},{\bf 1};{\bf 1}) \oplus \ldots]$
     \\
     $ $ & $({\bf 1},{\bf 1};{\bf 495}) \oplus ({\bf 2},{\bf 2};{\bf 220})$ & $[2 \, ({\bf 1},{\bf 1};{\bf 1},{\bf 1};{\bf 1}) \oplus ({\bf 1},{\bf 1};{\bf 3},{\bf 3};{\bf 1})$ \\
     $ $ & $ $ & $({\bf 1},{\bf 1};{\bf 1},{\bf 5};{\bf 1}) \oplus 
     ({\bf 1},{\bf 1};{\bf 5},{\bf 1};{\bf 1}) \oplus \ldots]$
     \\
     $ $ & $ $ & $[({\bf 2},{\bf 2};{\bf 2},{\bf 2};{\bf 1}) \oplus ({\bf 2},{\bf 2};{\bf 4},{\bf 4};{\bf 1}) \oplus \ldots ]$
     \\
     $ $ & $({\bf 2},{\bf 2};{\bf 12})$ & $({\bf 2},{\bf 2};{\bf 2},{\bf 2};{\bf 3})$
     \\
     \hline
\end{tabular}
}
\caption{Branching rules for $\textrm{SO}(8,8) \supset {\textrm{SL}(2)^{2} \times \textrm{SO}(6,6)} \supset {\textrm{SL}(2)^{4} \times \textrm{SO}(3)}$. The ellipsis in the bottom-right box denote additional non-singlets under $\textrm{SO}(3)$.}
\label{Table:SO(3)_branchings}
\end{center}
\end{table}

\subsubsection{Fluxes and tadpoles}

The set of type~IIB fluxes compatible with an O$5$-plane was presented in Table~\ref{Table:O5_fluxes}. In there, the $\textrm{SL}(4) \times \textrm{SL}(3)$ covariance of the O$5$-plane duality frame forced a splitting of the ${\bf 16}$ of $\textrm{SO}(8,8)$ of the form
\begin{equation}
\label{decomposition_16_O5-plane}
\begin{array}{ccc}
T_{M} \,=\, \left( \, T_{A} \,;\, T^{A} \,  \right)  & \rightarrow & \left( \, T_{i} \,,\, T_{\hat{a}}  \,,\, T_{8} \,;\, T^{i} \,,\, T^{\hat{a}} \,,\, T^{8} \,  \right) \ ,
\end{array}
\end{equation}
with $\hat{a}=1,3,5,7$ and $i=2,4,6$. This splitting followed from the O$5$-plane orientifold action in (\ref{sigma_O5}). A quick comparison with the $\textrm{SO}(3)$ splitting in (\ref{decomposition_16_SO(3)}) then yields $\hat{a}=(a,7)$. Using the $\textrm{SO}(3)$ invariant tensors in (\ref{SO(3)_inv_tensors}) one can construct the following set of $\textrm{SO}(3)$ invariant metric fluxes
\begin{equation}
\label{metric_flux_SO(3)}
\begin{array}{lclclc}
\omega_{ab}{}^{k}=\omega_{1} \, \epsilon_{ab}{}^{k}
& \hspace{3mm} , & \hspace{3mm} 
\omega_{7a}{}^{i}=\omega_{2} \, \delta_{a}^{i}
& \hspace{3mm} , & \hspace{3mm} 
\omega_{i7}{}^{a}=\omega_{3} \, \delta_{i}^{a} & , \\[2mm]
\omega_{ai}{}^{7}=\omega_{4} \, \delta_{ai}
& \hspace{3mm} , & \hspace{3mm} 
\omega_{aj}{}^{c}=-\omega_{5} \, \epsilon_{aj}{}^{c}
& \hspace{3mm} , & \hspace{3mm} 
\omega_{ij}{}^{k}=\omega_{6} \, \epsilon_{ij}{}^{k} & ,
\end{array}
\end{equation}
which are allowed by the O5-plane orientifold action and specify the isometry algebra of the group manifold $\textrm{G}$. More concretely, (\ref{X_commutators}) reduces to
\begin{equation}
\label{algebra_isometry_SO(3)}
\begin{array}{lclclc}
[X_a, X_b] = \omega_{1} \, \epsilon_{ab}{}^{k}\, X_k 
& \hspace{2mm} , & \hspace{2mm} 
[X_7, X_b] = \omega_{2} \, \delta_{bk}\, X_k
& \hspace{2mm} , & \hspace{2mm} 
[X_i, X_7] = \omega_{3} \, \delta_{ic} \, X_c & , \\[2mm]
[X_a, X_j] = -\omega_{5} \, \epsilon_{aj}{}^{c} \, X_c + \omega_{4} \, \delta_{aj} \, X_7
& \hspace{2mm} , & \hspace{2mm} 
[X_i, X_j] = \omega_{6} \, \epsilon_{ij}{}^{k} \, X_k & \hspace{2mm} .
\end{array}
\end{equation}
In addition to the metric fluxes, the $\textrm{SO}(3)$ invariant gauge fluxes are
\begin{equation}
\label{gauge_flux_SO(3)}
\begin{array}{c}
H_{abc}= h_{31}  \, \epsilon_{abc}
\hspace{3mm} , \hspace{3mm} 
H_{aij}= h_{32}  \, \epsilon_{aij} 
\hspace{3mm} , \hspace{3mm} 
F_{ijk}= -f_{31}  \, \epsilon_{ijk}
\hspace{3mm} , \hspace{3mm} 
F_{ia7}= f_{32}  \, \delta_{ia}
\hspace{3mm} , \hspace{3mm} 
F_{ibc}= f_{33}  \, \epsilon_{ibc} \ , \\[2mm]
F_{abij7}= -f_{5}  \, \delta_{ai} \, \delta_{bj}
\hspace{3mm} , \hspace{3mm} 
F_{abcijk7}= f_{7}  \, \epsilon_{abc} \, \epsilon_{ijk} \ ,
\end{array}
\end{equation}
so that one ends up with $6+7=13$ arbitrary flux parameters in total.\footnote{A proper antisymmetrisation of indices is understood in \eqref{metric_flux_SO(3)} and \eqref{gauge_flux_SO(3)}.} These $13$ fluxes are a subset of the $158$ embedding tensor components of half-maximal supergravity which are $\textrm{SO}(3)$ singlets (see Table~\ref{Table:SO(3)_branchings}). The rest are known as ``non-geometric'' fluxes (see \cite{Shelton:2005cf}) and their study goes beyond the scope of this work.

The type~IIB flux parameters in (\ref{metric_flux_SO(3)}) and (\ref{gauge_flux_SO(3)}) are restricted by the QC's of half-maximal supergravity in (\ref{QC_N=8}). An explicit computation yields
\begin{equation}
\label{QC_ww_O5}
\begin{array}{rclcrclcrcl}
\omega_{3} \, \omega_{4} + \omega_{5} \,(\omega_{5} +\omega_{6}) &=& 0 
& \hspace{3mm} , & \hspace{3mm}    
\omega_{4} \,(2\, \omega_{5} +\omega_{6}) &=& 0  \\[2mm]
\omega_{1} \, \omega_{3} - \omega_{2} \,(\omega_{5} +\omega_{6}) &=& 0 
& \hspace{3mm} , & \hspace{3mm}    
\omega_{3} \,(2\, \omega_{5} +\omega_{6}) &=& 0  
& \hspace{3mm} , & \hspace{3mm}    
\omega_{1} \, \omega_{4} &=& 0 \ ,\\[2mm]
\omega_{2} \, \omega_{4} - \omega_{1} \,(\omega_{5} +\omega_{6}) &=& 0
& \hspace{3mm} , & \hspace{3mm}    
\omega_{1} \, \omega_{3} - 2\, \omega_{2} \, \omega_{5} &=& 0  
\end{array}
\end{equation}
which originate from the nilpotency condition $(D^2=0)$ of the $D = d + \omega$ twisted exterior derivative on the internal space. Alternatively, (\ref{QC_ww_O5}) can be viewed as the Jacobi identity of the isometry algebra (\ref{algebra_isometry_SO(3)}). In addition, one also finds
\begin{equation}
\label{QC_DH3_O5}
\omega_{3} \, h_{32} \,\,=\,\, 0    
\hspace{5mm} , \hspace{5mm} 
2\,\omega_{2} \, h_{32} - \omega_{3} \, h_{31} \,\,= \,\,0   \ , 
\end{equation}
coming from $D H_{(3)}=0$ and reflecting the absence of NS$5$-branes, as well as
\begin{equation}
\label{QC_DF3_O5}
\omega_{2} \, f_{31} + (2\, \omega_{5} + \omega_{6}) \, f_{32} + 2 \, \omega_{3} \, f_{33}  \,\,=\,\, 0    
\hspace{4mm} , \hspace{4mm} 
\omega_{1} \, f_{31} - (2\, \omega_{5} + \omega_{6}) \, f_{33} - 2 \, \omega_{4} \, f_{32}  \,\,=\,\, 0  \ , 
\end{equation}
originating from $D F_{(3)}=0$ and reflecting the absence of O$5$/D$5$ sources different from the ones in (\ref{O5_location}). More concretely, the unrestricted O$5$/D$5$ sources entering (\ref{unrestrited_5_branes}) yield
\begin{equation}
\label{unrestrited_5_branes_example}
\left. DF_{(3)} \, \right|_{dy^{\hat{a}} \wedge \,dy^{\hat{b}}  \wedge \,dy^{\hat{c}} \wedge \,dy^{\hat{d}} } = 3\, \omega_{1} \, f_{32}  - 3\,   \omega_{2} \, f_{33} = J_{\textrm{O}5/\textrm{D}5} \ .
\end{equation}
Finally, there is a last condition of the form
\begin{equation}
\label{QC_DF5_O5}
3 \, \omega_{4} \, f_{5} - h_{31} \, f_{31} + 3 \, h_{32} \, f_{33} \,\, = \,\, 0 \ , 
\end{equation}
which originates from $DF_{(5)} - H_{(3)} \wedge F_{(3)}=0$ and reflects the absence of O$3$/D$3$ sources. The flux parameters must satisfy the algebraic conditions (\ref{QC_ww_O5})-(\ref{QC_DF3_O5}) and (\ref{QC_DF5_O5}) for them to specify a well-defined gauging of half-maximal supergravity.

\subsubsection{Exploring the landscape of Mkw$_{3}$ and AdS$_{3}$ flux vacua}

Let us now investigate the vacuum structure of the RSTU-model constructed in the type~IIB with O$5$ duality frame. The scalar potential of the half-maximal models is very lengthy (and not very enlightening) so we will not present it here explicitly. Still, it possesses some interesting features that we will exploit in order to explore its vacuum structure.
\begin{itemize}

\item The first feature is generic and follows from the general form of the scalar potential (\ref{V_N=8}) in half-maximal supergravity: it is quadratic in the embedding tensor (flux parameters) and involves high powers of the scalar fields.

\item The second property applies to the type~IIB with O$5$ duality frame we are discussing here. The scalar potential of the RSTU-model turns out to be invariant under non-compact $\textrm{SL}(2)$ transformations (scalings and shifts) on the complex scalars, of the form
\begin{equation}
\label{SL(2)_transformation_RSTU}
R' = \lambda_{R} \, R + c_{R} 
\hspace{5mm} , \hspace{5mm} 
S' = \lambda_{S} \, S
\hspace{5mm} , \hspace{5mm} 
T' = \lambda_{T} \, T + c_{T}
\hspace{5mm} , \hspace{5mm} 
U' = \lambda_{U} \, U  \ , 
\end{equation}
with the $\lambda$'s and $c$'s being real parameters, accompanied by the following transformation of the metric fluxes
\begin{equation}
\label{SL(2)_transformation_metric}
\begin{array}{c}
\omega_1' = \left( \dfrac{\lambda_R \lambda_S \lambda_T^3}{\lambda_U}\right)^{\frac12} \omega_1 
\hspace{5mm} , \hspace{5mm} 
\omega_2' = \lambda_T^2 \,  \omega_2
\hspace{5mm} , \hspace{5mm} 
\omega_3' = \dfrac{\lambda_T}{\lambda_S} \, \omega_3 \ , \\[3mm]
\omega_4' = \dfrac{\lambda_R}{\lambda_U} \, \omega_4
\hspace{5mm} , \hspace{5mm} 
\omega_5' = \left( \dfrac{\lambda_R \lambda_T}{\lambda_S \lambda_U}\right)^{\frac12} \omega_5
\hspace{5mm} , \hspace{5mm} 
\omega_6' = \left( \dfrac{\lambda_R \lambda_T}{\lambda_S \lambda_U} \right)^{\frac12} \omega_6 \ ,
\end{array}
\end{equation}
and of the gauge fluxes
\begin{equation}
\label{SL(2)_transformation_gauge}
\begin{array}{c}
h_{31}' = \left( \dfrac{\lambda_R \lambda_S \lambda_T^3}{\lambda_U^3}\right)^{\frac12} h_{31} 
\hspace{5mm} , \hspace{5mm} 
h_{32}' = \left( \dfrac{\lambda_R \lambda_T}{\lambda_S \lambda_U^3}\right)^{\frac12} h_{32} \ , \\[2mm]
f_{31}' = \dfrac{\lambda_R}{\lambda_S} \, f_{31}
\hspace{5mm} , \hspace{5mm}
f_{32}' = \left( \dfrac{\lambda_R \lambda_U \lambda_T^3}{\lambda_S}\right)^{\frac12} f_{32}
\hspace{5mm} , \hspace{5mm}
f_{33}' = \lambda_R \, \lambda_T \, f_{33}  \ , \\[2mm]
f_5' = \left( \dfrac{\lambda_R \lambda_T}{\lambda_S \lambda_U} \right)^{\frac12} \left[ \lambda_T \, f_5 + c_T \left( 2\omega_5 + \omega_6 \right) \right]
\hspace{5mm} , \hspace{5mm}
f_7' = \left( \dfrac{\lambda_R \lambda_T}{\lambda_S \lambda_U^3} \right)^{\frac12} \left[ \lambda_T f_7 + 3 \, c_T \, h_{32} \right] \ .
\end{array}
\end{equation}

\item The third feature, which actually can be inferred from (\ref{SL(2)_transformation_gauge}), is that the scalar potential does not depend on the scalar $\textrm{Re}R=B_{4}$. This explains why the transformation of the gauge fluxes in (\ref{SL(2)_transformation_gauge}) does not depend on the constant shift $c_{R}$ in (\ref{SL(2)_transformation_RSTU}).

\end{itemize}

\subsubsection*{Charting the landscape of flux vacua: the strategy}

The invariance of the scalar potential under the simultaneous transformations (\ref{SL(2)_transformation_RSTU}) and (\ref{SL(2)_transformation_metric})-(\ref{SL(2)_transformation_gauge}) allows us to set vacuum expectation values (VEV's) of the form
\begin{equation}
\label{GTTO}
\left\langle  R \, \right\rangle = i
\hspace{5mm} , \hspace{5mm}
\left\langle  S \right\rangle = \left\langle  u_{4} \right\rangle + i
\hspace{5mm} , \hspace{5mm}
\left\langle  T \right\rangle = i
\hspace{5mm} , \hspace{5mm}
\left\langle  U \right\rangle = \left\langle  u \right\rangle + i \ ,
\end{equation}
when searching for the extrema of $V$ without loss of generality. In other words, a given vacuum of the scalar potential can be brought to the form in (\ref{GTTO}) by applying a duality transformation at the expense of changing the flux configuration accordingly. Note that real shifts of $S$ and $U$ cannot be reabsorbed by a flux redefinition. In our scanning of critical points of the scalar potential we will restrict to vacua with
\begin{equation}
\left\langle  u_{4} \right\rangle = \left\langle  u \right\rangle = 0 \ .
\end{equation}
This amounts to search for vacua at the origin of the moduli space, which can afterwards be moved to a different location in field space by performing the duality transformations in (\ref{SL(2)_transformation_RSTU}) at the expense of changing fluxes as in (\ref{SL(2)_transformation_metric})-(\ref{SL(2)_transformation_gauge}).

\begin{table}[]
\begin{center}
\scalebox{0.82}{
\renewcommand{\arraystretch}{1.5}
\begin{tabular}{!{\vrule width 1.5pt}l!{\vrule width 1pt}c!{\vrule width 1pt}c!{\vrule width 1pt}cccccc!{\vrule width 1pt}cc!{\vrule width 1pt}ccc!{\vrule width 1pt}c!{\vrule width 1pt}c!{\vrule width 1.5pt}}
\Xcline{4-16}{1.5pt}
\multicolumn{3}{c!{\vrule width 1pt}}{}& \multicolumn{6}{c!{\vrule width 1pt}}{$\omega$} & \multicolumn{2}{c!{\vrule width 1pt}}{$H_{(3)}$} & \multicolumn{3}{c!{\vrule width 1pt}}{$F_{(3)}$} & \multicolumn{1}{c!{\vrule width 1pt}}{$F_{(5)}$} & \multicolumn{1}{c!{\vrule width 1pt}}{$F_{(7)}$}\\ 
\noalign{\hrule height 1.5pt}
     \hspace{5mm} ID & Type &  SUSY & $\omega_{1}$ & $ \omega_{2}$ & $  \omega_{3}$ & $\omega_{4}$ & $ \omega_{5}$ & $ \omega_{6}$ & $ h_{31}$ & $  h_{32}$ & $f_{31}$ & $f_{32}$ & $f_{33}$ & $ f_{5}$ & $  f_{7}$  \\ 
\noalign{\hrule height 1pt}
     $ \textbf{vac~1}^{\,\,\dagger} $ & \multirow{3}{*}{$\textrm{Mkw}_{3}$} & $\mathcal{N}=0,4$ & $ \kappa $ & $ \xi$ & $  0 $ & $ 0 $ & $ 0 $ & $ 0 $ & $ 0 $ & $  0 $ & $0$ & $ \kappa$ & $ -\xi$ & $ 0 $ & $  0 $  \\ 
     \cline{1-1}\cline{3-16} 
     $ \textbf{vac~2}^{\,\,\dagger} $ &  & $\mathcal{N}=0$ &  $ 0 $ & $ \kappa$ & $  0 $ & $0$ & $ 0$ & $ 0$ & $ 0$ & $  0$ & $0$ & $ 0$ & $ -\kappa$ & $ 0$ & $  0$  \\ 
     \cline{1-1}\cline{3-16}
     $ \textbf{vac~3}^{\,\,*} $ &  & $\mathcal{N}=0$ & $ 0$ & $ \kappa $ & $  \kappa$ & $0$ & $ 0$ & $ 0$ & $ 0$ & $  0$ & $0$ & $ 0$ & $ 0$ & $ 0$ & $  0$  \\ 
\noalign{\hrule height 1pt}
     $ \textbf{vac~4}^{\,\,* \,\dagger} $ & \multirow{2}{*}{${\textrm{AdS}_{3}}$} & $\mathcal{N}=4$ & $ 0$ & $ 0$ & $  0$ & $0$ & $ 0$ & $ \kappa$ & $ 0$ & $  0$ & $\pm\kappa$ & $ 0$ & $ 0$ & $ 0$ & $  -\kappa$ 
     \\ 
     \cline{1-1}\cline{3-16}
     $ \textbf{vac~5}^{\,\,*  \,\dagger} $ &  & $\mathcal{N}=0$ & $ 0$ & $ 0$ & $  0$ & $0$ & $ 0$ & $ \kappa$ & $ 0$ & $  0$ & $\pm\kappa$ & $ 0$ & $ 0$ & $ 0$ & $  \kappa$ 
     \\
\noalign{\hrule height 1pt}
     $ \textbf{vac~6}^{\,\,*} $ & \multirow{2}{*}{$\textrm{AdS}_{3}$} & $\mathcal{N}=3$ & $ \kappa$ & $ 0$ & $  0$ & $0$ & $ -\kappa$ & $ \kappa$ & $ 0$ & $  0$ & $\pm\kappa $ & $ 0$ & $ \mp\kappa$ & $ 0$ & $  -2\kappa$  \\ 
     \cline{1-1}\cline{3-16}
     $ \textbf{vac~7}^{\,\,*} $ &  & $\mathcal{N}=1$ & $ \kappa$ & $ 0$ & $  0$ & $0$ & $ -\kappa$ & $ \kappa$ & $ 0$ & $  0$ & $\mp\kappa $ & $ 0$ & $ \pm\kappa$ & $ 0$ & $  2\kappa$  \\
\noalign{\hrule height 1pt}
     $ \textbf{vac~8}^{\,\,* \,\dagger} $ & \multirow{2}{*}{${\textrm{AdS}_{3}}$} & $\mathcal{N}=1$ & $ 0$ & $ 0$ & $  0$ & $0$ & $ -\kappa$ & $ \kappa$ & $ 0$ & $  0$ & $\pm\kappa$ & $ 0$ & $ 0$ & $ 0$ & $  \kappa$  \\ 
     \cline{1-1}\cline{3-16}
     $ \textbf{vac~9}^{\,\,* \,\dagger} $ &  & $\mathcal{N}=0$ & $ 0$ & $ 0$ & $  0$ & $0$ & $ -\kappa$ & $ \kappa$ & $ 0$ & $  0$ & $\pm\kappa$ & $ 0$ & $ 0$ & $ 0$ & $ - \kappa$  \\ 
\noalign{\hrule height 1pt}
     $ \textbf{vac~10} $ & $\textrm{AdS}_{3}$ & $\mathcal{N}=0$ & $ 0$ & $ 2\kappa$ & $  \kappa$ & $0$ & $ 0$ & $ 0$ & $ 0$ & $  0$ & $\kappa$ & $ \pm\kappa$ & $ -\kappa$ & $ 0$ & $  \pm\kappa$  \\
     \hline 
     $ \textbf{vac~11}$ & $\textrm{AdS}_{3}$ & $\mathcal{N}=0$ & $ 0$ & $ 2\kappa$ & $  \kappa$ & $0$ & $ 0$ & $ 0$ & $ 0$ & $  0$ & $\kappa$ & $ \pm\kappa$ & $ -\kappa$ & $ 0$ & $  \mp\kappa$  \\
\noalign{\hrule height 1pt} 
     $ \textbf{vac~12}^{\,\,* \,\dagger} $ & \multirow{4}{*}{${\textrm{AdS}_{3}}$} & $\mathcal{N}=4$ & $ 0$ & $ 0$ & $  \mp\kappa$ & $\mp\kappa$ & $ \kappa$ & $ -2\kappa$ & $ 0$ & $  0$ & $\mp 2\kappa$ & $ 0$ & $ 0$ & $ 0$ & $  2\kappa$  \\
      \cline{1-1}\cline{3-16}
     $ \textbf{vac~13}^{\,\,* \,\dagger} $ &  & $\mathcal{N}=1$ & $ 0$ & $ 0$ & $  \mp\kappa$ & $\mp\kappa$ & $ \kappa$ & $ -2\kappa$ & $ 0$ & $  0$ & $\mp 2\kappa$ & $ 0$ & $ 0$ & $ 0$ & $  -2\kappa$  \\
     \cline{1-1}\cline{3-16}
     $ \textbf{vac~14}^{\,\,* \,\dagger} $ &  & $\mathcal{N}=0$ & $ 0$ & $ 0$ & $  \pm\kappa$ & $\pm\kappa$ & $ \kappa$ & $ -2\kappa$ & $ 0$ & $  0$ & $\mp 2\kappa$ & $ 0$ & $ 0$ & $ 0$ & $  2\kappa$  \\
     \cline{1-1}\cline{3-16}
     $ \textbf{vac~15}^{\,\,* \,\dagger} $ &  & $\mathcal{N}=0$ & $ 0$ & $ 0$ & $  \pm\kappa$ & $\pm\kappa$ & $ \kappa$ & $ -2\kappa$ & $ 0$ & $  0$ & $\mp 2\kappa$ & $ 0$ & $ 0$ & $ 0$ & $  -2\kappa$ \\
\noalign{\hrule height 1.5pt}
\end{tabular}}
\caption{Fluxes in the type~IIB with O$5$-planes duality frame producing a vacuum at the origin of moduli space, \textit{i.e.} $\left\langle R  \, \right\rangle = \left\langle S  \right\rangle = \left\langle T  \right\rangle = \left\langle U   \right\rangle = i$. The \textbf{vac~1} is Mkw$_{3}$ and generically non-supersymmetric, but becomes $\mathcal{N}=4$ when $\kappa = \pm \xi$. For the AdS$_{3}$ supersymmetric vacua, the $\,\mathcal{N}=p\,$ supersymmetry is realised as $\,\mathcal{N}=(p,0)\,$ or $\,\mathcal{N}=(0,p)\,$  depending on the upper/lower sign choice of the fluxes. For instance, taking $\kappa >0$ in \textbf{vac~4}, one finds $\,\mathcal{N}=(4,0)\,$ for $\,f_{31}=+\kappa\,$ and $\,\mathcal{N}=(0,4)\,$ for $\,f_{31}=-\kappa\,$. (The opposite happens if $\,\kappa<0$). The vacuum solutions with an asterisk $\,*\,$ satisfy the extra self-dual constraint (\ref{SD_condition}) and admit an embedding into maximal supergravity. Those with a dagger $\,\dagger\,$ satisfy the anti-self-dual condition (\ref{ASD_condition}). If both conditions are satisfied simultaneously then (\ref{KK_spectroscopy_condition}) holds. Finally, the full algebra (\ref{brackets_N=8}) induced by the fluxes activated in this table is presented in the Appendix~\ref{appendix:IIB_O5_flux_algebra}.}
\label{Table:flux_vacua_IIB}
\end{center}
\end{table}

The main advantage of searching for critical points at the origin of moduli space is that the extremisation conditions
\begin{equation}
\label{extremisation_cond}
\partial V \, \Big|_{\left\langle  R \right\rangle=\left\langle  S \right\rangle=\left\langle  T \right\rangle=\left\langle  U \right\rangle =i} \,\,=\,\, 0 \ ,
\end{equation}
become a set of algebraic equations which are quadratic on the flux parameters. In addition, there are the constraints (\ref{QC_ww_O5})-(\ref{QC_DF5_O5}) required by half-maximal supergravity which are also quadratic on the flux parameters. Therefore, putting together (\ref{extremisation_cond}) and (\ref{QC_ww_O5})-(\ref{QC_DF5_O5}), one is left with a complicated algebraic set of quadratic conditions on the flux parameters. This system can be solved in full generality with the help of the algebra software \textsc{Singular} \cite{DGPS}. The outcome is the set of $15$ vacua summarised in Table~\ref{Table:flux_vacua_IIB}.

Before moving to discuss some of the properties of the flux vacua in Table~\ref{Table:flux_vacua_IIB}, let us make a couple of observations about them: $i)$ They turn out to involve only metric fluxes $\,\omega\,$ together with gauge fluxes $F_{(3)}$ and $F_{(7)}$. $ii)$ They are found within an SO(3)-invariant sector of half-maximal supergravity which, as argued in Sections~\ref{sec:invariant_sectors} and \ref{Flux_vacua_IIA_O2}, is part of the larger $\mathbb{Z}_{2}^{*} \times \mathbb{Z}_{2}^{2}$ invariant sector describing type II reductions on co-calibrated G$_{2}$ orientifolds. Therefore, all the flux vacua presented in Table~\ref{Table:flux_vacua_IIB} must also extremise the scalar potential in the class of type~IIB reductions on co-calibrated G$_{2}$ orientifolds explicitly worked out in \cite{Emelin:2021gzx}. In order to prove this, let us first present the dictionary between the eight moduli in \cite{Emelin:2021gzx} and the dilatons in the eight-chiral model of Section~\ref{sec:eight-chiral_model}. The two sets of fields are related by
\begin{equation}
\label{moduli_dictionary_1}
e^{-\frac{\sqrt{7}}{4}v} = \left(\frac{A_{4}}{\mu_{4}}\right)^{\frac{1}{2}} \, \frac{\left( A_{1} \, A_{2} \, A_{3} \, \mu_{4} \right)^{\frac{1}{4}}}{\left( \mu_{1} \, \mu_{2} \, \mu_{3} \, A_{4} \right)^{\frac{1}{8}}}
\hspace{8mm} \textrm{ and } \hspace{8mm}
e^{\phi} = \frac{1}{\left(\mu_{1} \, \mu_{2} \, \mu_{3} \,  A_{4}\right)^{\frac{1}{2}}}  \ ,
\end{equation}
together with
\begin{equation}
\label{moduli_dictionary_2}
\tilde{s}_{1} = \frac{\mathcal{X}}{\mu_{1}}
\hspace{3mm} , \hspace{3mm}
\tilde{s}_{2} = \frac{\mathcal{X}}{\mu_{2}}
\hspace{3mm} , \hspace{3mm}
\tilde{s}_{3} = \frac{\mathcal{X}}{\mu_{3}}
\hspace{5mm} , \hspace{5mm}
\tilde{s}_{4} = A_{3} \, \mathcal{Y}
\hspace{3mm} , \hspace{3mm}
\tilde{s}_{5} = A_{1} \, \mathcal{Y}
\hspace{3mm} , \hspace{3mm}
\tilde{s}_{6} = A_{2} \, \mathcal{Y} \ ,
\end{equation}
where we have introduced two moduli-dependent quantities
\begin{equation}
\label{moduli_dictionary_3}
\mathcal{X}^{7} \equiv A_{4} \left(\frac{A_{1} \,A_{2} \,A_{3} \, \left(  \mu_{1} \,\mu_{2} \,\mu_{3}\right)^{3} }{\mu_{4}\,A_{4}^{3}}\right) 
\hspace{6mm} \textrm{ and } \hspace{6mm}
\mathcal{Y}^{14} \equiv A_{4}^{2} \left(\frac{A_{4}\,\mu_{4}^{5}}{\mu_{1} \,\mu_{2} \,\mu_{3} \left(A_{1} \,A_{2} \,A_{3}\right)^{5}} \right) \ .
\end{equation}
Starting from the scalar potential in eqs~$(3.44)$-$(3.46)$ of \cite{Emelin:2021gzx}, restricting the allowed fluxes therein\footnote{Our metric fluxes $\,\omega\,$ are related to the metric fluxes $\tau$ in \cite{Emelin:2021gzx} as $\,\tau_{mn}{}^{p} = -2 \,  \omega_{mn}{}^{p}$.} to the $\textrm{SO}(3)$-invariant ones in (\ref{metric_flux_SO(3)}) and (\ref{gauge_flux_SO(3)}), and using the moduli dictionary (\ref{moduli_dictionary_1})-(\ref{moduli_dictionary_2}) supplemented with the SO(3)-invariant identifications in (\ref{RSTU_identifications}), \textit{i.e.} $A_1=A_2=A_3\equiv A\,$ and $\,\mu_1=\mu_2=\mu_3\equiv\mu$, then the scalar potential we obtain using the embedding tensor formalism precisely matches the one from \cite{Emelin:2021gzx} upon use of the quadratic constraints (\ref{QC_ww_O5}) and (\ref{QC_DF3_O5}) required by half-maximal supersymmetry. The same holds for the normalisation of the kinetic terms of the moduli fields.

Equipped with a well-established correspondence between our three-dimensional setup and the top-down construction of \cite{Emelin:2021gzx}, we can take the expressions of the internal volume $\,\textrm{vol}_{7}\,$ (in Einstein frame) and the string coupling $\,g_{s}\,$ in terms of the moduli fields derived in \cite{Emelin:2021gzx}, and use (\ref{moduli_dictionary_1})-(\ref{moduli_dictionary_2}) to convert them to our moduli fields. The result is given by
\begin{equation}
\label{vol7&gs_definitions}
\textrm{vol}_{7} = e^{-\frac{\sqrt{7}}{4}v} = \left(\frac{A_{4}}{\mu_{4}}\right)^{\frac{1}{2}} \, \frac{\left( A_{1} \, A_{2} \, A_{3} \, \mu_{4} \right)^{\frac{1}{4}}}{\left( \mu_{1} \, \mu_{2} \, \mu_{3} \, A_{4} \right)^{\frac{1}{8}}}
\hspace{8mm} \textrm{ and } \hspace{8mm}
g_{s} =  e^{\phi} = \frac{1}{\left( \mu_{1} \, \mu_{2} \, \mu_{3} \, A_{4} \right)^{\frac{1}{2}}}  \ .
\end{equation}
We will come back to these two quantities when exploring phenomenological aspects of the flux vacua presented in Table~\ref{Table:flux_vacua_IIB}.

\subsubsection*{Supersymmetry and stability of flux vacua}

Table~\ref{Table:flux_vacua_IIB} contains the complete set of Mkw$_{3}$ ($V_{0} \equiv \left\langle V \right\rangle=0$) and AdS$_{3}$ ($V_{0} \equiv \left\langle V \right\rangle < 0$) flux vacua located at the origin of moduli space, \textit{i.e.}, $\left\langle R \right\rangle=\left\langle S \right\rangle=\left\langle T \right\rangle=\left\langle U \right\rangle=i$. For each vacuum, the complete set of gravitini masses -- from which we extract the number $\,\mathcal{N}\,$ of preserved supersymmetries -- is presented in Table~\ref{Table:flux_vacua_IIB_masses_gravitini}. The spectrum of scalar masses -- from which we assess the perturbative stability of the solutions -- is summarised in Table~\ref{Table:flux_vacua_IIB_masses}.

There are three Mkw$_{3}$ vacua which are generically non-supersymmetric. However, one of them -- labelled \textbf{vac~1} in Table~\ref{Table:flux_vacua_IIB} -- features an enhancement of supersymmetry to $\,\mathcal{N}=4\,$ when the flux parameters are adjusted to $\,\kappa = \mp \xi\,$ so that four gravitini become massless (see Table~\ref{Table:flux_vacua_IIB_masses_gravitini}). The remaining Mkw$_{3}$ vacua, labelled \textbf{vac~2} and \textbf{vac~3}, have the same gravitino (see Table~\ref{Table:flux_vacua_IIB_masses_gravitini}) and scalar (see Table~\ref{Table:flux_vacua_IIB_masses}) mass spectra but, as we will see in a moment, \textbf{vac~3} is embeddable into maximal supergravity while \textbf{vac~2} is not. On the other hand, amongst the twelve AdS$_{3}$ vacua we find examples preserving $\,\mathcal{N}=0,1,3,4\,$ supersymmetry. Curiously, there are four groups of AdS$_{3}$ vacua in which each member has a different amount of preserved supersymmetry, and therefore different gravitini masses, but they all have the same spectrum of scalar fluctuations. These four groups are \,\textbf{vac~4,5}, \,\textbf{vac~6,7}, \,\textbf{vac~8,9} \,and \,\textbf{vac~12,13,14,15}\, in Tables~\ref{Table:flux_vacua_IIB}, \ref{Table:flux_vacua_IIB_masses_gravitini} and \ref{Table:flux_vacua_IIB_masses}.

In order to assess the four-dimensional perturbative stability of the non-supersymmetric vacua, we have computed the scalar mass spectrum at each flux vacuum including \textit{all} the scalars in half-maximal supergravity (see Table~\ref{Table:flux_vacua_IIB_masses}). Surprisingly (at least to us), we find that only non-negative masses appear in the scalar spectra ensuring the perturbative stability of all the non-supersymmetric vacua within half-maximal supergravity. For those non-supersymmetric vacua that admit an embedding into maximal supergravity (marked with $*$ in Table~\ref{Table:flux_vacua_IIB}), unstable modes (tachyons) violating the Breitenlohner--Freedman (BF) bound for stability in AdS$_{3}$ \cite{Breitenlohner:1982bm} could still appear in the maximal theory which has a larger $\,\textrm{E}_{8(8)}/\textrm{SO}(16)\,$ scalar geometry. Moreover, the non-supersymmetric AdS$_{3}$ vacua in Table~\ref{Table:flux_vacua_IIB} could also decay non-perturbatively or due to the presence of tachyons in the tower of KK states. Regarding the first possibility, it would be interesting to explore whether some fake supersymmetry could protect the non-supersymmetric vacua belonging to the groups \textbf{vac~4,5}, \,\textbf{vac~8,9} \,and \,\textbf{vac~12,13,14,15}. These non-supersymmetric vacua have an almost identical supersymmetric partner obtained by flipping the sign of some flux parameters. Note that this is no longer the case for \textbf{vac~10,11} which do not possess a supersymmetric partner. Regarding the higher-dimensional stability of the non-supesymmetric vacua, a precise study of KK masses is, in general, out of computational reach. However, such a KK spectrometry analysis can be carried out for those vacua in Table~\ref{Table:flux_vacua_IIB} marked simultaneously with $\,* \,\dagger\,$ (as we explain below). We leave the exploration of these stability issues for the future.

\subsubsection*{Embedding into maximal supergravity and KK spectrometry}

The embedding tensor configurations associated with vacua marked with an asterisk $\,*\,$ in Table~\ref{Table:flux_vacua_IIB} satisfy the extra QC's in (\ref{QC_N=16_extra}) and specify a consistent gauging of maximal supergravity. More concretely, the first condition in (\ref{QC_N=16_extra}) trivialises in our type IIB with O$5$ duality frame. The second condition in (\ref{QC_N=16_extra}) is non-trivial and, by virtue of the self-dual (SD) nature of $\,[\Gamma^{M_{1} \ldots M_{8}}]_{\mathcal{\dot{A} \, \dot{B}}}\,$, amounts to compute the self-dual part of the $\,\textrm{SO}(8,8)\,$ eight-form $\,\theta_{[M_{1}M_{2}M_{3}M_{4}} \,\, \theta_{M_{5}M_{6}M_{7}M_{8}]}\,$. The result is just a single condition
\begin{equation}
\label{SD_condition}
\theta_{[M_1 M_2 M_3 M_4} \, \theta_{M_5 M_6 M_7 M_8]} \Big|_{\textrm{SD}} = 0 \,\,\,\,\,\, \Longrightarrow \,\,\,\,\,\, \omega_{1} \, f_{32}  -  \omega_{2} \, f_{33} = 0 \ .
\end{equation}
A direct comparison with (\ref{unrestrited_5_branes_example}) then shows that, as expected, an embedding into maximal supergravity is possible whenever the net charge of unrestricted O$5$/D$5$ sources of the type in (\ref{O5_location}) vanishes, namely, $\,J_{\textrm{O}5/\textrm{D}5}=0$. These were precisely the sources causing the breaking of supersymmetry from maximal to half-maximal.

\begin{table}[t]
\label{Table:flux_vacua_IIB_spectrum_gravitini}
\begin{center}
\scalebox{0.9}{
\renewcommand{\arraystretch}{1.5}
\begin{tabular}{!{\vrule width 1.5pt}c!{\vrule width 1pt}c!{\vrule width 1pt}c!{\vrule width 1pt}cccccc!{\vrule width 1pt}cc!{\vrule width 1pt}ccc!{\vrule width 1pt}c!{\vrule width 1pt}c!{\vrule width 1.5pt}}
\noalign{\hrule height 1.5pt}
     ID & Gravitini spectrum  \\
\noalign{\hrule height 1pt}
     $ \textbf{vac~1} $ & $g^{-2} m^2_{3/2} = \left[\frac{(\kappa \pm \xi)^{2}}{16}\right]_{(3)} , \left[\frac{9(\kappa \pm \xi)^{2}}{16}\right]_{(1)}$  \\[2mm] 
     \hline 
     $ \textbf{vac~2} $ & \multirow{2}{*}{$g^{-2} m^2_{3/2} = \left(\frac{9 \kappa^{2}}{16}\right)_{(2)} , \left(\frac{\kappa^{2}}{16}\right)_{(6)}$}  \\ 
     \cline{1-1} 
     $ \textbf{vac~3} $ &  \\ 
\noalign{\hrule height 1pt}
     $ \textbf{vac~4} $ & $m^2_{3/2} L^{2} = 1_{(4)} , 9_{(4)}$
     \\[1mm] 
    \hline
     $ \textbf{vac~5} $ &  $m^2_{3/2} L^{2} = 9_{(4)} , 25_{(4)}$ 
     \\[1mm]
\noalign{\hrule height 1pt}
     $ \textbf{vac~6} $ & $m^2_{3/2} L^{2} = 1_{(3)} , 9_{(4)} , 25_{(1)}$  \\[1mm] 
     \hline 
     $ \textbf{vac~7} $ & $m^2_{3/2} L^{2} = 1_{(1)} , 9_{(4)} , 25_{(3)}$  \\[1mm]
\noalign{\hrule height 1pt}
     $ \textbf{vac~8} $ & $m^2_{3/2} L^{2} = 1_{(1)} , 9_{(1)} , 25_{(3)}  , 49_{(3)}$ \\[1mm] 
     \hline 
     $ \textbf{vac~9} $ &  $m^2_{3/2} L^{2} = 9_{(4)} , 25_{(4)}$ \\[1mm]
\noalign{\hrule height 1pt}
     $ \textbf{vac~10} $ & $m^2_{3/2} L^{2} = 9_{(3)} , 49_{(3)} , 81_{(1)}  , 169_{(1)}$   \\[1mm]
     \hline 
     $ \textbf{vac~11} $ & $m^2_{3/2} L^{2} = 25_{(6)} , 49_{(1)} , 225_{(1)}$   \\[1mm]
\noalign{\hrule height 1pt} 
     $ \textbf{vac~12} $ &  $m^2_{3/2} L^{2} = 1_{(4)} , 9_{(1)} , 25_{(3)}$  \\[1mm]
     \hline 
     $ \textbf{vac~13} $ & $m^2_{3/2} L^{2} = 1_{(1)} , 9_{(4)} , 49_{(3)}$   \\[1mm]
     \hline
     $ \textbf{vac~14} $ &  $m^2_{3/2} L^{2} = 9_{(7)} , 25_{(1)}$  \\[1mm]
     \hline
     $ \textbf{vac~15} $ &  $m^2_{3/2} L^{2} = 9_{(1)} , 25_{(7)}$ \\[1mm]
\noalign{\hrule height 1.5pt}
\end{tabular}}
\caption{Gravitini masses at the type~IIB with O$5$-planes vacua of Table~\ref{Table:flux_vacua_IIB}. The subscript in $n_{(s)}$ denotes the multiplicity of the mass $n$ in the spectrum. For the AdS$_{3}$ vacua we have normalised the gravitino spectrum using the AdS$_{3}$ radius $L^2=-2/V_{0}$. Note that all the normalised gravitini masses are integer-valued for the AdS$_{3}$ vacua. A ``massless" gravitino has $\,m_{3/2}^{2}=0\,$ in Mkw$_{3}$ and $\,m^2_{3/2} L^{2} =1\,$ in AdS$_{3}$. The number $\,\mathcal{N}\,$ of such gravitini correspond to the number of preserved supersymmetries at the corresponding vacuum in Table~\ref{Table:flux_vacua_IIB}.}
\label{Table:flux_vacua_IIB_masses_gravitini}
\end{center}
\end{table}

Although not necessary to have an embedding into maximal supergravity, it is also interesting to evaluate whether the anti-self-dual (ASD) part of the $\,\textrm{SO}(8,8)\,$ eight-form $\,\theta_{[M_{1}M_{2}M_{3}M_{4}} \,\, \theta_{M_{5}M_{6}M_{7}M_{8}]}\,$ vanishes. This implies two additional conditions which are purely geometrical since they only involve metric fluxes
\begin{equation}
\label{ASD_condition}
\theta_{[M_1 M_2 M_3 M_4} \, \theta_{M_5 M_6 M_7 M_8]} \Big|_{\textrm{ASD}} = 0 \,\,\,\,\,\, \Longrightarrow \,\,\,\,\,\, 
\begin{array}{r}
\omega_{2} \, \omega_{3}  = 0   \\[2mm]
\omega_{2} \, \omega_{4} - 2  \, \omega_{1} \, \omega_{5} = 0
\end{array}
 \ .
\end{equation}
The embedding tensors associated with the flux vacua marked with a dagger $\,\dagger\,$ in Table~\ref{Table:flux_vacua_IIB} satisfy (\ref{ASD_condition}). Lastly, if both (\ref{SD_condition}) and (\ref{ASD_condition}) hold, the corresponding embedding tensor satisfies
\begin{equation}
\label{KK_spectroscopy_condition}
\theta_{[M_1 M_2 M_3 M_4} \, \theta_{M_5 M_6 M_7 M_8]} = 0 \ .
\end{equation}
As explained in \cite{Hohm:2017wtr}, a generalised Scherk--Schwarz (gSS) ansatz with twist matrices that obey the so-called \textit{section constraint} of the $\textrm{O}(8,8)$ enhanced double field theory (DFT) can only reproduce gaugings whose embedding tensor satisfies (\ref{KK_spectroscopy_condition}). It would then be interesting to understand the role of the geometric conditions (\ref{ASD_condition}) in the $\textrm{O}(8,8)$-DFT context. Finally, reformulating the flux models in Table~\ref{Table:flux_vacua_IIB} which obey (\ref{KK_spectroscopy_condition}) within the $\textrm{O}(8,8)$-DFT context, namely, constructing the gSS twist matrix generating both the gauge and the metric fluxes, would allow for a precise KK spectrometry analysis of the corresponding AdS$_{3}$ vacua along the lines of \cite{Malek:2019eaz,Malek:2020yue,Eloy:2020uix,Eloy:2023acy,Eloy:2024lwn}.

\begin{table}[t]
\label{Table:flux_vacua_IIB_spectrum}
\begin{center}
\scalebox{0.9}{
\renewcommand{\arraystretch}{1.5}
\begin{tabular}{!{\vrule width 1.5pt}c!{\vrule width 1pt}c!{\vrule width 1pt}c!{\vrule width 1pt}cccccc!{\vrule width 1pt}cc!{\vrule width 1pt}ccc!{\vrule width 1pt}c!{\vrule width 1pt}c!{\vrule width 1.5pt}}
\noalign{\hrule height 1.5pt}
     ID & Scalar spectrum  \\ 
\noalign{\hrule height 1pt}
     $ \textbf{vac~1} $ & $ g^{-2} \, m^2 =  0_{(30)}, \left(\frac{\kappa^2}{16} \right)_{(9)}, \left(\frac{\kappa^2}4\right)_{(9)} ,
     \left(\frac{\xi^2}4\right)_{(9)}  , 
     \left(\frac{9\kappa^2}{16} \right)_{(1)}, \left[ \frac{\left( \kappa - 2\xi \right)^2}{16} \right]_{(3)}, \left[ \frac{\left( \kappa + 2\xi \right)^2}{16} \right]_{(3)}  $  \\[2mm] 
     \hline 
     $ \textbf{vac~2} $ & \multirow{2}{*}{$g^{-2} \, m^2 =  \left( \frac{\kappa^2}{4} \right)_{(15)} , 0_{(49)} $}  \\ 
     \cline{1-1} 
     $ \textbf{vac~3} $ &  \\ 
\noalign{\hrule height 1pt}
     $ \textbf{vac~4} $ & \multirow{2}{*}{$\begin{array}{ccl}
        m^2 L^2 &=&  8_{(19)}, \ 0_{(45)}  \\[-2mm]
         \Delta &=&  4_{(19)}, \ 2_{(45)} 
     \end{array}$}
     \\ 
    \cline{1-1} 
     $ \textbf{vac~5} $ &   
     \\
\noalign{\hrule height 1pt}
     $ \textbf{vac~6} $ & \multirow{2}{*}{$\begin{array}{ccl}
     m^2 L^2 &=&  8_{(10)}, \ 4_{(18)} , \ 0_{(36)}  \\[-2mm]
      \Delta &=&  4_{(10)}, \ (1+\sqrt{5})_{(18)} , \ 2_{(36)}
     \end{array}$}  \\ 
     \cline{1-1} 
     $ \textbf{vac~7} $ &   \\
\noalign{\hrule height 1pt}
     $ \textbf{vac~8} $ & \multirow{2}{*}{$\begin{array}{ccl}
     m^2 L^2 &=& 24_{(10)}, \ 8_{(25)}, \ 0_{(29)} \\[-2mm]
      \Delta &=& 6_{(10)}, \ 4_{(25)}, \ 2_{(29)} \end{array}$} \\ 
     \cline{1-1} 
     $ \textbf{vac~9} $ &   \\ 
\noalign{\hrule height 1pt}
     $ \textbf{vac~10} $ & $\begin{array}{ccl}
     m^2 L^2 &=&  80_{(3)}, \ 48_{(9)}, \ 24_{(4)}, \ 8_{(7)}, \ 0_{(41)} \\[-2mm]
     \Delta &=&  10_{(3)}, \ 8_{(9)}, \ 6_{(4)}, \ 4_{(7)}, \ 2_{(41)}
     \end{array}$   \\[1mm]
     \hline 
     $ \textbf{vac~11} $ & $\begin{array}{ccl} 
     m^2 L^2 &=& 48_{(15)}, \ 8_{(13)}, \ 0_{(36)} \\[-2mm]
     \Delta &=& 8_{(15)}, \ 4_{(13)}, \ 2_{(36)}
     \end{array}$   \\[1mm]
\noalign{\hrule height 1pt} 
     $ \textbf{vac~12} $ & \multirow{4}{*}{$\begin{array}{ccl}
     m^2 L^2 &=&  15_{(8)}, \ 8_{(19)}, \ 3_{(8)}, \ 0_{(29)} \\[-2mm]
     \Delta &=&  5_{(8)}, \ 4_{(19)}, \ 3_{(8)}, \ 2_{(29)}
     \end{array}$}  \\
     \cline{1-1} 
     $ \textbf{vac~13} $ &    \\
     \cline{1-1}
     $ \textbf{vac~14} $ &    \\
     \cline{1-1}
     $ \textbf{vac~15} $ &  \\
\noalign{\hrule height 1.5pt}
\end{tabular}}
\caption{Scalar masses at the type~IIB with O$5$-planes vacua of Table~\ref{Table:flux_vacua_IIB}. The subscript in $n_{(s)}$ denotes the multiplicity of the mass $n$ in the spectrum. For the AdS$_{3}$ vacua we have normalised the scalar spectrum using the AdS$_{3}$ radius $L^2=-2/V_{0}$. In addition, for each AdS$_3$ vacuum, the conformal dimension $\Delta$ of the would-be CFT$_{2}$ dual operators are also indicated. These correspond to the larger root of $m^2 L^2 = \Delta (\Delta -2)$. Note that all but \textbf{vac~6} and \textbf{vac~7} AdS$_{3}$ solutions come along with integer-valued $\Delta$'s.}
\label{Table:flux_vacua_IIB_masses}
\end{center}
\end{table}

\subsubsection{On the conformal dimension of dual operators and scale separation}
\label{Delta's&scale-separation}

In the context of type~IIA supergravity, the supersymmetric and scale-separated AdS$_4$ vacua of DGKT \cite{DeWolfe:2005uu} come along with a peculiar feature: the conformal dimensions of the would-be operators dual to the closed string sector are integer-valued \cite{Conlon:2021cjk} (see also \cite{Apers:2022zjx,Apers:2022tfm,Quirant:2022fpn,Ning:2022zqx}). More concretely, the authors of \cite{Conlon:2021cjk} showed that, independently of the details of the compactification, the conformal dimensions of the low-lying scalar primaries of the CFT$_3$ dual to supersymmetric AdS$_4$ DGKT flux vacua on general Calabi--Yau three-folds (CY$_3$'s) are always integer-valued. When considering instead non-supersymmetric configurations (see \textit{e.g.} \cite{Quirant:2022fpn}), integer-valued conformal dimensions are not guaranteed but, in light of the AdS swampland conjecture \cite{Ooguri:2016pdq}, one might argue that such vacua are unstable and so not suitable for applying the AdS/CFT correspondence. More general type~IIA flux models including not only gauge but also metric fluxes have been extensively investigated in (not so) recent years (see \textit{e.g.} \cite{Dibitetto:2011gm,Derendinger:2004jn,Camara:2005dc,Villadoro:2005cu,Dall'Agata:2005fm,Derendinger:2005ph,DeWolfe:2005uu,Andriot:2022way,Andriot:2022yyj,Grana:2006kf,Caviezel:2008ik,DallAgata:2009wsi,Dibitetto:2012ia,Dibitetto:2014sfa,Derendinger:2014wwa,Danielsson:2014ria,Andriot:2016ufg,Font:2019uva} for an incomplete list). In this case, the conformal dimensions of the would-be CFT$_3$ operators dual to the light scalars defining the corresponding AdS$_4$ vacua cease to be (only) integer-valued. Differently from the type~IIA DGKT-like vacua, such models with metric fluxes do not admit scale separation and one might then wonder whether integer-valued conformal dimensions are actually related to scale separation in four dimensions.  Similar questions on supersymmetric AdS$_{4}$ flux vacua with scale separation and integer-valued conformal dimensions of would-be dual operators have been investigated in the mirror-symmetric type~IIB context of \cite{Plauschinn:2022ztd}. When considering one-loop corrections to the complex-structure moduli space of the mirror-dual of DGKT, it was shown in \cite{Plauschinn:2022ztd} that the conformal dimensions of the would-be CFT$_3$ dual operators get also corrected and no longer take integer values. Only in the strict large complex structure regime where the aforementioned corrections are ignored, the integer-valued conformal dimensions of \cite{Conlon:2021cjk,Apers:2022zjx,Apers:2022tfm} are recovered. In addition, ref.~\cite{Plauschinn:2022ztd} also considered supersymmetric AdS$_{4}$ vacua in type IIB models including non-geometric fluxes \cite{Shelton:2005cf}. For these more exotic supersymmetric vacua, the conformal dimensions of the would-be dual operators were (numerically) determined to be non-integer-valued.

The above story about simultaneously having scale-separated AdS vacua and integer-valued conformal dimensions, as in DGKT, seems to change when moving to the context of AdS$_{3}$/CFT$_{2}$. In this context, various examples of $\mathcal{N}=1$ supersymmetric AdS$_{3}$ flux vacua have been constructed upon compactification of type~IIA \cite{Farakos:2020phe} and type~IIB \cite{Emelin:2021gzx} supergravity on (co-calibrated) $\textrm{G}_{2}$ orientifolds. For the parametrically scale-separated type IIA flux vacua of \cite{Farakos:2020phe}, it was noticed in \cite{Apers:2022zjx} that the light single-trace operators of the would-be dual CFT$_2$ have non-integer (actually irrational) conformal dimensions. On the other hand, the type~IIB flux vacua of \cite{Emelin:2021gzx} allow for integer-valued (and rational) conformal dimensions but do not exhibit a parametrically-controlled hierarchy between the size of the internal space and AdS$_{3}$. Therefore, and differently from what happens in four dimensions, the phenomenon of having integer-valued conformal dimensions seems to be in tension with having scale-separated AdS flux vacua in three dimensions. In the rest of this section we will argue that the type IIB vacua labelled as \textbf{vac~10} and \textbf{vac~11} in Table~\ref{Table:flux_vacua_IIB} -- and which are non-supersymmetric -- come along with integer-valued conformal dimensions of would-be dual operators and, at the same time, allow for parametrically-controlled scale separation.

\subsubsection*{AdS$_{3}$ vacua: on integer-valued conformal dimensions}

The complete scalar mass spectra of the type~IIB with O$5$ flux vacua of Table~\ref{Table:flux_vacua_IIB} are reported in Table~\ref{Table:flux_vacua_IIB_masses}. Quite surprisingly, and with the exception of \textbf{vac~6,7}, the rest of AdS$_3$ vacua therein exhibit all integer-valued conformal dimensions for the would-be low-lying CFT$_2$ dual operators. This observation places itself within the on-going discussion on the relation between (non-)supersymmetric AdS flux vacua and their CFT potential description, and the attempts of understanding what makes the aforementioned conformal dimensions to be integer-valued. By looking at Table~\ref{Table:flux_vacua_IIB}, one observes that the groups \,\textbf{vac~4,5} \,and\, \textbf{vac~12,13,14,15}\, possess a member with $\,\mathcal{N}=4\,$ supersymmetry. This large amount of supersymmetry might help to understand the integer nature of the $\Delta$'s at those groups of AdS$_{3}$ vacua. However, the groups \,\textbf{vac~8,9}\, and \,\textbf{vac~10,11}\, have members with smaller $\,\mathcal{N}=1\,$ or $\,\mathcal{N}=0\,$ supersymmetry, respectively, making the appearance of integer-valued $\Delta$'s more puzzling in these cases. Finally, let us also highlight that all the AdS$_{3}$ vacua with integer-valued $\Delta$'s come along with a non-zero contribution of the metric fluxes to the scalar potential (which is positive for \textbf{vac~10,11} and negative for the rest), thus corresponding to type~IIB reductions on non Ricci-flat manifolds. It would be interesting to understand the ultimate reason behind the spectral structures in Table~\ref{Table:flux_vacua_IIB_masses}.

\subsubsection*{AdS$_{3}$ vacua: on scale separation}

Let us now look at some phenomenological aspects of the set of type~IIB vacua with a single O$5$-plane listed in Table~\ref{Table:flux_vacua_IIB}. In particular, we will deal with the phenomenon of scale separation between the AdS$_{3}$ external space and the internal seven-dimensional space. To this end we will resort to the more standard picture of having the VEV's of the scalar fields as a function of the flux parameters. This amounts to undo the duality transformation (\ref{SL(2)_transformation_RSTU})-(\ref{SL(2)_transformation_gauge}) that we implemented to go to the origin of moduli space when extremising the potential. The result is summarised in Table~\ref{Table:fields_vacua_IIB}.

\begin{table}[t!]
\begin{center}
\scalebox{0.88}{
\renewcommand{\arraystretch}{1.7}
\begin{tabular}{!{\vrule width 1pt}c!{\vrule width 1pt}c!{\vrule width 1pt}c!{\vrule width 1pt}c!{\vrule width 1pt}c!{\vrule width 1pt}c!{\vrule width 1pt}c!{\vrule width 1pt}c!{\vrule width 1pt}c!{\vrule width 1pt}c!{\vrule width 1pt}}
\Xcline{3-10}{1.5pt}
\multicolumn{2}{c!{\vrule width 1pt}}{}& \multicolumn{2}{c!{\vrule width 1pt}}{$\left\langle R \, \right\rangle$} & \multicolumn{2}{c!{\vrule width 1pt}}{$\left\langle S \right\rangle$} & \multicolumn{2}{c!{\vrule width 1pt}}{$\left\langle T \right\rangle$} & \multicolumn{2}{c!{\vrule width 1pt}}{$\left\langle U \right\rangle$} \\ 
\noalign{\hrule height 1.5pt}
     ID & $V_{0} \equiv \left\langle V \right\rangle$ & $ B_4$ & $  A_4$ & $u_4$ & $\mu_4$ & $B$ & $A$ & $u$ & $\mu$ \\ 
\noalign{\hrule height 1pt}
     $ \textbf{vac~1} $ & $0$ & $  $ & $ -\frac{f_{33}}{\omega_2} A$ & $ 0 $ & $ \frac{\omega_1}{f_{32}} \mu $ & $  $ & $  $ & $ 0 $ & $  $
     \\[1mm] 
     \hline 
     $ \textbf{vac~2} $  &  $ 0$ & $  $ & $    -\frac{f_{33}}{\omega_2} A $ & $ 0 $ & $  $ & $  $ & $  $  & $ 0$ & $  $\\[1mm] 
     \hline
     $ \textbf{vac~3} $  & $ 0 $ & $  $ & $  $ & $ 0 $ & $ \frac{\omega_2}{\omega_3} A^{-1} $ & $  $ & $  $  & $ 0 $ & $  $\\[1mm] 
\noalign{\hrule height 1pt}
     $ \textbf{vac~4} $  & \multirow{2}{*}{$ -\frac{g^2}{32} \frac{\omega_6^6}{f_{31}^2 f_7^2}$} & $  $ & $ -\frac{f_{31}^2 f_7}{\omega_6^3} \mu_4 $ & \multirow{2}{*}{$ 0 $} & $  $ & \multirow{2}{*}{$ \frac{f_5}{\omega_6} $} & $ - \frac{f_7}{\omega_6} \mu $ & \multirow{2}{*}{$ 0 $} & $  $
     \\[1mm]
    \cline{1-1}\cline{4-4}\cline{8-8}
     $ \textbf{vac~5} $ &  & $  $ &  $ \frac{f_{31}^2 f_7}{\omega_6^3} \mu_4  $ &  & $  $ &  & $  \frac{f_7}{\omega_6} \mu $ &  & $  $
     \\[1mm]
\noalign{\hrule height 1pt}
     $ \textbf{vac~6} $  & \multirow{2}{*}{$ - \frac{g^2}{2} \frac{\omega_6^6}{f_{31}^2 f_7^2} $} & $  $ & \multirow{2}{*}{$  \frac{ \omega_1 f_{31}^2}{\omega_6^3} \mu^{-1}$} & \multirow{2}{*}{$ 0 $} & $ -\frac{2\omega_1}{f_7} \mu^{-1} $ & \multirow{2}{*}{$ -\frac{f_5}{\omega_6} $} & $ -\frac{f_7}{2\omega_6} \mu $ & \multirow{2}{*}{$ 0 $} & $  $
     \\[1mm] 
     \cline{1-1}\cline{6-6}\cline{8-8}
     $ \textbf{vac~7} $  &  & $  $ &  &  & $ \frac{2\omega_1}{f_7} \mu^{-1} $ &  & $ \frac{f_7}{2\omega_6} \mu $ &  & $  $ \\[1mm]
\noalign{\hrule height 1pt}
     $ \textbf{vac~8} $  & \multirow{2}{*}{$ -\frac{g^2}{32} \frac{\omega_6^6}{f_{31}^2 f_7^2}$} & $  $ & $ \frac{f_{31}^2 f_7}{\omega_6^3} \mu_4 $ & \multirow{2}{*}{$ 0 $} & $  $ & \multirow{2}{*}{$ -\frac{f_5}{\omega_6} $} & $ \frac{f_7}{\omega_6} \mu $ & \multirow{2}{*}{$ 0 $} & $  $
     \\[1mm] 
     \cline{1-1}\cline{4-4}\cline{8-8}
     $ \textbf{vac~9} $ &  & $  $ & $ -\frac{f_{31}^2 f_7}{\omega_6^3} \mu_4 $ &  & $  $ &  & $ -\frac{f_7}{\omega_6} \mu $ &  & $  $
     \\[1mm] 
\noalign{\hrule height 1pt}
     $ \textbf{vac~10} $ & \multirow{2}{*}{$-\frac{g^2}{32} \frac{\omega_3^6 f_{33}^6}{f_{31}^2 f_{32}^6 f_7^2}$} & $  $ & $ -\frac{f_{31} (f_{32}^3 f_7)^{\frac12}}{\omega_3^2 f_{33}} $ & \multirow{2}{*}{$ 0 $} & $ \frac{\omega_3 f_{33}^2}{f_{31} (f_{32}^3 f_7)^{\frac12}} $ & $  $ & $-\frac{(f_{32}^3 f_7)^{\frac12}}{\omega_3 f_{33}}$ & \multirow{2}{*}{$0$} & $ \left( \frac{f_{32}}{f_7} \right)^{\frac12} $
     \\[2mm]
     \cline{1-1}\cline{1-1}\cline{4-4}\cline{6-6}\cline{8-8}\cline{10-10}
     $ \textbf{vac~11} $ &  & $  $ & $  -\frac{f_{31} (-f_{32}^3 f_7)^{\frac12}}{\omega_3^2 f_{33}} $ &  & $ \frac{\omega_3 f_{33}^2}{f_{31} (-f_{32}^3 f_7)^{\frac12}} $ & $  $ & $ -\frac{(-f_{32}^3 f_7)^{\frac12}}{\omega_3 f_{33}} $ &  & $ \left(- \frac{f_{32}}{f_7} \right)^{\frac12} $
     \\[2mm]
     \noalign{\hrule height 1pt} 
     $ \textbf{vac~12} $ & \multirow{4}{*}{$-2g^2\frac{\omega_5^6}{f_{31}^2 f_7^2}$} & $ $ & $ \frac{f_{31} f_7}{4 \omega_3 \omega_5} \mu $ & \multirow{4}{*}{$0$} & $ \frac{2\omega_5^2}{\omega_3 f_{31}} \mu$ & $ $ & $\frac{f_7}{2\omega_5} \mu$ & \multirow{4}{*}{$0$} & $  $
     \\[1mm]
     \cline{1-1}\cline{4-4}\cline{6-6}\cline{8-8} 
     $ \textbf{vac~13} $  &  & $ $ & $-\frac{f_{31} f_7}{4 \omega_3 \omega_5} \mu $ &  & $\frac{2\omega_5^2}{\omega_3 f_{31}} \mu$ & $ $ & $-\frac{f_7}{2\omega_5} \mu$ &  & $  $
     \\[1mm]
     \cline{1-1}\cline{4-4}\cline{6-6}\cline{8-8}
     $ \textbf{vac~14} $ &  & $ $ & $-\frac{f_{31} f_7}{4 \omega_3 \omega_5} \mu $ &  & $-\frac{2\omega_5^2}{\omega_3 f_{31}} \mu$ & $ $ & $\frac{f_7}{2\omega_5} \mu $ &  & $  $
     \\[1mm]
    \cline{1-1}\cline{4-4}\cline{6-6}\cline{8-8}
     $ \textbf{vac~15} $ &  & $ $ & $ \frac{f_{31} f_7}{4 \omega_3 \omega_5} \mu $ &  & $-\frac{2\omega_5^2}{\omega_3 f_{31}} \mu$ & $ $ & $-\frac{f_7}{2\omega_5} \mu $ &  & $  $ \\[1mm]
\noalign{\hrule height 1.5pt}
\end{tabular}}
\caption{Type~IIB fields as a function of the fluxes after undoing the ``go to the origin" process. The empty boxes in the table indicate that the corresponding field is not fixed by the extremisation of the scalar potential $V$. For a given vacuum solution, the fluxes appearing in the table must be considered as independent. Those which do not appear are either zero or, otherwise, can be expressed in terms of the independent ones. In particular: $\omega_5 = -\omega_6\,$ and $\,f_{33} = -\omega_1 f_{31}/\omega_6\,$ at \,\textbf{vac~6,7}\,; $\,\omega_5 =- \omega_6\,$ at \,\textbf{vac~8,9}\,; $\,\omega_2 = - 2 \, \omega_3 f_{33}/f_{31}\,$ at \,\textbf{vac~10,11}\,; $\,\omega_4 =\omega_5^2/\omega_3\,$ and $\,\omega_6 = -2 \, \omega_5\,$ at \,\textbf{vac~12,13,14,15}. These additional relations are necessary for the QC's (\ref{QC_ww_O5})-(\ref{QC_DF3_O5}) and (\ref{QC_DF5_O5}) of half-maximal supergravity to hold.}
\label{Table:fields_vacua_IIB}
\end{center}
\end{table}

The four dilatons in the RSTU-model set the volumen of the internal space (in Einstein frame) and the string coupling in (\ref{vol7&gs_definitions}) to be
\begin{equation}
\label{vol7&gs_RSTU}
\textrm{vol}_{7} = \left(\frac{A_{4}}{\mu_{4}}\right)^{\frac{1}{2}} \, \frac{\left( A^{3} \, \mu_{4} \right)^{\frac{1}{4}}}{\left( \mu^{3} \, A_{4} \right)^{\frac{1}{8}}}
\hspace{10mm} \textrm{ and } \hspace{10mm}
g_{s} = \frac{1}{\left( \mu^{3} \, A_{4} \right)^{\frac{1}{2}}}  \ .
\end{equation}
Importantly, we observe that \textbf{vac~10} and \textbf{vac~11} in Table~\ref{Table:fields_vacua_IIB} stabilise all the dilatons in the RSTU-model so that the two quantities in (\ref{vol7&gs_RSTU}) become well-defined and fully determined by the flux parameters.\footnote{The axions $\,B\,$ and $\,B_4\,$ are potentially stabilised by other, \textit{e.g.} quantum, effects.} In particular, imposing
\begin{equation}
\label{vac_10_signs_constraints}
\pm f_{32} \, f_{7} > 0
\hspace{6mm} , \hspace{6mm} 
\omega_{3} \, f_{31} >0
\hspace{6mm} , \hspace{6mm} 
\omega_{3} \, f_{33} < 0
\hspace{6mm} \textrm{ and } \hspace{6mm} 
f_{31} \, f_{33}  < 0
\end{equation}
for the solutions to be physically acceptable, one finds
\begin{equation}
\label{L&vol7_vac_10,11}
L^2 = \frac{64}{g^2} \frac{(f_{32}^3 f_7)^2}{(\omega_3 f_{33})^6} \, f_{31}^2
\hspace{8mm} , \hspace{8mm}  
(\textrm{vol}_{7})^{8} = \mp    \frac{(\omega_{3} f_{31})^{5} (f_{32}^3 f_7)^5 f_{7}^{2}}{(\omega_3 f_{33})^{13} \, \omega_{3}^{6}}  \ ,
\end{equation}
with the upper (lower) choice of sign for the \textbf{vac~10} (\textbf{vac~11}) solution and where, as before, $\,L^2 = - 2/V_{0}\,$ denotes the AdS$_3$ radius. In addition, the string coupling is given by
\begin{equation}
\label{gs_vac_10,11}
g_s^{2} = \mp \frac{(\omega_3 f_{33}) f_{7}^{2} \, \omega_{3}^{2}}{(\omega_3 f_{31})(f_{32}^3 f_7)} \ .
\end{equation}
Therefore, \textbf{vac~10} and \textbf{vac~11} come along with five flux parameters $\,(f_{31},f_{32},f_{33},f_{7},\omega_{3})\,$ which are free and can be tuned to set $\,L$, $\,\textrm{vol}_{7}\,$ and $\,g_{s}\,$ independently at any desired value. Still there is the tadpole cancellation condition (\ref{unrestrited_5_branes_example}) which, when evaluated at \textbf{vac~10} and \textbf{vac~11}, becomes
\begin{equation}
\label{tadpole_vac_10,11}
J_{\textrm{O}5/\textrm{D}5} = 6\,  (\omega_3 f_{33}) \, \left(\frac{f_{33}}{f_{31}}\right)  > 0 \ ,
\end{equation}
requiring the presence of O5-planes in the compactification scheme contributing negatively to the scalar potential. Since the number of O5-planes in a type IIB orientifold reduction is fixed by the orientifold action, the tadpole condition (\ref{tadpole_vac_10,11}) introduces an upper bound for the flux combination entering its right-hand side. However, this fact does not spoil the (independent) parametric control over $\,L\,$, $\,\textrm{vol}_{7}\,$ and $\,g_{s}\,$ at \textbf{vac~10} and \textbf{vac~11}. This becomes more clear when inverting the relations (\ref{L&vol7_vac_10,11})-(\ref{tadpole_vac_10,11}) to obtain
%
%
%
\begin{equation}
\label{flux_from_scales}
\begin{array}{rclcrcll}
|f_{7}| &=& \dfrac{2^3 \, g_{s}^{\frac{1}{2}} \,(\textrm{vol}_{7})^{2}}{(gL)}
& \hspace{5mm} , & \hspace{5mm}
|f_{31}| &=& \dfrac{2^{10} \, 3 \,(\textrm{vol}_{7})^{4}}{g_{s} \, (gL)^{3} \, J_{\textrm{O}5/\textrm{D}5}} & , \\[6mm]
\left| \dfrac{f_{32}^{2}}{\omega_{3}}\right|^{3} &=& \dfrac{J_{\textrm{O}5/\textrm{D}5}^{2} \, (g L)}{2^{5} \, 3^{2} \, g_{s}^2}
& \hspace{5mm} , & \hspace{5mm}
|\omega_{3} \, f_{33}^{2}| &=& \dfrac{2^{9} \, (\textrm{vol}_{7})^{4}}{ g_{s} \, (gL)^{3}} & .
\end{array}
\end{equation}
As a result, arbitrary values of $\,(gL\,,\,\textrm{vol}_{7}\,,\,g_{s}\,,\,J_{\textrm{O}5/\textrm{D}5})\,$ can be obtained by appropriately tuning the four flux combinations in the l.h.s of (\ref{flux_from_scales}).\footnote{One can, for example, fix $\,\omega_{3}\,$ at wish, and then set the values of $\,(gL\,,\,\textrm{vol}_{7}\,,\,g_{s}\,,\,J_{\textrm{O}5/\textrm{D}5})\,$ in (\ref{flux_from_scales}) by adjusting the gauge fluxes $\,(f_{31},f_{32},f_{33})\,$ and $f_{7}$. Therefore, the metric flux $\,\omega_{3}\,$ still represents an arbitrary flux parameter. In summary, the non-zero metric fluxes turned on at \textbf{vac~10,11} (see Table~\ref{Table:fields_vacua_IIB}) are $\,\omega_{3}\,$ and $\,\omega_2 = - 2 \, \omega_3 f_{33}/f_{31}$. With only these two metric fluxes activated, the isometry algebra (\ref{algebra_isometry_SO(3)}) simplifies to
\begin{equation}
\label{algebra_isometry_vac10,11}
[X_7, X_b] = \omega_{2} \, \delta_{bk}\, X_k
 \hspace{5mm} ,  \hspace{5mm} 
[X_i, X_7] = \omega_{3} \, \delta_{ic} \, X_c \ ,
\end{equation}
which corresponds to a $2$-step solvable algebra with degenerate Killing--Cartan metric.} In this sense, \textbf{vac~10} and \textbf{vac~11} come along with enough flux parameters to achieve a weakly-coupled ($\,g_{s} \ll 1\,$) and scale-separated regime while keeping $\,J_{\textrm{O}5/\textrm{D}5} =\textrm{fixed}$. It is worth stressing that we define \textit{scale separation} as the existence of a parametric hierarchy  between the AdS$_{3}$ radius and the characteristic size of the internal space, \textit{i.e.} $\,m_P^{-1} (g L) \gg \left( \textrm{vol}_7^s \right)^{\frac17}$, where $\,\textrm{vol}_7^s = g_s^{\frac74} \textrm{vol}_7\,$ is the seven-dimensional internal volume in the string frame and $\,m_P = g_s^{-2} \, \textrm{vol}_7^s\,$ denotes the 3D Planck constant (in string units) \cite{Farakos:2020phe,Petrini:2013ika}. To provide an example, let us parameterise the fluxes as
\begin{equation}
f_{7}= 2 \, \alpha^{2} \, \beta \, N^{3}
\hspace{3mm} , \hspace{3mm}
f_{31} = 2 \, \alpha \, \beta^{2}
\hspace{3mm} , \hspace{3mm}
f_{32} = \alpha \, \beta \, N^{5}
\hspace{3mm} , \hspace{3mm}
f_{33} = - \alpha^{2}
\hspace{3mm} , \hspace{3mm}
\omega_{2} = \alpha
\hspace{3mm} , \hspace{3mm}
\omega_{3} = \beta^{2} \ ,
\end{equation}
with $\,\alpha,\beta,N \in \mathbb{N}\,$ so that the fluxes are integer-valued, and with $\,N\,$ playing the role of a scaling parameter which is to be taken arbitrarily large. This flux configuration realises the scale-separated and weakly-coupled regime as
\begin{equation}
g_{s} = N^{-6}
\hspace{3mm} , \hspace{3mm}
\textrm{vol}_{7}^s = 2^{\frac{3}{2}} \, \alpha \, \beta^{\frac{1}{2}} \, N^{\frac32}
\hspace{3mm} , \hspace{3mm}
m_P^{-1} (g L)  = 2^{\frac72} \alpha^{-1} \beta^{-\frac12}\, N^{\frac92}
\hspace{3mm} , \hspace{3mm}
J_{\textrm{O}5/\textrm{D}5} = 3 \, \alpha^{3} \ .
\end{equation}
If $\alpha=1$ then the compactification scheme will include the number $J_{\textrm{O}5}$ of O5-planes located at the fixed points of the orientifold involution, but also a number $J_{\textrm{D}5}$ of D5-branes sitting on top of them such that $\,J_{\textrm{O}5/\textrm{D}5} \equiv J_{\textrm{O}5}-J_{\textrm{D}5} = 3$.

\section{Summary and future directions}
\label{sec:Conclusions}

In this work we have identified which $\,\mathcal{N}=8\,$ (half-maximal) supergravities in three dimensions arise from type~II orientifold reductions including background gauge and metric fluxes, O$p$-planes and D$p$-branes. Demanding half-maximal supersymmetry only allowed for a single type of coincident O$p$-planes/D$p$-branes, \textit{e.g.} type~IIA with O$2$/D$2$ or type~IIB with O$5$/D$5$. Relying on group-theoretical arguments, we presented the dictionary between M/string theory fluxes and the embedding tensor of the three-dimensional half-maximal supergravity that results from the compactification. This was done for all the possible type~IIA and type~IIB duality frames (see Section~\ref{sec:Type_II_orientifolds}). Formulating the various type~II flux models within the embedding tensor formalism \cite{Nicolai:2001ac,deWit:2003ja} allowed us to initiate a systematic study of their vacuum structure. This also gave us the possibility to compute the complete scalar mass spectrum within half-maximal supergravity and to establish connections with the Swampland Program, for example, with the AdS conjecture \cite{Ooguri:2016pdq}.

We then focused on the particular type~IIB with O$5$ duality frame and considered its $\textrm{SO}(3)$ invariant sector, deriving what we dubbed the RSTU-model. Making a combined use of the method put forward in \cite{Dibitetto:2011gm} and of algebraic geometry techniques, we investigated the landscape of the RSTU-model. We found a surprisingly rich structure of vacua consisting of Mkw$_3$ and AdS$_3$ solutions with and without supersymmetry (see Table~\ref{Table:flux_vacua_IIB}). In all the vacua, the masses of all the scalar fields in the coset geometry (\ref{scalar_geometry_N=8}) of half-maximal supergravity turned out to be non-negative definite, establishing the perturbative stability of all the vacua (irrespective of supersymmetry) within half-maximal supergravity. Some of these vacua -- in particular the ones marked with $\,*\,$ in Table~\ref{Table:flux_vacua_IIB} -- can be embedded into maximal supergravity since the corresponding embedding tensors satisfy the self-duality condition (\ref{SD_condition}). In other words, they are vacua of a maximal $\,\mathcal{N}=16\,$ supergravity in three dimensions. In the absence of supersymmetry, it would be important to compute the masses of all the scalar fields in the $\textrm{E}_{8(8)}/\textrm{SO}(16)$ scalar geometry of the maximal theory in order to assess the stability of such solutions. In any case, for non-supersymmetric vacua like \textbf{vac~10} and \textbf{vac~11} in Table~\ref{Table:flux_vacua_IIB} which cannot be embedded into maximal supergravity due to the need of O$5$/D$5$ sources to cancel the flux-induced tadpole (\ref{tadpole_vac_10,11}), we have established their perturbative stability within half-maximal supergravity. It would be interesting to investigate mechanisms for such vacua to decay in light of the AdS conjecture \cite{Ooguri:2016pdq}. Besides being perturbatively stable within half-maximal supergravity, \textbf{vac~10} and \textbf{vac~11} display two interesting properties. Firstly, they provide examples of AdS vacua with scale separation between AdS$_3$ and the internal space within the regime of validity of the supergravity approximation and in a classical type IIB context. Secondly, and despite being non-supersymmetric AdS$_{3}$ vacua involving non Ricci-flat internal spaces, all the would-be dual CFT$_{2}$ operators turned out to have integer-valued conformal dimensions. These results motivate a systematic exploration of the landscape of vacua of the various half-maximal flux models presented in Section~\ref{sec:Type_II_orientifolds}. It would also be interesting to have a (real) superpotential formulation thereof along the lines of \cite{deIaOssa:2019cci,VanHemelryck:2022ynr} and \cite{Farakos:2020phe,Emelin:2021gzx}.

There are some directions we would like to investigate in the future. For example, it would be interesting to explore lower-supersymmetric scenarios arising from the simultaneous presence of O$p$-planes (and D$p$-branes) of various types, \textit{e.g.}, type~IIA with three types of O$4$-planes breaking supersymmetry down to $\,\mathcal{N}=2$, or type~IIA with O$2$/O$6$-planes to capture the $\,\mathcal{N}=1\,$ models of \cite{Farakos:2020phe,Farakos:2023nms,Farakos:2023wps}. Having a superpotential formulation of the simple flux models presented in this work would provide a good starting point to construct less supersymmetric models. These less supersymmetric models would have new terms in the superpotential enriching the dynamics of the moduli fields and, perhaps, opening the possibility to realise the phenomenon of fixed-point annihilation \cite{Kaplan:2009kr,Faedo:2019nxw} whose top-down realisation is still lacking.\footnote{The phenomenon of fixed-point annihilation is holographically described by the merging (and subsequent disappearance) of two AdS vacua in field space as a consequence of the tuning of a parameter in the superpotential. This phenomenon does not take place in the RSTU-model of Section~\ref{Section:RSTU_model}, as it can be deduced by looking at Table~\ref{Table:fields_vacua_IIB}.} In addition to this, and for the flux models satisfying the algebraic condition (\ref{KK_spectroscopy_condition}), it would be interesting to reformulate them in an $\textrm{O}(8,8)$-DFT context. This would open up the possibility to carry out a systematic study of KK modes at the corresponding AdS$_{3}$ flux vacua along the lines of \cite{Eloy:2020uix,Eloy:2023acy,Eloy:2024lwn}. Tracking the masses of the KK modes while moving the AdS$_{3}$ vacua in field space by tuning the flux parameters would allow us to perform quantitative tests of the distance conjecture of \cite{Ooguri:2006in}. Another interesting question has to do with the smearing of the sources in the AdS$_{3}$ flux vacua we have presented. We are assuming that smeared O$p$-planes are a consistent ingredient of flux compactifications and that there exists a full-fledged solution in string theory that is described by such an approximation at low energies. Building upon the results in, \textit{e.g.}, \cite{Junghans:2020acz,Junghans:2023yue,Emelin:2022cac,Emelin:2024vug}, it would be interesting to take into account the backreaction of such O$p$-planes (order by order in $g_s$) and see whether it can always be considered as a small perturbation around the smeared solution. Due to the non-linear character of the backreaction near the O$p$-planes, a conclusive answer can not be given yet. Finally, following a more conventional top-down approach, it would be interesting to further investigate higher-dimensional aspects -- like compactness of the internal space -- at the three-dimensional flux vacua we have presented in this work. It would also be interesting to investigate a possible connection between these three-dimensional flux vacua and the extensive list of four-dimensional solutions already existing in the literature (see \textit{e.g.} \cite{Dibitetto:2011gm,Derendinger:2004jn,Camara:2005dc,Villadoro:2005cu,Dall'Agata:2005fm,Derendinger:2005ph,DeWolfe:2005uu,Andriot:2022way,Andriot:2022yyj,Grana:2006kf,Caviezel:2008ik,DallAgata:2009wsi,Dibitetto:2012ia,Dibitetto:2014sfa,Derendinger:2014wwa,Danielsson:2014ria,Andriot:2016ufg,Font:2019uva}). We hope to come back to these and other related issues in the near future.

\section*{Acknowledgements}

We thank Fotis Farakos and Gabriel Larios for interesting conversations. We also gratefully acknowledge email correspondence and discussions with George Tringas, Thomas Van Riet and Vincent Van Hemelryck. This work is supported by the Spanish national grant MCIU-22-PID2021-123021NB-I00.

\appendix

\section{Spinorial representations and $\Gamma$-matrices of $\textrm{SO}(8)_{\textrm{R}}$}
\label{App:Gamma_conventions}

In this appendix we collect our conventions for the set of $\Gamma$-matrices of $\textrm{SO}(8)_{\textrm{R}}$ (we adopt conventions in Appendix~A of \cite{Guarino:2013gsa}). 
In the Majorana representation they are of the form
\begin{equation}
\left[\Gamma^{\mathcal{I}}\right]_{\mu}{}^{\nu} = 
\left(\begin{matrix}
0 & \left[\gamma^\mathcal{I}\right]_{\cal{A} \dot{\cal B}} \\
\left[\overline{\gamma}^{\mathcal{I}}\right]^{\dot{\cal A} \cal{B}} & 0
\end{matrix}\right)  \ ,
\end{equation}
where $\mathcal{I}=1,\ldots,8$ is the vector index of $\textrm{SO}(8)_{\textrm{R}}$ and $\mathcal{A} \,(\dot{\mathcal{A}})=1,\ldots,8$ is the left-handed (right-handed) Majorana-Weyl index. Majorana indices are lowered/raised using the invariant matrix $\mathcal{C} \equiv \mathcal{C}_{\mu\nu}$ and its inverse $\mathcal{C}^{-1} \equiv \mathcal{C}^{\mu\nu}$, with
\begin{equation}
\mathcal{C}_{\mu\nu} = \left(\begin{matrix}
\mathbb{I}_{{\cal A}{\cal B}} & 0 \\
0 &  \mathbb{I}_{\dot{\cal A}\dot{\cal B}}
\end{matrix}\right) \ .
\end{equation}
This implies that left-handed and right-handed Majorana-Weyl indices are trivially raised and lowered. Lastly, using the above set of $\textrm{SO}(8)_{\textrm{R}}$ invariant matrices, $\Gamma^{(p)}$-forms can be introduced as
\begin{equation}
\label{Gamma-forms}
\left[ \Gamma^{\mathcal{I}_{1} \cdots \mathcal{I}_{p}} \right] = \Gamma^{[\mathcal{I}_{1}} \cdots \Gamma^{\mathcal{I}_{p}]} \, \mathcal{C} \ .
\end{equation}

All the results in the main text have been derived using the following set of $\left[\gamma^\mathcal{I}\right]_{\cal{A} \dot{\cal B}}$ matrices
\begin{equation}
\begin{array}{lcll}
\gamma^1 = i\sigma_2 \otimes i\sigma_2 \otimes i\sigma_2 & \hspace{5mm} ,  & \hspace{5mm} \gamma^2 = \mathbb{I}_2 \otimes \sigma_3 \otimes i\sigma_2 & , \\[2mm]
\gamma^3 = \mathbb{I}_2 \otimes \sigma_1 \otimes i\sigma_2 & \hspace{5mm} ,  &  \hspace{5mm} \gamma^4 = \sigma_1 \otimes i\sigma_2 \otimes \mathbb{I}_2 & , \\[2mm]
\gamma^5 = \sigma_3 \otimes i\sigma_2 \otimes \mathbb{I}_2 & \hspace{5mm} ,  &  \hspace{5mm} \gamma^6 =  i\sigma_2 \otimes \mathbb{I}_2 \otimes \sigma_1 & , \\[2mm]
\gamma^7 = i\sigma_2 \otimes \mathbb{I}_2 \otimes \sigma_3 & \hspace{5mm} ,  &  \hspace{5mm} \gamma^8 = \mathbb{I}_2 \otimes \mathbb{I}_2 \otimes \mathbb{I}_2 & ,
\end{array}
\end{equation}
in terms of the Pauli matrices $\sigma_{1,2,3}$. It then follows that $\overline{\gamma}^{\mathcal{I}} = \left(\gamma^{\mathcal{I}}\right)^T$ and the following Clifford algebra holds
\begin{equation}
\left[\gamma^{\mathcal{I}}\right]_{{\cal A}\dot{\cal A}} \, \left[\overline{\gamma}^{\mathcal{J}}\right]^{\dot{\cal A}{\cal B}} + \left[\gamma^{\mathcal{J}}\right]_{{\cal A}\dot{\cal A}} \, \left[\overline{\gamma}^{\mathcal{I}}\right]^{\dot{\cal A}{\cal B}} = 2 \, \delta^{\mathcal{I} \mathcal{J}} \, \delta_{{\cal A}}^{{\cal B}} \ .
\end{equation}
Vector indices of $\textrm{SO}(8)_{\textrm{R}}$ are then trivially raised/lowered too. Finally, the set of $\textrm{SO}(8)_{\textrm{R}}$ invariant $\gamma^{(p)}$-forms entering (\ref{Fermion-Shifts}) are straightforwardly extracted from (\ref{Gamma-forms}). They are of the form
\begin{equation}
\begin{array}{rlll}
p=4 :  & \hspace{5mm}  \left[\gamma^{\mathcal{IJKL}}\right]_{{\cal A}{\cal B}} =  \gamma^{\mathcal{[I}} \bar{\gamma}^{\mathcal{J}} \gamma^{\mathcal{K}} \bar{\gamma}^{\mathcal{L]}}
&,&
\left[\gamma^{\mathcal{IJKL}}\right]_{{\dot{\cal A}}{\dot{\cal B}}} =  \bar{\gamma}^{\mathcal{[I}} \gamma^{\mathcal{J}} \bar{\gamma}^{\mathcal{K}} \gamma^{\mathcal{L]}}  \ , \\[3mm] 
p=3 : & \hspace{5mm} \left[ \gamma^{\mathcal{IJK}} \right]_{{\cal A}{\dot{\cal B}}} = \gamma^{\mathcal{[I}} \bar{\gamma}^{\mathcal{J}} \gamma^{\mathcal{K}]} \ , \\[3mm]
p=2 : & \hspace{5mm} \left[ \gamma^{\mathcal{IJ}} \right]_{\dot{\cal A}{\dot{\cal B}}} = \bar{\gamma}^{[\mathcal{I}} \gamma^{\mathcal{J}]} \ .
\end{array}
\end{equation}

\section{Gauge algebra for the type~IIB with O$5$ vacua in Table~\ref{Table:flux_vacua_IIB}}
\label{appendix:IIB_O5_flux_algebra}

The embedding tensors giving rise to the type~IIB with O$5$ flux vacua in Table~\ref{Table:flux_vacua_IIB} generically activate $\,73\,$ (out of the $120$) generators in the gauge algebra (\ref{brackets_N=8}). These are generators of the form 
\begin{equation}
X^{AB} 
\hspace{5mm} , \hspace{5mm}
X_{m}{}^{i}
\hspace{5mm} , \hspace{5mm}
X_{\hat{a}}{}^{\hat{b}}
\hspace{5mm} \textrm{ and } \hspace{5mm}
X^{8}{}_{A} \ ,
\end{equation}
where $\,A=(m,8)\,$ and $\,m=(i,\hat{a})$. The gauge algebra they span is described by the following non-trivial commutation relations induced by the fluxes
\begin{equation}
\begin{array}{rcll}
\left[ X^{ij} , X^{k {\hat c}} \right]  & = &  - 2 \, \tilde{\omega}_{\hat{c}\hat{d}}{}{}^{ij} \, X^{k}{}_{\hat{d}}  &  , \\[2mm]
\left[ X^{ij} , X^{k8} \right]  & = & - 2 \, \omega_{ij}{}^{h} \, X^{kh} - 2 \tilde{F}_7 \epsilon^{ijh}  X^{k}{}_{h}  &  , \\[2mm]
\left[ X^{ij} , X^{{\hat{c}}{\hat{d}}} \right]  & = &  2 \, \tilde{\omega}_{\hat{c}\hat{e}}{}{}^{ij} \, X^{{\hat d}}{}_{\hat e} - (c \leftrightarrow d) & , \\[2mm]
\left[ X^{i\hat{a}} , X^{j{\hat b}} \right]  & = & -2 \, \tilde{\omega}_{\hat{a}\hat{c}}{}{}^{ij}  X^{\hat{b}}{}_{\hat{c}}-2 \, \tilde{\omega}_{\hat{a}\hat{b}}{}{}^{ih}  X^{j}{}_{h}  & , \\[2mm]
\left[ X^{ij} , X^{{\hat c}8} \right]  & = & -2 \, \omega_{ij}{}{}^{k} X^{{\hat c}k} + 2 \, \tilde{\omega}_{\hat{c}\hat{d}}{}{}^{ij}  X^{8}{}_{\hat{d}}  & , \\[2mm]
\left[ X^{i{\hat a}} , X^{k8} \right]  & = & - 2 \, \omega_{{\hat b} i}{}^{\hat a} X^{k}{}_{\hat b}  - 2\, \tilde{\omega}_{\hat{a}\hat{b}}{}{}^{ik}  X^{8}{}_{\hat b} + 2  \tilde{F}_{(3)i\hat{a}\hat{b}} X^{k}{}_{\hat b} & , \\[2mm]
\left[ X^{i\hat{a}}, X^{{\hat{c}} 8} \right] & = & -2 \, \omega_{{\hat b} i}{}^{\hat a} X^{\hat{c}\hat{b}} + 2\, \tilde{\omega}_{\hat{a}\hat{c}}{}{}^{ij}  X^{8}{}_{j}  + 2  \tilde{F}_{(3)i\hat{a}\hat{d}} X^{\hat c}{}_{\hat d} & , \\[2mm]
\left[ X^{i8}, X^{k 8} \right] & = & - 2 \, \omega_{ik}{}^{h} X^{8h} + 2  \tilde{F}_0 X^{ik} - 2 \tilde{F}_7 \epsilon^{ikh} X^{8}{}_{h} & , \\[2mm]
\left[ X^{i8}, X^{\hat{c}\hat{d}} \right] & = & -2 \, \omega_{{\hat b} i}{}^{\hat c} X^{\hat{d}\hat{b}} + 2\tilde{F}_{(3)i\hat{c}\hat{b}} X^{\hat d}{}_{\hat b} - (c \leftrightarrow d) & , \\[2mm]
\left[ X^{i8}, X^{\hat{c}8} \right] & = & -2 \, \omega_{{\hat b} i}{}^{\hat c} X^{8\hat{b}} - 2 \tilde{F}_0 X^{\hat{c}i} + 2 \tilde{F}_{(3)i\hat{c}\hat{b}} X^{8}{}_{\hat b} & , \\[2mm]
\left[ X^{{\hat a}8}, X^{\hat{c}8} \right] & = &  2 \tilde{F}_0 X^{{}\hat{a}\hat{c}} + 2 \tilde{F}_{(3)i\hat{a}\hat{c}} X^{8}{}_{i} & 
\end{array}
\end{equation}
and
\begin{equation}
\begin{array}{rcll}
\left[ X_{i}{}^{j} , X^{k 8} \right]  & = &  - 2 \, \omega_{jh}{}{}^{i} \, X^{k}{}_{h}   &  , \\[2mm]
\left[ X_{i}{}^{8} , X^{kh} \right]  & = & - 2 \, \omega_{kj}{}^{i} \, X^{h}{}_{j}  - (k \leftrightarrow h) &  , \\[2mm]
\left[ X_{i}{}^{8} , X^{k8} \right]  & = &  - 2 \, \omega_{kh}{}{}^{i} \, X^{8}{}_{h} - 2 \tilde{F}_0 X^{k}{}_{i}  & , \\[2mm] 
\left[ X_{\hat{a}}{}^{i} , X^{k8} \right]  & = & - 2 \, \omega_{\hat{a}i}{}{}^{\hat b} X^{k}{}_{\hat b}  & , \\[2mm]
\left[ X_{{\hat a}}{}^{i} , X^{\hat{c}8} \right]  & = & - 2 \, \omega_{{\hat a} i}{}^{\hat d} X^{\hat c}{}_{\hat d}  & , \\[2mm]
\left[ X_{\hat{a}}{}^{\hat b}, X^{k 8} \right] & = &  2 \, \omega_{{\hat a} j}{}^{\hat b} X^{k}{}_{j}  & , \\[2mm]
\left[ X_{\hat a}{}^{8}, X^{k h} \right] & = & - 2 \, \omega_{\hat{a}k}{}^{\hat b} X^{h}{}_{\hat b} - (k \leftrightarrow h) & , \\[2mm]
\left[ X_{\hat a}{}^{8}, X^{k\hat{c}} \right] & = &- 2 \, \omega_{{\hat a} k}{}^{\hat d} X^{\hat{c}}{}_{\hat d} - 2 \, \omega_{{\hat a} h}{}^{\hat c} X^{k}{}_{h} & , \\[2mm]
\left[ X_{\hat a}{}^{8}, X^{k8} \right] & = & -2 \, \omega_{{\hat a} k}{}^{\hat b} X^{8}{}_{\hat{b}} - 2 \tilde{F}_0 X^{k}{}_{\hat a}  & , \\[2mm]
\left[ X_{{\hat a}}{}^{8}, X^{\hat{c}8} \right] & = & 2 \, \omega_{{\hat a} i}{}^{\hat c} X^{8}{}_{i} - 2 \tilde{F}_0 X^{\hat{c}}{}_{\hat a} + 2 \delta_{\hat a}{}^{\hat c} \tilde{F}_0 X^{8}{}_{8} & , \\[2mm]
\end{array}
\end{equation}
where we have introduced the flux-dependent quantities
\begin{equation}
\begin{array}{lcll}
\tilde{\omega}_{\hat{a}\hat{b}}{}{}^{ij} \equiv \frac{1}{2!}\epsilon^{\hat{a}\hat{b}\hat{c}\hat{d}} \, \epsilon^{ijk} \, \omega_{\hat{c}\hat{d}}{}{}^{k}
& \hspace{5mm} , & \hspace{5mm}
\tilde{F}_7 \equiv \frac{1}{3!\,4!}\epsilon^{\hat{a}\hat{b}\hat{c}\hat{d}} \epsilon^{ijk}  F_{\hat{a}\hat{b}\hat{c}\hat{d}ijk} & , \\[4mm]
\tilde{F}_{(3)i\hat{a}\hat{b}} \equiv \frac{1}{2!} \epsilon^{\hat{a}\hat{b}\hat{c}\hat{d}} F_{i\hat{c}\hat{d}}
& \hspace{5mm} , & \hspace{5mm}
\tilde{F}_0 \equiv \frac{1}{3!} \epsilon^{ijk} F_{ijk}  & .
\end{array}
\end{equation}


\bibliographystyle{JHEP}
\bibliography{references}

\end{document}